# Antiferroelectric Oxide Thin-Films: Fundamentals, Properties, and Applications

Yangyang Si[a,1], Tianfu Zhang[a,1], Chenhan Liu[b,1], Sujit Das,[c*] Bin Xu[d], Roman G Burkovsky[e,*], Xian-Kui Wei[f,g,*] Zuhuang Chen[a,*]

[a] *School of Materials Science and Engineering, Harbin Institute of Technology, Shenzhen, 518055, P.R. China*

[b] *Micro- and Nano-scale Thermal Measurement and Thermal Management Laboratory, School of Energy and Mechanical Engineering, Nanjing Normal University, Nanjing, 210023 P. R. China*

[c] *Material Research Centre, Indian Institute of Science, Bangalore, 560012, India*

[d] *Jiangsu Key Laboratory of Thin Films, School of Physical Science and Technology, Soochow University, Suzhou, 215006, China*

[e] *Peter the Great Saint-Petersburg Polytechnic University, 29 Politekhnicheskaya, 195251, St. Petersburg, Russia*

[f] *College of Chemistry and Chemical Engineering, Xiamen University, Xiamen 361005, P.R. China*

[g] *Innovation Laboratory for Sciences and Technologies of Energy Materials of Fujian Province (IKKEM), Xiamen 361005, China*

[1] Equal contribution.

[*] Corresponding authors

*Email address:* zuhuang@hit.edu.cn (Z. H. Chen), sujitdas@iisc.ac.in (S. Das), roman.burkovsky@gmail.com (R. G. Burkovsky), xkwei@xmu.edu.cn (X.-K. Wei),

**Abstract**

Antiferroelectrics have received blooming interests because of a wide range of potential applications in energy storage, solid-state cooling, thermal switch, transducer, actuation, and memory devices. Many of those applications are the most prospective in thin film form. The antiferroelectric ordering in thin films is highly sensitive to a rich set of factors, such as lattice strain, film thickness, surface and interface effects as well as film stoichiometry. To unlock the full potential of these materials and design high-quality thin films for functional devices, a comprehensive and systematic understanding of their behavior is essential. In conjunction with the necessary fundamental background of antiferroelectrics, we review recent progress on various antiferroelectric oxide thin films, the key parameters that trigger their phase transition and the device applications that rely on the robust responses to electric, thermal, and optical stimuli. Current challenges and future perspectives highlight new and emerging research directions in this field. It is hoped that this review can boost the development of antiferroelectric thin-film materials and device design, stimulating more researchers to explore the unknowns together.

## 1. Introduction

Ferroic materials are among the most attractive research topics as they offer the intriguing ability to manipulate their physical properties such as polarization, magnetization, and strain states using external fields [1-3]. Among these systems, antiferroelectrics (AFEs) have been underexplored for a long time despite the intriguing properties and rich structural transition phenomena, because their practical benefits were less accessible than those in ferroelectrics (FEs) [4] and their functioning mechanisms were perceived much less transparent than those in antiferromagnets [5]. With the slowing down on the development of conventional microelectronics, the merits of AFE thin films gradually appear in the public vision for their exceptional performance in various electronic devices. This includes electrocaloric [6] and electrostrain devices [7], energy storage capacitors [8], memories [9], thermal switches [10], and photovoltaics [11], which promote the prosperity of present AFE thin films research. As a result, AFE thin films have become an important branch of ferroic materials due to their diverse range of properties that are tunable between different states.

At the current state of the art, the understanding of the bulk AFEs is now well systematized and the standing questions, if not yet answered, are now identified solidly [12]. However, this is not the case in AFE thin films, which strongly differ from the bulk materials [13] and whose properties and functional behaviors are more of a mystery [14-16]. One typical example is that associated with the miniaturization of the microelectronic devices, the tendency of size effect from bulk to thin film makes the structural order parameters even more complex in the AFE, which is manifested by the deviation of cationic displacement from the antiparallel shifts as documented in prototypical AFE $PbZrO_3$ [17]. Pertinent to structure diversities in AFE thin films, this gives rise to abundant exotic phenomena, including the unusual structural phase transition [18] and complex ferrielectric-like behavior with non-zero remnant polarization [16, 19]. A spring of recent experimental and theoretical findings have produced a plethora of diverse and occasionally conflicting accounts, which clearly demand a unified description to navigate the field. This is even more significant on the application aspect, which is not yet clear about which of those are going to flourish in the near future. This highlights the necessity of updating our conventional understanding of AFEs through carrying out systematic thin-film studies. In addition to this, thin-film epitaxy also provides additional strategies to effectively tune the structural and physical properties in AFEs [20, 21]. Moreover, for many device applications, AFE is preferred in thin film form. However, most previous articles, including reviews, mainly focus on bulk ceramics [22-24], and a comprehensive guideline is still missing on establishing a detailed structure-property relationship and deterministic realization of desirable physical properties in thin-film AFEs, which strongly hinders their practical applications.

This review aims to address the exiting gaps by providing a comprehensive understanding of AFE thin films from the fundamental physics, engineering strategies, and application perspectives. We start with an introduction of the definition and research history, fundamental mechanisms, and characteristics of AFEs. Subsequently, we delve into the engineering strategies employed in thin films, such as surface and interface effects, chemistry and doping modifications, domain and octahedral tilt engineering, strain and orientation effects, and emphasize how they work on AFE thin films. The materials include the most widely explored lead-based and lead-free perovskite crystals, fluorite AFEs with emphasis either on their nanotechnology integration or on their otherwise interesting properties that were not broadly covered in other reviews. In terms of applications, our goal is to provide a clear and accessible explanation of the core concept, followed by a discussion on how that concept can be effectively implemented in thin films. We also highlight the key advantages and challenges associated with the application of these ideas in thin films. Finally, we conclude the review by outlining the key scientific challenges and opportunities in the field. We hope that reexamining AFE materials from the thin film perspective will update our conventional understanding on relevant physics in the bulk and deliver a comprehensive insight to future study of AFEs, both for fundamental research and practical applications.

### 1.1. Definition of antiferroelectricity

Antiferroelectricity was first proposed by Kittel in 1951, which is comparable to the concept of antiferromagnetism [25]. According to the respective model, AFEs consist of sublattices of ions with antiparallel spontaneous polarization (*i.e.*, having dipoles with "↑↓↑↓" arrangement). Despite that, none of the known AFE oxides is found to conform to this collinear two-sublattice model, the model highlights the essence of individual sub-unit cells for the definition of antiferroelectricity. A modern definition of antiferroelectricity is not totally unambiguous, but usually one expects at least two attributes in an AFE material. The first attribute of antiferroelectricity is the ability to be switched from a nonpolar state to a polar state by electric field below its own breakdown strength, giving rise to the emblematic double hysteresis loop. The second attribute is the presence of dipoles that are aligned in antiparallel, leading to unit-cell doubling or structural modulation, and thus averagely compensate each other. It is important to note that there are cases where materials exhibit the first attribute without the second, such as FEs with pinned antiparallel domain configurations [26], or a FE at the high-temperature paraelectric (PE) state but with antiparallel polar nanoregions [27]. Both examples are typically not classified as AFE materials. However, there is an exception. $(Hf_{1-x}Zr_x)O_2$ is commonly seen as an AFE when nonpolar-polar switching can be induced by an external electric field, but antiparallel dipolar displacements in the nonpolar phase have not yet been clearly identified [28, 29]. The second attribute can also be present without the presence of the first one [30], which is also not considered as an AFE. In many well-established AFE samples, the suppressed FE soft mode and related high-temperature dielectric anomaly credit as the vital ingredient of antiferroelectricity [31], this includes classic perovskite AFE oxides such as $PbZrO_3$, $PbHfO_3$, $NaNbO_3$, and $AgNbO_3$.

### 1.2. Timeline of notable events

Key advancements in the field of AFEs are presented in Fig. 1, with a focus on their chronological development over time. This figure provides a comprehensive overview of the significant milestones and breakthroughs in AFE research, highlighting the key findings and discoveries that have shaped our understanding of the materials [7, 9-11, 25, 30-46]. The first AFE, $PbZrO_3$, is discovered in the early 1950s [47]. Analogous to FEs, $PbZrO_3$ also undergoes a structural phase transition to a high-symmetry phase above a transition temperature ($T_A$), and a sharp dielectric peak can be observed around $T_A \approx 230$ °C. Therefore, $PbZrO_3$ was considered as a $BaTiO_3$-type FE material at the very beginning [48]. However, anomalous superstructure and physical properties revealed by X-ray diffraction, volume expansion, field-dependent dielectric constant and specific heat suggest that $PbZrO_3$ is not FE, but probably the first AFE material ever known [49, 50]. Soon after that, a series of AFE systems such as $PbHfO_3$ [51], $AgNbO_3$ [52], H-bonded ADP [53], and liquid crystals [54] were reported one after another. Because externally applied electric field can induce an AFE-to-FE phase transition, devices such as high energy-storage capacitors [34] and large displacement transducers [55] were demonstrated in AFE-based materials. At the same time, many endeavors were made to understand the AFE phase transition from both the theoretical and experimental perspectives. By examining changes of dielectric constant of $PbZrO_3$ in the pressure-temperature phase diagram, Samara suggested the existence of two independent soft phonon modes: a FE soft mode determining the dielectric anomaly and an AFE soft mode accounting for the phase transition [36]. Soon after, complex electric field-temperature phase diagrams about $PbZrO_3$ and $PbHfO_3$ were reported [56, 57], suggesting a sensitivity of AFE structure to an external stimulus. These studies constitute the initial recognition about the AFE bulks.

As the research interest migrates from bulk to thin film in 1990s, a series of unprecedented findings refresh our conventional understanding of AFEs. In 1998, a thickness driven AFE-to-FE phase transition was first reported in $PbZrO_3$ thin films [58]. Subsequent to its initial observation, this behavior has been consistently observed in various AFE thin films, and it is commonly attributed to surface and/or strain effects [19, 39, 59]. In addition to the size effect, recent research has unveiled new AFE systems in multiferroic $BiFeO_3$ -based [41], lead-based $PbSnO_3$ [60], fluorite-structure $Hf_xZr_{1-x}O_2$ thin films [44], and quasi-2 dimensional francisite $Cu_3Bi(SeO_3)_2O_2Cl$ [30]. These discoveries provide an alternative avenue to explore the origin of antiferroelectricity and the interplay of competing order parameters in these materials, which were previously known for their FE properties. Meanwhile, recent advances in thin-film synthesis and characterization techniques have also facilitated the development of novel devices, including above-bandgap photo-voltages [61], four-state random access memories [9], negative capacitance [45, 62], thermal switches [10, 46] and pulsed power capacitors [63]. Alongside the discovery of new material systems, there is a growing diversity of perspectives regarding the mechanisms underlying antiferroelectricity. Proposed mechanisms include flexoelectric coupling with single soft polar optic mode [31], multiple soft (zone-boundary and zone-center) modes coupling [43], and order-disorder transition scenarios [64], which will be discussed in detail in the following section.

### 1.3. Mechanisms of antiferroelectricity

To understand the origin of antiferroelectricity, one of the key issues is to find out the free energy difference between different phases by considering their dependence on temperature, pressure, electric field, and epitaxial strain. For example, if the free energy of a polar phase is slightly higher than that of the AFE phase, the former can be stabilized by an applied electric field via adjusting the energy landscape (Fig. 2a and 2b). When considering cooling-induced phase transition under a field-free condition, the emerged AFE phase can be understood as an interruption of an imminent FE phase transition at or slightly above the FE critical temperature $T_0$, by a structural phase transition at $T_A$ ($T_A \geq T_0$). This interruption can be modeled by a repulsive interaction between polarization $P$, and structural order parameter $\xi$, that spontaneously emerges at $T_A$.

Taking $PbZrO_3$ as an example, there are at least three modes softening near the phase transition with decreasing temperature [31, 42]. One is the polar mode at the Γ point of Brillouin zone center, which gives birth to the dielectric anomaly. However, before the softening of this polar mode at Curie temperature $T_0$, two other modes interrupt this process, prevent the softening of the polar mode and dominate the PE to AFE structural phase transition at $T_A$, *i.e.*, the $\Sigma$ mode with wave vector $\boldsymbol{q}_{\Sigma} \equiv \frac{2\pi}{a_c}(\frac{1}{4}\ \frac{1}{4}\ 0)$ and the $R$ mode with wave vector $\boldsymbol{q}_R \equiv \frac{2\pi}{a_c}(\frac{1}{2}\ \frac{1}{2}\ \frac{1}{2})$ ($a_c$ is the pseudocubic unit cell), which control the antiparallel Pb displacements and antiphase octahedral rotations, respectively. Therefore, it can be addressed that the structural order parameter $\xi$ is correlated with the $\Sigma$ and $R$ modes and has a repulsive interaction with the polar-mode-related polar order parameter $P$.

Considering the even function nature of the free energy expansion related to $\xi$ and $P$, the power of each individual term must be an even number and the interaction between the two order parameters must include at least a bi-quadratic coupling term $P^2\xi^2$. When the interaction is repulsive, an increase in $P$ and $\xi$ would elevate the free energy. This requires a positive coefficient, controlling the strength of the repulsive interaction. As a result, the repulsion is represented by the bi-quadratic term with a positive coefficient and the free energy model can be given by [65, 66]:

$$F(P,\xi)= \frac{A}{2}(T\text{-}T_0)P^2+\frac{B}{2}P^2\xi^2+F_A(\xi) \qquad (1)$$

wherein $A > 0$, and the coefficient $B > 0$ controls the repulsive biquadratic coupling between the polarization, $P$, and structural order parameter, $\xi$. The transition at $T_A$ implies softening of the lattice modes associated with the order parameter $\xi$, which is incapsulated in the term $F_A(\xi)$.

For the low-temperature phase, where the structural order parameter acquires a spontaneous value of $\xi_0$, the susceptibility, defined as $\chi=1/\frac{\partial^2 F}{\partial P^2}$, can be described as [65]:

$$\chi=1/(A(T\text{-}T_0)+B\xi_0^2) \qquad (2)$$

which corresponds to the AFE-type anomaly below the transition if $A(T\text{-}T_0)+\eta\xi_0^2$ increases on cooling below the transition temperature. This is possible as $\xi_0$ increases with decreasing temperature, which dominates the behavior of this term. The AFE phase transition can be seen as a competition among PE, FE and AFE phases with a temperature dependent free energy change as schematically shown in Fig. 2a. During PE-AFE phase transition around $T_A$, a FE phase would emerge briefly under electric field or under the influence of defects and impurities [67, 68]. From the electric-field dependent free energy landscape, the AFE state occupies the minimum energy position at zero electric field, while the two FE states of opposite polarization direction take unstable or metastable energy positions close to AFE state. When an external electric field is applied, the interaction between the polarization and external electric field can lift the otherwise degenerate states with different polarization directions and thus one polar state becomes preferential in energy, leading to the characteristic AFE-FE phase transition, as depicted in Fig. 2b.

In experiments, lattice dynamics was studied to probe the soft mode behaviors to better understand the AFE origin. So far, several different viewpoints have been proposed. Inelastic X-ray scattering and Brillouin light scattering experiments suggested that the phase transition in $PbZrO_3$, the most widely studied AFE crystal, was driven by flexoelectric coupling with polar soft mode at Brillouin zone center ($\Gamma$ point), which gives rise to a missed incommensurate structural modulation [31]. However, subsequent experiments found that the anomalous flexoelectric coupling was not observed in AFE materials, which mildly questioned the flexoelectric effect as the origin of antiferroelectricity [69]. In parallel, multimode $\Sigma$, $R$, and $S$ trilinear coupling was proposed to account for the AFE phase transition [42, 43]. In addition, few other approaches have also been reported: a first-principles-based model with only $\Sigma$ and $R$ modes actively involved [70], bi-linearly [71] and dipole-dipole [72] coupled displacements as a possible origin of the AFE lattice instability, *etc.* The atomic scale picture for AFE $\Sigma$ and $R$ distortion modes is presented in Fig. 2c [73]. Specifically, the $\Sigma$ mode outweighs the $R$ distortion mode, which is associated to the antiphase octahedral rotation and results in an antiferrodistortive (AFD) character of oxygen displacement. Instead, the $S$ mode maintains a mixed feature of the AFE and AFD modes [74]. With selection of different but dominant soft-mode coupling in AFE, one can obtain different structure symmetries as summarized in Fig. 2d by generalized gradient approximation method [75]. This figure illustrates many possible structures, and two dominant ones are *Pbam*-symmetric AFE phase and *R*3*c*-symmetric FE phase. For these two phases, the ab-initio calculations yield very similar energies [75].

**2. Phase transitions in antiferroelectrics**

Despite ongoing debates on the fundamental basis of AFEs, there is a consensus regarding their macroscopic performance and characteristics. In this section, we provide a comprehensive overview of the fundamental characteristics exhibited by AFE materials.

## 2.1. Temperature-driven transitions

Similar to FEs, AFEs also exhibit a temperature-dependent dielectric response. This indicates an anomaly in the dielectric constant near the AFE transition point. For the high temperature PE phase of bulk AFEs, the dielectric constant follows the *Curie-Weiss* law [76]:

$$\varepsilon = C/(T-T_0) \qquad (3)$$

where $C$ is the Curie constant, $\varepsilon$ is the dielectric permittivity and $T_0$ is the Curie temperature. As temperature decreases to the transition temperature $T_A$, the dielectric permittivity reaches a high value and then drops steeply at $T < T_A$.

Regarding various material systems, *e.g.*, $NaNbO_3$ and $PbZrO_3$, one may usually see different dielectric anomaly and hysteresis as a function of temperature during the heating and cooling cycles [47, 77, 78]. It is certain that the dielectric anomaly and hysteretic behavior indicate the presence of different structural phases during the cyclic measurement of the physical properties, suggesting the first order transition. However, for the low temperature transition between AFE and FE in $NaNbO_3$, a broad dielectric response and negligible thermal hysteresis indicate the deviation from first order transition [77]. This deviation can be attributed to the relaxor feature of the low temperature FE phase [77]. Therefore, we could not conclude whether the AFE transition is first order or not without considering the specific structures from both sides of the transition. In this context, recent research findings on phase-transition frustration near the tricritical point of $Pb(Zr_{1-x}Ti_x)O_3$ are noteworthy [79]. This is evidenced by competition of the first- and second-order transitions near $T_C$, where the birefringence exhibits a first-order transition feature during heating and a second-order transition feature during cooling. In other word, the AFE transition may also arise from a kind of frustrated phase-transition order near the Curie temperature. Certainly, this requires future verification by a series of experimental and theoretical studies on different systems.

The physical property in thin films usually show differences from that in bulks due to the additional constraints on electrical and mechanical boundary conditions. Taking the prototypical AFE $PbZrO_3$ as an example, as shown in Fig. 3a, thin films exhibit a modified phase transition order with diffuse characteristics, while bulk $PbZrO_3$ undergoes a sharp first-order phase transition from an AFE phase to a PE phase at a critical temperature of ∼232 °C. In addition, the transition temperature of $PbZrO_3$ thin film is found to be higher than that of bulk, which effectively broadens the temperature region of the AFE phase [14, 16]. Based on temperature dependent X-ray diffraction, Kniazeva *et al.* [18] reported an emergent in-plane *M* reflection, which coexists with AFE superstructure diffractions during temperature increasing. This manifests a new phase as an intermediate AFE state before the AFE to PE phase transition [18]. In contrast, Dufour *et al.* [80] recently re-examined this phase transition but revealed another FE-like state that emerges in a large temperature window (100K). From their results, transmission electron microscopy (TEM) revealed the atomic-scale high temperature polar structure, which is further supported by atomistic simulations. Overall, the phase transition in thin films behaves drastically distinct from that in the bulk owing to complex competition between the FE and AFE orders, and many aspects of this behavior remain open questions.

### 2.2. Electric field induced transitions

In traditional FEs, polarization hysteresis loops are of great interest in microelectronic applications, for example, in information storage media [81]. In contrast to FEs, AFEs exhibit a null macroscopic polarization owing to their mutually compensated dipoles at the sublattice scale. At a small electric field, the polarization in the AFEs exhibits a linear dielectric behavior. However, once the electric field surpasses a critical value, denoted as $E_{AF}$, the antiparallel dipoles within the materials undergo a switching process to minimize the electrostatic energy, resulting in the completion of the AFE-FE transition. This transition is accompanied by a sharp switching current, similar to that observed in FEs. It is important to note that most of the induced FE phases are metastable, which is evidenced by a back-switching of the FE phase to the prior AFE phase when the applied electric field $E$ is smaller than another critical field $E_{FA}$ ($E_{FA}$ < $E_{AF}$). However, there are also exceptions in lead free $NaNbO_3$ [82] and some doped AFE systems [83], in which the field-induced FE phase would remain stable after withdrawing the electric field. The reversible AFE-FE phase transition process generates the double hysteretic behavior (Fig. 3b), which is considered as a fingerprint of AFE materials. Along with minimized loop area, the steep AFE-FE transition under a strong electric field highlights the superiority of AFEs in the application of pulse power capacitor and digital displacement transducers [23, 38].

Another typical characteristic of AFEs is the “double butterfly” shaped capacitance -voltage ($C$-$V$) curves or dielectric constant-electric field ($\varepsilon$-$E$) curves (Fig. 3c). With increasing the external DC electric field, the dielectric constant rises sharply near the AFE-FE transition point, reaching the first peak value at the critical field of $E_{AF}$. In the reverse process, a similar behavior can also be observed at $E_{FA}$. It should be noted that the phase transition points differ between the AFE-FE (rising electric field, $E_{AF}$) and the FE-AFE (decreasing electric field, $E_{FA}$) transitions, which yield dielectric hysteresis. The “double butterfly”-shaped curve displays four dielectric peaks in the bi-directional measurement. In contrast, the FEs only exhibit two dielectric peaks, and the application of a DC electric field generally leads to a decrease of the dielectric constant.

The field-dependent dielectric constant $\varepsilon$, which is related to the slope of the $P$-$E$ curve, is expressed by [84]:

$$\varepsilon(E) = \frac{dD}{dE} \approx \frac{dP}{dE} \qquad (4)$$

From this equation, we can consider the dielectric constant as the ability for polarization change under electric field. AFEs are nonlinear materials that exhibit distinct polarization switching during the electric-field-induced AFE-FE phase transition. Therefore, when the DC bias field is below the critical field $E_{AF}$, the dielectric constant increases with increasing electric field [85]. The increasing of dielectric constant is associated to the antipolar dipole reorientation. For FEs, the dielectric constant would decrease when the DC electric field increases as the film would transform toward a single domain state, resulting a decrease in the number of domain walls [86]. It should be noted that the decreasing trend of dielectric constant in FE rely on the DC field direction. When the DC field is in the opposite direction of pre-poled polarization, the dielectric constant will also increase due to the domain switching between two opposite directions. The different behaviors in the dielectric constant with applied DC field between AFEs and FEs have been widely verified by experiment [85, 86] and utilized to distinguish the AFE phase [87, 88]. The increase of dielectric constant when subjected to an electric field can greatly enhance the energy storage density, making AFEs as excellent candidates for the application of dielectric energy-storage capacitors.

Distinct from the dipole switching without symmetry changes in FEs, the switching of antiparallel dipoles in AFEs is accompanied with a centrosymmetric to a non-centrosymmetric structural transition. More precisely, AFEs usually undergo a transformation from an AFE structure to a FE structure under a sufficiently high electric field. Since the lattice parameters of the induced FE phase are larger than those of the original AFE cell, the volume expansion would lead to a big jump in the electro-strain response (Fig. 3d) [89]. For instance, recent studies have demonstrated that the highly textured (001)-oriented $PbZrO_3$ thin films can provide a strain up to 1% [7], which is several times larger than that of FE thin films, such as $Pb(Zr_xTi_{1-x})O_3$. The field-induced ultra-large strain makes AFE materials superior in the fabrication of actuators.

The electric field induced AFE to FE switching has been extensively studied in bulk AFEs, often explained through the two-state energy competition. However, how the antipolar dipoles rotate as transient state remains poorly understood. In recent report on thin films, a ferrielectric-like state has been identified as an intermediate state before reaching the final FE phase [15]. The stepwise dipole switching geometry sheds light on the potential manipulation of multistate polarization for information storage capability and tunability of dielectric and strain properties. Further discussion on this topic will be presented in section 4.

### 2.3. Phase transition dynamics

Compared with FE materials, there have been relatively few studies investigating the phase transition dynamics in AFEs, resulting in limited knowledge regarding the kinetics of polarization switching during the AFE-FE phase transition. In particular, the most concerned one is how fast the polarization responds to an external stimulus, which directly determines the performance of the AFE devices. In practice, the width of switching current *versus* time is considered as a good indicator for the phase-transition time. As previously reported in AFE $(Pb_{0.89}La_{0.06}Sr_{0.05})(Zr_{0.95}Ti_{0.05})O_3$ thin films [90], the time dependent switching current and polarization as a function of bias voltage are presented (Fig. 3e), revealing that the polarization switching time is on the order of nanosecond (ns). Furthermore, by referring to the nucleation-limited theory, where the time for nucleation is considered as the time for a complete switching, fitting results of the (001)-textured AFE $Pb_{0.97}La_{0.02}(Zr_{0.95}Ti_{0.05})O_3$ films show that the characteristic switching time is as short as 3 ns [91], which is comparable to that of FE films. Moreover, it is known that the switching time can be influenced by the applied electric field, temperature, electrode area, and film thickness [92, 93] and there is a great possibility that the intrinsic dynamics of AFE switching can be as fast as switching in FE thin films, such as $BiFeO_3$, reaching picosecond (ps) scale [94].

In FE thin films, a scaling relationship between coercive field $E_c$ and film thickness $t$, *i.e.*, the semi-empirical Janovec–Kay–Dunn (JKD) scaling law, wherein $E_c \sim t^{-2/3}$, was frequently observed [95, 96]. Very recently, a JKD-type scaling behavior is also reported in AFE $ZrO_2$ thin films, wherein a decrease in the film thickness from 20 nm to 5.4 nm results in an increase in critical fields for both nonpolar-to-polar ($E_{AF}$) and polar-to-nonpolar ($E_{FA}$) transitions [97]. First-principles calculations suggest that an increase in tetragonality of the ultrathin films contributes to an increase of energy barrier between non-polar tetragonal phase and polar orthorhombic phase, leading to the increase in critical electric fields. In perovskite-type AFE thin films, similar critical field scaling relationship had been reported [98], while the underlying mechanism is underexplored and need further exploration.

Benefit from the critical field scaling, AFE thin films exhibit much higher $E_{AF}$ and $E_{FA}$ (hundreds of kV/cm) [16] than bulk AFEs (tens of kV/cm)[99], which potentially boosts the phase transition speed [92]. Therefore, AFE thin films with high-density charge release offer advantages in high-speed chip modules and pulsed energy storage capacitors [100]. Additionally, investigating the field driven structural evolution path is also important for understanding the AFE-FE transition. Previously, piezoelectric measurements and in-situ X-ray diffraction studies were conducted to analyze the domain reorientation and structural transformation in bulk AFEs [101, 102]. By employing in-situ TEM imaging techniques, it has been demonstrated that the AFE structure undergoes a series of intermediate states. These states involve the AFD-ferrodistortive transition of oxygen octahedra, resulting in the emergence of a unique cycloidal polarization order and enabling the presence of a spontaneous polarization [103, 104]. Until now, the complete understanding of phase transition process and the field-induced FE structure in AFE thin films remains elusive. It is believed that advanced characterization techniques, such as time-resolved X-ray diffraction [105], and optical pump-based detection [106], which have been successfully employed to investigate the switching kinetics of FE films, can provide further insights into the AFE-FE transition in AFE films

### 2.4. Fatigue in antiferroelectric thin films

Electrical fatigue is manifested as the gradual decrease of switchable polarization under repetitive switching cycles [107]. Like FEs, AFE materials also suffer from fatigue problem, which has been one of the major problems hindering their practical applications. In general, the fatigue performance of bulk or thin film AFEs is directly influenced by the sample quality. Over the past decades, several models have been proposed to explain the fatigue mechanism in AFEs. One widely accepted model is that, during the AFE-FE phase transition, a significant strain induced by an external field leads to the formation of microcracks, which has been observed in fatigued AFE ceramics [108]. Another popular model is based on domain wall pinning [108]. During the AFE-FE phase transition, the ionic defects and trapped charges, such as oxygen vacancies, can impede the domain switching and domain wall motion. This gradually reduces the amount of reversible domains and subsequently diminishes the switchable polarization in AFE polycrystalline films [109]. Additionally, the local phase decomposition model has been proposed, suggesting a correlation with the local heat generated during the switching event in polycrystalline films [110].

Although a universally applicable fatigue mechanism for AFEs has not been established yet, various effective methods have been proposed to mitigate the electrical fatigue. For example, previous studies have reported that AFE $Pb_{0.99}Nb_{0.02}(Zr_{0.85}Sn_{0.13}Ti_{0.02})_{0.98}O_3$ polycrystalline thin films grown on the $LaNiO_3$-buffered $Pt/Ti/SiO_2/Si$ substrates displayed improved fatigue properties in comparison to those grown on $Pt/Ti/SiO_2/Si$ directly [111]. The improved behavior is attributed to the fact that the oxide bottom electrode can serve as a sink for oxygen vacancies, resulting in a reduced number of oxygen vacancies at the film-electrode interface [111]. Additionally, chemical doping has been demonstrated as an effective approach to alleviate the fatigue problem. For instance, in Sr-doped $PbZrO_3$ polycrystalline films, the polarization reduction was only 28% after $10^8$ cycling, whereas pure $PbZrO_3$ polycrystalline films experienced a polarization loss of over 50% under the same cycling conditions (Fig. 3f) [112]. Overall, the fatigue mechanism in AFE thin films is still not fully understood, and there is a lack of research on epitaxial AFE thin films to supplement the understandings.

## 3. Engineering antiferroelectric order in thin film form

In recent decades, research based on epitaxial oxide thin films has significantly advanced our understanding of the physics, structures, and functionalities in AFE materials. Firstly, compared to bulk materials, thin films offer more control methods, such as size scaling effect, strain engineering, and orientation control, to manipulate the AFE ordering. Secondly, the development of thin-film growth techniques enables flexible configurations of layers and architectures as well as device miniaturization. Lastly, thin films have been demonstrated to have an improved breakdown field strength, thereby facilitating the AFE-FE phase transitions (for instance, the room-temperature critical field $E_{AF}$ for undoped bulk $PbZrO_3$ could be even higher than the breakdown field). Thus, the AFE research in thin film form is essential and holds great promise.

However, in contrast to the much more developed theory about FE films, the theory for controlling the functional behavior of AFE films in terms of specific demands is still in its development stage [113]. Currently, a comprehensive understanding that addresses the specific requirements for manipulating the functional properties of AFE thin films is still lacking. Although AFE films share some properties and control principles with FE films, they have distinct characteristics. To systematically organize these aspects, we propose an eight-corner paradigm scheme, as illustrated in Figure 4.

### 3.1. Size effect

Size effects in ferroic materials have been extensively studied over the years, and these studies have revealed the presence of a critical thickness necessary to stabilize the primary ferroic order parameter and thus the device functionality [114, 115]. Similar to FEs, knowledge about the AFE order and its size effect is essential for both fundamental understanding to the nature of antiferroelectricity and for creating deterministic performance in functional devices [116]. In 1998, Ayyub *et al.* first reported the size effect in AFEs in polycrystalline $PbZrO_3$ thin films grown on Si substrates. By reducing the film thickness from 900 nm to 100 nm, the AFE behavior is replaced by FE behavior, and whether it is induced by residual strain or the self-biasing effect at the interface remains an open question [58, 117]. Subsequently, in 2007, a thickness driven AFE-FE phase transition was reported in $PbZrO_3/PbZr_{0.8}Ti_{0.2}O_3$ multilayers. The electric measurements have revealed that the multilayers undergo a thickness-dependent transition from a mixed AFE-FE behavior to a purely FE behavior. This indicates that the $PbZrO_3$ sub-layer undergoes a structural transition from an orthorhombic AFE phase to rhombohedral FE phase when the $PbZrO_3$ sub-layer thickness decreases to ~10 nm [118]. Further, X-ray diffraction studies suggest that the substrate strain plays a crucial role in driving the AFE-to-FE transition of the $PbZrO_3$ layers. In 2011, Chaudhuri *et al.* [19] reported that a FE rhombohedral structure could be stabilized by substrate-induced strain in ultrathin $PbZrO_3$ epitaxial films grown on (001) $SrRuO_3$-buffered $SrTiO_3$ substrates. The contribution from FE structure leads to an increase in the remnant polarization of the $PbZrO_3$ thin films as the film thickness decreases, resulting in a mixed feature of FE and AFE behavior in the AFE films, as illustrated in Fig. 5a [19].

To understand the size effect in AFE thin films, phenomenological theory approach has been applied by considering the mechanical stress and surface piezoelectric effect. As Bratkovsky and Levanyuk suggested, the surface of a FE can be regarded as a defect coupled to the order parameter, resulting in the phase transition smearing in thin film [119]. In 2007, Eliseev and Glinchuk took the size effects into account in AFE thin films and reported that the surface piezo-effect can break the inversion symmetry near the surface and generate a built-in electric field that is inversely proportional to the film thickness [120]. With decreasing thickness, the built-in electric field progressively increases until it reaches a critical value and induces the AFE-FE phase transition. In 2014, Mani *et al.* conducted meticulous first principles-based simulations to thoroughly investigate the size effect in AFEs. They examined the energy stability between AFE and FE phases in $PbZrO_3$ films as a function of film thickness under various mechanical and electrical boundary conditions [116]. Surprisingly, below a critical thickness (about 6 nm), the $PbZrO_3$ films would develop a FE tetragonal phase for films under short-circuit boundary condition even in the absence of epitaxial strain. According to Mani *et al.* [116], the origin of this size-driven AFE-FE transition is attributed to the intrinsic surface effect that stabilizes the FE phase by removing energy costly short-range interactions between head-to-tail dipoles. This theoretical model has been supported by recent experiment. Xu *et al.* demonstrated the intrinsic size effect in lead-free AFE $NaNbO_3$ free-standing membranes, revealing the emergent ferroelectricity as the thickness decrease and coexistence of FE and AFE orders at an intermediate film thickness [121]. By removing the mechanical-clamping contribution from the substrate, it delivered more intrinsic behavior of AFEs and proposed the distortion of surface structure, rather than substrate-induced strain, as the origin of AFE size effect [121].

The mixed feature of FE and AFE are commonly seen in AFE thin films, and its microstructures exhibit complex constitutions due to the close energy barrier of the polar and antipolar phases. On the one hand, Recent report on $PbZrO_3$ thin films discovered the antipolar structure remains locally uncompensated. In this case, the coexistence of different structure periods, *i.e.*, ↑↑↓↓ and ↑↑↓, are reported and correlated with thickness scaling [122]. Both antipolar structures transform to FE structure with decreasing thickness below 25 nm [122]. This ferrielectric structure exists on $SrTiO_3$ substrates and disappear on $KaTaO_3$ substrates, which can be attributed to the epitaxial compressive strain that may induced the fluctuation of the ferrielectric structure as theoretically predicted by DFT calculations [123]. On the other hand, the presence of wide-range FE rhombohedral phase has also been reported in $PbZrO_3$ thin films grown on $SrRuO_3$-buffered $SrTiO_3$ substrates [124]. This FE structure is energetically favorable below certain critical thickness due to surface effect [116] or strain effect [124], and typically exists near the film-bottom electrode heterointerface.

The size scaling induced mixed FE/AFE feature can also be found in $HfO_2$-based thin films [125-127]. As suggested by DFT calculations, due to the impact of surface energy, the monoclinic, orthorhombic and tetragonal structure are energetically favorable in sequence as the thickness decreases [128]. By varying the film thickness, the remnant polarization has a maximum at an intermediate thickness where the polar orthorhombic structure dominant the FE behavior, while decreasing (increasing) the thickness would lower the remnant polarization [127], and the more energetic-favorable AFE tetragonal (PE monoclinic) phase would dominant the electrical behavior [126-128]. In contrast with $HfO_2$ thin films, it is widely acknowledged that $ZrO_2$ thin films predominantly persist AFE tetragonal structure [129, 130]. As the thickness decrease below 3 nm, Cheema *et al.* [129], recently reported an AFE-FE phase transition that occurs, and the orthorhombic FE structure can be stabilized down to 5 Å [129]. The ferroelectricity in ultrathin films is promising for highly scaled next-generation Si electronics beyond 2D limit.

### 3.2. Strain engineering

For epitaxial growth of thin films, the in-plane lattice constant of thin films is expected to be identical to that of the substrate when the film is thinner than a specific critical thickness value. In this case, the misfit strain in the film can be given as:

$$\varepsilon=(a_s-a_f)/a_s \qquad (5)$$

where $a_s$ and $a_f$ are the lattice constants of the substrate and film, respectively. To achieve high-quality, coherently-strained epitaxial films, the key factor is to keep the film thickness thinner than the critical thickness [131]. Otherwise, the strain tends to relax with the assistance of interfacial misfit dislocations. Epitaxial strain has been recognized as an effective tool to modify the crystal and domain structures, to tune the physical properties, and even to create new emergent functionalities in ferroic oxide thin films [132, 133].

Based on first principles calculations, it is anticipated that epitaxial strain can disturb the balance of phase stability between the AFE and FE phases, thereby inducing the AFE-FE phase transition. For bulk $PbZrO_3$, calculations have shown a small energy difference of only ~1 meV/f.u. between the AFE *Pbam* and FE *R*3*c* phases [134]. The calculations further indicate that the FE phase is stabilized under compressive epitaxial strain, while the AFE *Pbam* phase is favored under tensile strains (Fig. 5b). For lead-free AFE $AgNbO_3$ thin films, first-principles calculations revealed that two phase transitions (from *Pm* to *Cm* and from *Cm* to *Amm*2) would occur at the range of compressive strain and tensile strain, respectively. Additionally, compressive strain can enhance the stability of AFD and, subsequently, enhance the antiferroelectricity [135]. In $NaNbO_3$ thin films, an energy-versus-misfit strain phase diagram has been constructed, revealing three phases (monoclinic *Cc*, orthorhombic $Pca2_1$, and orthorhombic $Pmc2_1$ symmetries) corresponding to different strain regimes [136]. In addition to the strain-induced structural phase transition, the strain can also alter the critical electric fields ($E_{FA}$ and $E_{AF}$) for AFE-FE phase transition and thus tune the energy storage properties [137].

It is important to emphasize that most of our current understanding regarding the influence of epitaxial strain on the stability of different phases in AFEs has been derived from zero-Kelvin density functional theory (DFT) calculations. However, these findings may not accurately reflect the actual behavior at finite temperatures. Alternative approaches, such as phenomenological Landau approach, which are based on empirical free energy parametrizations and have been extensively used for analyzing strain effects in FE films [138], have not yet been widely utilized in the study of AFE films.

In polycrystalline AFE oxide thin films grown on semiconducting substrates (*e.g.*, Si), the thermal and growth strain would play a crucial role in determining the free energy of AFE phase and FE phase, altering the structure stability and consequent electrical behaviors. Thermal strain originates from the mismatch of the thermal expansion coefficients between the film and the underlying substrate during the cooling process after film deposition [139, 140]. The thermal strain and stress can be given by [139]:

$$\epsilon_{thermal}=\int_{T_{RT}}^{T_D}(\alpha_{film}-\alpha_{substrate})\,dT \qquad (6)$$

$$\sigma_{thermal}=\int_{T_{RT}}^{T_D}\frac{(\alpha_{film}-\alpha_{substrate})E_{film}}{1-v_{film}}\,dT \qquad (7)$$

where $\alpha$, $v$ and E represent the thermal expansion coefficient, Poison's ratio and elastic modulus, respectively. For instance, in Nd-doped $BiFeO_3$ thin films grown on Pt/Si substrate, a coexistence of AFE, FE and PE phases was reported, while bulk counterpart only exhibits single phase AFE. It is proposed that clamping due to mismatch in thermal expansion coefficient with the substrate suppresses the formation of AFE phase by forming FE and PE phases [141]. Growth strain, originating from the coalescence of the nucleating grains and densification during crystallization, is another important factor that influences the structure stability [140, 142]. In $Hf_{0.5}Zr_{0.5}O_2$ thin films, a giant in-plane tensile strain, induced by coalescence of the nucleating grain, are proposed to enhance the ferroelectricity by transforming the nonpolar tetragonal phase to polar orthorhombic phase [142].

### 3.3. Orientation control

Functional oxides generally exhibit significant anisotropy in crystallography, and their physical properties may differ significantly with the change in crystallographic orientation [143]. For epitaxial thin films, their orientation is determined by the underlying substrates and their performance can be enhanced by selecting suitable substrate orientation [144]. In conventional FE systems, it has been reported that manipulating the energy barrier via orientation control is an effective approach to switch the electric dipoles. For example, *Xu et al.* demonstrated that the coercive field in 20 nm $PbZr_{0.2}Ti_{0.8}O_3$ thin films can be reduced from ~350 kV/cm for films on $SrTiO_3$ (001) substrate to ~150 kV/cm on $SrTiO_3$ (111) substrate [144]. For AFEs, taking $PbZrO_3$ as an example, it is widely accepted that the AFE phase has an orthorhombic structure (*Pbam*, $a^-a^-c^0$), which transforms into a FE rhombohedral (*R*3*c*, $a^-a^-c^-$) phase with the polarization along the [111] direction under a sufficiently strong electric field [102]. By applying an electric field along different crystallographic orientations in AFEs, the critical electric field for AFE-FE phase transition and maximum polarization varies significantly. In a recent study, Chu *et al.* reported that the saturated polarization and critical transition field undergo dramatic changes with varying orientations, from 30.54 to 38.91 $\mu C/cm^2$ and from 650 to 870 kV/cm as the AFE $PbHfO_3$ thin films were grown on (100)-, (111)- and (110)-oriented $SrTiO_3$ substrates (Fig. 5c) [145].

In $PbZrO_3$, by using a first-principles-based effective Hamiltonian approach, Ponomareva *et al.* predicted that the phase transition pathway of $PbZrO_3$ also depends on the crystallographic orientation with respect to the applied electric field [146]. Notably, they discovered the emergence of three new polar phases: *Cc*, *Ima*2 and *I*4*cm* are obtained under electric field along the [111], [110] and [001] directions, respectively. None of these newly identified phases exhibit the commonly acknowledged rhombohedral structure with the *R*3*c* symmetry. Moreover, the field-induced phase transition into these new polar phases is predicted to generate extremely large strain of up to 3.6%, which holds significant technological significance [146].

It is interesting to note that the above results of pioneering effective Hamiltonian studies disagree with experimental observations by Fesenko *et al.* [56] in $PbZrO_3$ single crystals. In their experiments, applying an electric field along a specific direction triggered a series of transitions to various FE-like phases, where the polarization did not necessarily align along the field. The underlying mechanism for this discrepancy between experiment and theory thus requires further investigation and clarification.

### 3.4. Domain engineering

FE and AFE phases in single crystals are usually composed of domains with different orientation of polar/antipolar dipoles to minimize the depolarization field energy and/or elastic energy [147, 148]. Manipulating functional domains and domain walls in ferroic materials has been a recent focus, resulting in the discovery of intriguing phenomena, including charged and chiral walls, domain-wall conductivity, topological polar structures and more [149-153]. In AFEs like $PbZrO_3$, the orthorhombic AFE structure results in 60°, 90°, 120° and 180° domains [154]. Although the structural geometry of AFE domains has been widely studied, only recent years, the atomic scale understanding about AFE domain and domain boundary were reported. In 2013, polar translational antiphase boundary was initially discovered in $PbZrO_3$ [65]. By connecting two 180° domains with a $\pi$ phase shift, the dipoles in the domain wall are parallel, making the antiphase boundary polar. Viewing the 2D polar domain walls in non-polar matrix as information carrier, the AFEs exhibit possibility for usage in high-density non-volatile memory devices.

In $PbZrO_3/BaZrO_3/SrTiO_3$ thin films, it was found that the polar domain walls (180° walls) correlated strongly with the interfacial strain [155]. By mediating the interfacial compressive strain using $BaZrO_3$ buffer layer, the polar domain walls can be abundantly formed near the interface. The preferential creation of polar domain wall paves the way for domain wall engineering, presenting opportunities for further practical applications. Apart from the polar domain walls, domain engineering is another necessary step towards the practical use of the AFEs [65, 156], *e.g.*, shape memory induced by the AFE-to-FE phase transition [157]. Recent studies have also shown that reducing hysteresis with slim hysteresis loops can be achieved by breaking the long-range ordered domains into nanometer scale, thereby introducing nanoscale heterogeneity in AFEs *via* chemical doping [158], which is particularly important for their applications in energy storage devices.

Recent results in $PbZrO_3/SrRuO_3/SrTiO_3$ (001) heterostructures have shown stimulating, but rather puzzling results by an unusual experimental approach. Using the Bragg peak shape analysis combined with X-ray nanoscopy where a nano-focused X-ray beam is used, it has been shown that individual domains are usually not larger than about 13 nm [159]. This occurs regardless of film thickness, in contrast to the so-called Kittel law, *i.e.,* a square root dependence of domain size on thickness which has been reported for diverse ferroic systems [160, 161]. In the shape analysis of Bragg peaks, the domain size is probed as the length at which the AFE lattice is coherent with itself, *i.e.*, the lattice coherence length. This coherence can be easily broken by translational boundaries, which do not change the domain orientation but only displace domain lattices with respect to each other. Therefore, it is likely that the estimated domain size of 13-nm can characterize the length scale of the densest kind of domain packing – the scale of translational domains. In bulks, it has been shown that those translational domains pack themselves into superdomains, which are in turn separated by curved 90° domain walls and more exotic junctions, such as cloverleaf patterns [148]. In films this has not yet been thoroughly studied.

### 3.5. Defect engineering

Defects are common in materials and are inevitable during the sample preparation process. In traditional recognition, the defects in materials are deleterious to the physical properties and will greatly reduce the device reliability. Nevertheless, recent findings report that defect engineering, an intentional and purposeful introduction of defects, can enhance the material properties and even create emergent functionalities [14, 162]. One paradigm is that the presence of oxygen vacancies, leading to deviation from the ideal stoichiometry, can cause changes in vibrational frequency of the FE and AFE mode, resulting in the presence of intermediate FE phase [68]. In addition, by introducing point-defect via decreasing oxygen pressure, it has been revealed that a FE-like switching can be obtained in $PbZrO_3$ thin films at room temperature [163]. This FE-like switching can be attributed to the pinning effect from point-defect, which can effectively delay the AFE-FE phase transition. Pertinent to the atom occupancy, the anti-site defects, manifested by a false positional occupation, provides a new approach to break the inversion symmetry and altering the electric property [164]. For instance, the AFE-FE phase transition can directly be stimulated in nonstoichiometric $Pb_{1.2}ZrO_{3+\delta}$ thin films (Fig. 5d) [14]. Defect-controlled electrocaloric effect (ECE) was also reported in $PbZrO_3$ thin films [165]. The gradual increase of lead volatilization results in asymmetric interface nucleation. As a result of defect-related switching dynamics under different frequency of the same electric field, a large adiabatic temperature change of 30 K from positive ECE can be achieved at 100 Hz while a temperature change of -10 K due to negative ECE was obtained at 10 kHz [165]. One can see that the defect engineering opens new possibilities to design and tailor functionality of AFE thin films.

### 3.6. Doping effect

Chemical doping is an efficient way to modulate the material properties. Chemical doping has been widely used to mediate the competitive orderings in functional oxide systems and induce intriguing phenomena including enhanced piezoelectricity near the morphotropic phase boundary (MPB), high energy storage performance, metal-insulator transition, and high-temperature superconductivity [8, 166-169]. For PZT system, compositional doping can stabilize either the AFE or FE order, and this stabilization can be rather successfully interpreted in terms of adjusting the *Goldschmidt* tolerance factor $t$ [170]:

$$t=\frac{R_A+R_O}{\sqrt{2}(R_B+R_O)} \qquad (8)$$

where $R_A$, $R_B$, and $R_O$ refer to the radii of A- and B-site cations as well as of oxygen anions. Upon remaining within the same solid solution system, the value of *Goldschmidt* tolerance factor can be a good indicator to evaluate the FE and AFE stability. In AFEs, a smaller A-site dopant or bigger B-site dopant can stabilize the AFE phase. On the one hand, $Sr^{2+}$ [112] and $La^{3+}$ [171] can stabilize the AFE phase and bring a higher $E_{AF}$ than undoped $PbZrO_3$, on the other hand, the B-site dopant with a smaller radius, such as $Sn^{4+}$ and $Ti^{4+}$ [172], has been employed to decrease the $E_{AF}$ and minimize the hysteresis loop area. Moreover, chemical doping would induce complex phase diagrams, resulting in the identification of abundant structural phases, *e.g.*, the incommensurately modulated structure in $(Pb,La)(Zr,Sn,Ti)O_3$ system [173]. This would give rise to AFE-to-relaxor FE phase transition and to boost the energy storage performance [174].

Apart from the tolerance factor that reflects the arrangements of ions in perovskite structures, the properties of perovskite oxides are intricately linked to the average electronegativity. Halliyal *et al.*, found that besides tolerance factor $t$, the electronegativity difference $X$ is another important parameter that determine the stability of the perovskite $ABO_3$ structure [175]. The average electronegativity difference is given by:

$$X=(X_{AO}+X_{BO})/2 \qquad (9)$$

where $X_{AO}$ ($X_{BO}$) is the electronegativity difference between the A(B) cation and the oxygen anion. This average difference value in electronegativity between cations and anions indicates the polarity of the chemical bond. A larger average electronegativity value signifies the stronger the polarity and a closer proximity to an ionic bond, while a smaller average electronegativity value implies the weaker the polarity and the closer proximity to a covalent bond. Previous studies on $CaZrO_3$-$NaNbO_3$ [176], $AgNbO_3$-$CaZrO_3$ [177], $NaNbO_3$-$SrZrO_3$ [178] suggest that reducing the tolerance factor t and increasing the electronegativity difference $X$ favor antiferroelectricity.

In fluorite $HfO_2$-based thin films, different dopants can result in either FE or AFE behavior depend on the dopant size and doping content [44]. For the famous $Hf_{1-x}Zr_xO_2$ solid solution, the inherent AFE nature in $ZrO_2$ thin films enables the $Hf_{1-x}Zr_xO_2$ thin films to exhibit antiferroelectricity over a wide range of composition compared to other dopants [28, 44, 179]. Apart from the AFE in $Hf_{1-x}Zr_xO_2$ thin films, the AFE behavior is usually discovered in doped $HfO_2$ thin films with dopant size smaller than Hf, while the dopants with bigger size only result in FE behavior [44]. In Si-doped $HfO_2$ thin films where Si has a smaller radius than Hf, the AFE behavior can be obtained with increasing the doping content of Si [127]. Similar behavior can also be discovered in Al-doped $HfO_2$ thin films [180], where Al is also smaller than Hf. The AFE behavior is strengthened with Al content increase to 8.5%, and a PE-FE-AFE phase transition can be realized during increasing the Al content [180]. First principles study suggested that the dopant size can change the metal-oxygen bonding to cause energy differences [181]. As a result, the smaller dopants tend to stabilize the nonpolar tetragonal phase, which can be seen as the origin of antiferroelectricity [181]. Introducing Si and Al dopant into FE $Hf_{0.5}Zr_{0.5}O_2$ thin films can also assist FE to AFE phase transition, which can enhance the energy storage and efficiency without increasing the thermal budget of $Hf_{1-x}Zr_xO_2$ [182].

### 3.7. Superlattice and interface control

Functional oxide materials usually display delicate interaction between spin, orbital and lattice order parameters, and results in competing ground states that are sensitive to extrinsic stimulus. With the advancement of thin film deposition technique at atomic-scale precision, such as laser molecular beam epitaxy, artificial control of heterointerfaces and superlattices between different oxides give chance to introduce variant boundary conditions for bring a plethora of emergent phenomena, for example, two-dimensional electron gas [183], polar vortices/skyrmions [152, 184], and improper ferroelectricity [185]. Considering the energy vicinity between ferroelectricity and antiferroelectricity, interfacial engineering could bring unprecedent behaviors between two counterpart phase diagrams. Emergent AFE behavior has been discovered in several artificial superlattice systems. In 2002, the first one is observed in $KNbO_3/KTaO_3$ 1/1 superlattices (samples in which each $KNbO_3$ and $KTaO_3$ sub-layer are 1 unit cell thick, respectively), while bulk $KTaO_3$ and $KNbO_3$ are FE and PE, respectively, the superlattice appears AFE based on an increase in the dielectric constant with applied DC bias [87]. The B-site cation modulation imposed by the superlattice is considered as the origin of antiferroelectricity [87]. Soon afterwards, a similar AFE-type behavior has been demonstrated in $SrTiO_3/BaZrO_3$ superlattices, where both constituent phases are PE in bulk. Those superlattices with large periodicities of 20/20 and 38/38 exhibit FE ordering, while for superlattices with smaller periodicities of 7/7 and 15/15 [88], the strain-induced ferroelectricity in $SrTiO_3$ and the layer-spacing-dependent coupling are proposed to be the leading mechanism for producing AFE-like behavior [88].

Very recently, using high-throughput second-principles calculations, Aramberri *et al.* further verified the existence of an AFE-like behavior in artificial electrostatically engineered FE/PE superlattices [186]. It is found that the $(PbTiO_3)_4/(SrTiO_3)_4$ superlattice presents an overall antipolar 180° domain structure at zero field and undergoes a field-induced phase transition into a polar state for fields of a few megavolts per centimeter. Moreover, the superlattices can be tailored to address specific needs by means of epitaxial strain, layer thicknesses, and dielectric stiffness. In this proposed model, superior energy storage performance is obtained [186]. Apart from the emergent antiferroelectricity in para- and ferro-electric materials, the AFE behavior can also be derived from multiferroic $BiFeO_3$ nanolayers with delicate interface engineering. This is evidenced by identification of antipolar phases in $BiFeO_3/La_{0.7}Sr_{0.3}MnO_3$, $BiFeO_3/La_{0.4}Bi_{0.6}FeO_3$, and $BiFeO_3/TbScO_3$ superlattices [187-190]. These studies demonstrate that interfacial control offers an effective strategy for the design of new AFE materials for versatile applications.

The interfacial control is also an effective engineering strategy to tune AFE properties in fluorite-structure $Hf_{1-x}Zr_xO_2$ thin films [191]. In 2019, Park *et al.* [192], reported the $HfO_2/ZrO_2$ nanolaminates and superlattices with various layering combinations and thicknesses. In that work, the increasing single layer thickness of $ZrO_2$ can stabilize the AFE behavior due to the energetically favorable tetragonal structure in pure $ZrO_2$. The starting layer with $HfO_2$ can strengthen the field-induced ferroelectricity from AFE thin films. This can be attributed to the low monoclinic phase fraction in the starting layer of $HfO_2$. In comparison, $ZrO_2$ with tetragonal matrix displays reduced energy barrier for monoclinic phase formation and reduces the field-induced FE behavior [191, 192]. TiN capping layer is commonly used as a mechanical confinement to engineer the FE/AFE behavior in $Hf_{1-x}Zr_xO_2$ films [44, 140]. In $ZrO_2$ thin films, with introducing the mechanical stress induced by the TiN capping layer, transformation from PE amorphous phase to AFE tetragonal phase can be realized [193], while etching the TiN layer away leads to the transition from the AFE tetragonal phase to FE orthorhombic phase [193]. Additionally, the interfacial engineering is also reported in $ZrO_2$ thin films with sub-nanometer $HfO_2$ and $TiO_2$ interfacial layer [194]. Additionally, the interfacial engineering is also reported in $ZrO_2$ thin films with sub-nanometer $HfO_2$ and $TiO_2$ interfacial layer [194]. In that work, the $HfO_2$ interfacial layers are reported to boost the formation of (111)-texture orthorhombic $ZrO_2$ to enhance the ferroelectricity, while the $TiO_2$ interfacial layers favor the formation of (110)-textured tetragonal $ZrO_2$ to support antiferroelectricity. Moreover, such interfacial engineering strategy is post-annealing-free and thus is beneficial to the CMOS process integration which need a low thermal budget [194].

### 3.8. Octahedral tilt control

Since most of the prospective AFE thin films are of perovskite structure, it is possible to affect their state through the connectivity of octahedral tilt systems across the substrate-film interface. Originally, it was envisioned by ab-initio modeling in LSMO/$SrTiO_3$ heterostructures [195] and developed in few other works [196, 197]. The approach is being rather extensively investigated nowadays in the context of $SrRuO_3$/$SrTiO_3$ functional interfaces [198, 199]. The essence is that the substrate's octahedral tilt pattern can propagate into the film to a small but considerable depth.

For AFEs, this can have, at least in principle, a solid impact on the ability to tune the material properties. The link between antiferroelectricity and octahedral tilt patterns has been the focus of research by theoretical groups [200], which flourished recently in various systems [201]. This theory is expected to clarify its connection to the experiments in the future. Single-crystal diffuse scattering experiments suggest that $a^-a^-c^0$ octahedral tilts are attractive to incommensuration and antiferroelectricity in $PbHfO_3$, at least at high temperatures [202]. Those distortions form simultaneously in a triggered incommensurate transition in pure material. While in Sn-doped $PbHfO_3$ crystals, the octahedral tilts form earlier on cooling than the incommensuration, and the octahedral tilts seem to aide in moving the AFD susceptibility maximum from zone center towards the incommensurate position [203]. This, again, suggests a collaborative nature of the coupling between octahedral tilts and antiferroelectricity. The picture gets complicated by molecular dynamics simulations by Mani *et al.* [70], who try simulations with and without the coupling between octahedral tilts and antiferroelectricity. When the coupling is turned off, both types of distortions appear at much higher temperatures as compared to the case when they are coupled. The coupling in those calculations is, therefore, of strange character. On the one hand, it is attractive in the sense of joining the AFE and AFD transitions into one, on the other hand, it is somehow repulsive in the sense of stabilizing the high-symmetry phase to lower temperatures. This shows that understanding the coupling between AFE and AFD orders is yet to be achieved.

In the meantime, it is necessary to highlight another complication on the way of octahedral control of AFEs. Presently, conventional high-quality epitaxial lead-based AFE films were synthesized mostly on highly mismatched buffer layers and single-crystal substrates [16, 18]. The pseudocubic lattice comparison of common perovskite AFE oxides, electrodes, and substrates at room temperature is drawn in Fig. 6. The commonly used substrates and electrode layers, such as $SrTiO_3$ and $SrRuO_3$, differs by near -3% from Pb-based AFEs. The films are, therefore, strongly relaxed through misfit dislocations [124] and this strain relaxation should likely impact the film stronger than the octahedral tilt connectivity effect. To overcome this obstacle, one needs to envision the newly developed perovskite electrodes/substrates with large lattice constants, like La-doped $BaSnO_3$ electrode and $BaZrO_3$ single crystal substrates [204, 205]. An exceptional work is related to the $PbZrO_3$ thin film grown on $BaZrO_3$-buffered $SrTiO_3$. The dislocations were created at the interface between $BaZrO_3$ and $SrTiO_3$ away from $PbZrO_3$, and the influence from $SrTiO_3$ as well as the potentially unchained octahedral connectivity effect has not been explored [155]. Alternatively, the situation can be more promising with lead-free AFE perovskites, where lattices match better with mainstream substrates, like $SrTiO_3$.

## 4. Materials

Through tracing the key guidelines about tailoring the AFE properties in thin films, we find that abundant structure and property diversities emerge in conventional AFE thin films, and a set of new AFE phenomena are developed. Understanding the existing AFE systems (Table 1) and its thin-film types are important for the development of AFE materials and devices. In this section, various lead-based and lead-free AFE oxide materials are reviewed, with particular emphasis on those from the thin film perspective, trying to grasp both the material science and physical perspectives.

### 4.1. Lead-based Perovskites

### 4.1.1. $PbZrO_3$

$PbZrO_3$ is the most investigated material in the field of AFEs. It undergoes a PE to AFE phase transition around 230°C and its AFE orthorhombic unit cell contains 8 formula units (*i.e.*, pseudocubic unit cells), where Pb ions displace antiparallelly along $[110]_{pc}$ direction (Fig. 7a). As an archetype AFE materials, much about it has become clear, from its structure details to practical applications in the bulk form. However, when it comes to the thin-film form, the story about its physical properties becomes richer and more complex.

During the PE-AFE phase transition, which is essentially related to the origin of antiferroelectricity, an unusually high transition temperature is observed in epitaxial $PbZrO_3$ thin films. This is manifested by the higher $T_A$ for dielectric anomaly and annihilation of 1/4 superstructure diffractions as the $PbZrO_3$ film is grown on $SrTiO_3$ substrate, which increases by 40 K in comparison with bulk $PbZrO_3$ [16]. Moreover, second-harmonic generation and hysteresis measurement detected the polar structure at room temperature and the incipient ferroelectricity above the transition temperature. This indicates there is a mixed feature of ferroelectricity and antiferroelectricity of $Pb^{2+}$ displacement in $PbZrO_3$ thin films, resulting in the fluctuation of uncompensated polarization as shown in Fig. 7b and 7c. Obviously, epitaxial integration has a huge impact on property of the materials. In FE thin films, epitaxial strain alone has shown great influence on the transition temperature [206]. Indications on the importance of such an effect in $PbZrO_3$ films are reported by earlier studies performed at Max Planck Institute, where Raman studies show that the presence of mechanical stress may lead to the presence of FE phase in $PbZrO_3/SrRuO_3/SrTiO_3$ heterostructures [124]. However, the $PbZrO_3/SrRuO_3/SrTiO_3$ heterostructures are rather far from being simply strained because of the large lattice mismatch of ~6%, which is relaxed almost completely by forming dislocations near the $PbZrO_3/SrRuO_3$ interface. Considering that the AFE lattice distortion is inhomogeneous, a more complex strain distribution is possibly developed inside $PbZrO_3$ films. This may either arise from the surface effect where the surface energy is dominant in straining the thin films, or from intra-domain strain that helps to minimize the material micro twists near the interface [18].

Recently, a question of the ground state of $PbZrO_3$ has attracted attention due to the rather odd discovery of ferrielectric state having lower energy than the well accepted AFE state at 0 K, according to first principles calculations [123]. Similar structures have been also experimentally discovered in $PbZrO_3$ thin films. Atomic scale TEM revealed the "↑↑↓" type $Pb^{2+}$ displacements and ferrielectric region inside the conventional AFE domains (with "↑↑↓↓" type $Pb^{2+}$ displacements), resulting in the coexistence of polar and nonpolar blocks [122]. To a certain degree, this further complicates our understanding on the following aspects. On the one hand, similar Pb displacement of "↑↑↓" feature has been observed at the $\boldsymbol{R}_{\text{II-1}}$-type translation boundary, which is featured as a displacement vector $\boldsymbol{R}[2\bar{1}n]$ with a $\pi/2$ phase shift for the separated but neighboring unit cells [207]. On the other hand, a long-range and periodic structure of such has not been observed in experiment so far. Therefore, this cannot be regarded as a pure phase but only as a local structure. Taking the discovered ferrielectric-like structure into account, it offers chances to achieve ultra-fine grid of polar entities within the nonpolar matrix, which can be used for multistate information storage. The control of domain states at nanometer scale (~10 nm) has been realized in FEs [208], and previous study has also revealed the preferential creation of polar translational boundary by interface engineering in $PbZrO_3$ thin films [155]. On the other hand, analogous to superior electrical conductivity identified at charged FE domain walls, one may also expect potential presence of conductivity difference between nanoscale polar regions and mesoscale non-polar regions. This may enable a flexible reading scheme in memory application if charged polar domains can be eventually realized and controlled at the nanoscale.

The first complete electric field-temperature phase diagram of bulk $PbZrO_3$ was proposed by Fesenko in 1978 (Fig. 7d) [56]. Besides the AFE orthorhombic phase and high-field FE rhombohedral phase, other intermediate low-field FE phases were predicted from the birefringence and dielectric characterization. However, how an AFE transforms into a FE under an electric field stimulus remains an open question, which relates to issues like the sample quality, the long recording time against the short phase-transition time and thus the difficulty of separating the intermediate phases at a narrow transient field regime. Until recently, the in-situ observation for AFE structural evolution is achieved in $PbZrO_3$ thin films [15]. With increasing electric field, the 1/4 superstructure diffraction damps, while weak 1/8 and strong 3/8 superstructure diffraction spots emerge. This indicates that a new type of AFE dipole modulation develops from fourfold to eightfold periodicity. By simulating the diffraction intensity with different dipole arrangement, the experimental results fit well with the "↑↑↑↓↑↑↓" dipole arrangement, revealing the stepwise dipole rotations during the AFE-FE phase transitions as shown in Fig. 7h-j. Interestingly, the intermediate state with an eightfold dipole periodicity exhibits another kind of ferrielectric feature, which covers the threefold ("↑↑↓") ferrielectric periodicity. This suggests a strong correlation between different dipole periodicity for the stepwise AFE-FE phase transitions and shed light on the electric field control of those nanodomains for memory device.

In fact, due to the low energy barrier between the AFE and FE phases, *e.g.*, in the $PbZrO_3$ system [209], the polarization/phase structure can be easily tuned by external stimuli, such as electric field, temperature, pressure, chemical doping, *etc*. Owing to the structure sensitivity, many new structures have been discovered recently in $PbZrO_3$-based crystals and thin films, all of which share intrinsic unit-cell-scale heterogeneity. Now it becomes possible to distinguish a few quite different groups of those.

First, the originally attractive incommensurate phases have shown a few interesting structural details. Contrary to the originally conceived transverse modulation waves, those phases appear now as modulated structures with certain degree uncompensated polarization. More surprisingly, the structure is accompanied with a longitudinal component in the modulation wave. Most of the above results were obtained in chemically (Ti and Sn) modified $PbZrO_3$. One of the most stunning discoveries is observation of the devil's staircase in electron diffraction [210]. Essentially, the modulation period for the transverse polarization wave is fixed at a series of commensurate positions. However, this may change in an abrupt manner upon when the external conditions are varied, which is even more generous in epitaxial thin films. This can be attributed to the synergistic effect of combined variables, *e.g.*, strain gradient and boundary conditions, which play as additional stimulus for flourish of the commensurate structure. For example, in $Pb_{0.97}La_{0.02}Zr_{0.95}Ti_{0.05}O_3$ epitaxial thin films, the phase stability can be controlled by thickness-related strain effect, presenting a depth dependent FE to incommensurate/commensurate AFE phase transition [211].

Second, the even more exciting structures are those with short-period modulations, especially in thin films. Actually, there is no direct observation of pure ferrielectric structure, only its fragments, like abundant but irregular appearance at the anti-phase domain walls and field-induced ferrielectric-like phase, where it sets the lone-displacement motion of more complicated "↑↑↑↑↓↑↑↓" structure. Model understanding of these short-period phases belongs to a different realm than that of the original incommensurate phases, despite they both have transverse lead ion displacements as the main feature. It is tempting to view the original incommensuration as weakly inhomogeneous polarization, possibly intermixed with other degrees of freedom [31]. On the other hand, the short-period modulations are obviously away from the ideal situation and need other perspectives for the description. One interesting approach that has been suggested [15] to simulate the displacement patterns in $PbZrO_3$ films is to use the Ising model with competing interactions [212], herein the coefficients are determined by first-principles calculations and can be slightly adjusted to account for defects in the material.

The third group consists of exotic structures, primarily manifested by FE cycloidal phase, which is activated by electron beam irradiation [213]. The instability under electron beam makes it difficult to capture the real ground state to some degree but provides opportunities to observe the pathway of electron beam induced phase transition. Different from the previous two groups, the main distortion takes place in the oxygen octahedral sublattices, while the anti-polar displacements of $Pb^{2+}$ ions in the original AFE phase are preserved [103]. Attempting to disentangle the mechanisms behind its formation leads to the appreciation of a stunning hierarchy of potentially realizable metastable phases, which can be, in principle, stabilized under various unusual conditions.

From these recent works, one can see that that the experimental progress goes far ahead of the theoretical one. Most identified structures are not directly accessible in first-principles-based modeling, and considerable breakthroughs are still needed on the theoretical aspect. Thin film specifics of $PbZrO_3$ makes the mission of theory even more complicated. For example, recent X-ray diffraction experiments on $PbZrO_3/SrRuO_3/SrTiO_3$ (001) heterostructures indicate a curiously frustrated state of the AFE domains there [18]. Instead of tilting by a fixed angle to comply with mechanical compatibility for neighboring domains, as observed in ordinary FEs, it is found that most of domains are non-tilted and thus should bear additional inhomogeneous mechanical stresses. Obviously, that is a huge complication to the theoretical modeling, where the $PbZrO_3$ films were commonly considered as only homogeneously strained [134] or not strained at all [214]. Unusual specifics of $PbZrO_3$ thin films, such as emergent transition order [16, 18], field-induced heterophase states with ferrielectric-like feature [15], and increased AFE transition temperature [14, 16], need to be disentangled in conjunction with appropriately modified theory that takes into account those important specifics.

#### 4.1.2. $PbHfO_3$

$PbHfO_3$ (PHO) is often considered as the second model of AFEs and is commonly overshaded by the research on $PbZrO_3$. Compared with the dielectric behavior and superstructure line from X-ray photography of $PbZrO_3$, PHO is recognized as AFE soon after the AFE study of $PbZrO_3$ [51]. The room temperature AFE phase of PHO has the same phase symmetry as that of $PbZrO_3$, while PHO has an even more complex phase diagram. The initial understanding of PHO is the isostructure of $PbZrO_3$ with *Pbam* symmetry. The electric field-temperature phase diagram of bulk PHO was given by *Fesenko* and *Balyunis*, suggesting similar multiphase transition versus electric field as occurred in $PbZrO_3$ [57]. However, from a modern perspective, PHO is remarkable in demonstrating that the incommensurate phase can be formed by antiferrodistortive soft mode [202]. The analysis of phase transitions in PHO driven by temperature and hydrostatic pressure unveils a quite different story from $PbZrO_3$ [215]. With temperature increasing, two intermediate phases are found to exist in two sequential temperature ranges below the transition temperature. The pressure increase can obviously affect the phase stability by altering the Pb-ion displacements and oxygen octahedral distortion. However, regarding the number of intermediate incommensurate phases, one or two, a consensus has not been reached so far [202, 216].

### 4.2. Lead-free perovskites

Traditional lead based AFE materials, such as $PbZrO_3$, still occupy most of the share in relevant research. However, these materials contain heavy metal lead, which is toxic to the environment and human body. Therefore, lead-free AFE materials with a perovskite structure such as $NaNbO_3$, $AgNbO_3$, $(Bi_{0.5}Na_{0.5})TiO_3$, rare-earth element doped $BiFeO_3$ thin films have attracted growing interest in recent years.

#### 4.2.1. $NaNbO_3$

Among lead-free AFEs, $NaNbO_3$ is particularly interesting due to the low cost of raw materials and rich structural motifs in the phase diagram. Being the same as $PbZrO_3$, $NaNbO_3$ was initially considered as one of the new FE materials [33, 217], since double hysteresis loop is even more difficult to be obtained in pure $NaNbO_3$. Because of the easy volatilization of Na at high temperature during sample preparation, it is very difficult to grow high-quality and stoichiometric $NaNbO_3$ samples [218]. To successfully verify its AFE characteristics, additional elements are usually introduced to stabilize the AFE structure, *e.g.,* $SrZrO_3$-$NaNbO_3$ solid solutions [219]. On this basis, a series of structural phase transitions as a function of temperature have been unveiled:

$$C \xrightarrow{639\,^oC} T_2 \xrightarrow{575\,^oC} T_1 \xrightarrow{519\,^oC} S \xrightarrow{479\,^oC} R \xrightarrow{359\,^oC} P \xrightarrow{-100\,^oC} N$$

At high temperature, $NaNbO_3$ is PE with a cubic structure and space group of $Pm3m$ ($C$ phase). It undergoes a phase transition to another PE phase while with a tetragonal structure and space group of $P4/mbm$ ($T_2$ phase) at a transition temperature of 639 °C. With further temperature decreasing, several orthorhombic phases appear, including $Cmcm$ ($T_1$ phase) at 575 °C, $Pmnm$ ($S$ phase) at 519 °C, $Pmnm$ ($R$ phase) at 479 °C, and to *Pbcm* (*P* phase) at 359 °C [220, 221]. These different orthorhombic phases are distinguished by two factors, *i.e.*, the atomic displacement of $Nb^{5+}$ and the oxygen octahedral rotation patterns.

Among these orthorhombic phases, the *P* and *R* phases are AFE, while S and $T_1$ phases are PE [220, 222]. At -100 °C, the $NaNbO_3$ transits into a rhombohedral FE *R*3*c* phase. Temperature-dependent dielectric permittivity of a pure $NaNbO_3$ ceramic shows that the largest dielectric anomaly occurs at the *P*-to-*R* phase transition (Fig. 8a) [223]. However, recent studies argued that the most stable phase near room temperature is not the AFE *P* phase, but rather another orthorhombic FE phase with space group $Pca2_1$ (*Q* phase) [224]. Furthermore, coexistence of FE *Q* and AFE *P* phases was reported, making the ground-state structure of $NaNbO_3$ more controversial [219, 225]. According to first-principles calculations, the energy difference between the AFE and FE phases is remarkably small (~1.4 meV/f.u.) [224, 226]. Due to the small energy difference between the competing phases, the structure of $NaNbO_3$ can be easily tuned by chemical composition, defects, electric field, temperature, epitaxial strain, film thickness and grain size. For instance, it was found that tensile-strained $NaNbO_3$ epitaxial films with thickness of 250 nm are a FE *Q* phase rather than AFE at room temperature [224, 227]. Therefore, $NaNbO_3$ has been dubbed as the most complex perovskite system [224].

#### 4.2.2. $AgNbO_3$

$AgNbO_3$ is another important lead-free AFE material that has attracted considerable interest in the field of energy storage and conversion. Similar to $NaNbO_3$, $AgNbO_3$ also exhibits a complex structural phase diagram. In 1958, Francombe *et al.* first successfully synthesized $AgNbO_3$ ceramics and several dielectric anomalies associated with sequential structural phase transitions were identified [52]. Based on the temperature dependent structural and dielectric measurements, the authors suggested that $AgNbO_3$ is possibly an AFE at room temperature. However, the evidence of a small remnant polarization, obtained from hysteresis loops and the pyroelectric effect, indicated the presence of weak ferroelectricity [52, 228], which was further confirmed in subsequent experiments [229]. With the advancement of structural characterization techniques, it is generally believed that $AgNbO_3$ undergoes a series of at least five reversible structural phase transitions (Fig. 8b) [230], described as follows:

$$M_1 \xrightarrow{67\ ^{\circ}\mathrm{C}} M_2 \xrightarrow{267\ ^{\circ}\mathrm{C}} M_3 \xrightarrow{353\ ^{\circ}\mathrm{C}} O(O_1, O_2) \xrightarrow{387\ ^{\circ}\mathrm{C}} T \xrightarrow{579\ ^{\circ}\mathrm{C}} C,$$

where both $M_2$ and $M_3$ phases are AFE, the orthorhombic *O* phase, tetragonal *T* phase and cubic *C* phase are PE. So far, the exact phase symmetry of the room-temperature $M_1$ phase is still under debate. The transitions among the *M* phases are related to cation displacements, while the high-temperature $M_3 \leftrightarrow O$, $O \leftrightarrow T$, and $T \leftrightarrow C$ phase transitions have been attributed to octahedral-tilting instabilities.

In 2007, the first complete double hysteresis loop with a large polarization of 52 $\mu C/cm^2$ was obtained by Fu *et al.* in $AgNbO_3$ ceramics, where electric field is applied up to 220 kV/cm [231]. However, a slim FE hysteresis loop with a small remnant polarization was also observed under a weak electric field of 80 kV/cm [231], leading to controversy in the structure of $M_1$ phase. In 2009, using a combination of high-resolution X-ray diffraction, neutron total scattering, electron diffraction, and X-ray absorption fine-structure measurements, Levin *et al.* reported that the $M_1$ phase exhibits orthorhombic structure with centrosymmetric *Pbcm* space group [232]. They further demonstrated that partial ordering of Nb displacements induced by coupling between $(a^-b^-c^-)/(a^-b^-c^+)$ octahedral-tilting and local polar Nb displacements dominate the structural behavior of the *M*-phase polymorphs [232]. However, the centrosymmetric structure is not consistent with the presence of remnant polarization. Later in 2011, by using convergent-beam electron diffraction, neutron and synchrotron powder diffraction and first principles calculations, Yashima *et al.* revealed that $AgNbO_3$ is FIE at room temperature and has a non-centrosymmetric orthorhombic structure with space group of $Pmc2_1$ [233]. In contrast to the nonpolar *Pbcm*, the atomic displacements along the *c* axis in the FIE phase are antiparallel aligned but with different magnitudes, leading to a net polarization at zero field in $AgNbO_3$ [233]. Overall, despite of some disputes in structure, its macroscopic features are basically the same and all $M_1$, $M_2$ and $M_3$ phases show the characteristic double hysteresis loops, which become slimmer with increasing temperature (Fig. 8c) [234]. Compared with bulk, $AgNbO_3$ thin films are poorly studied mainly because of the difficulty in synthesizing high-quality and phase-pure samples [235, 236]. Although both AFE and FE loops are observed, it is hard to distinguish whether this behavior is intrinsic or not. An important reason lies in the fact that impurity phases, mainly arising from decomposition of $AgNbO_3$, are prone to appear during the film deposition [237].

#### 4.2.3. $(Bi_{0.5}Na_{0.5})TiO_3$

$(Bi_{0.5}Na_{0.5})TiO_3$ (BNT) is first found by Smolensky *et al.* in 1960 [238] to be a rhombohedral perovskite FE with space group of $R3c$ at room temperature, and exhibits an AFE behavior at temperature between 200 °C and 320 °C (Fig. 8d) [239]. BNT has received considerable attention in recent years because of its potential interest as a lead-free piezoelectric ceramic. In the structure of BNT, the A site is occupied by $Bi^{3+}$ and $Na^+$ ions with equal concentration to balance the electroneutrality [240, 241]. Similar to $Pb^{2+}$, lone-pair electrons in the outermost $6s^2$ layer of Bi can hybridize with the $6p$ vacant orbital or $O^{2-}$ orbital to produce strong electron cloud polarization [242]. In 1974, AFE-like double hysteresis loops were first found in BNT-$SrTiO_3$ solid solutions at a high temperature of 170 °C [37]. Later, it is reported that BNT undergoes rhombohedral ($R3c$, $a^-a^-a^-$) → tetragonal ($P4bm$, $a^0a^0c^+$) → cubic phase transitions with increasing temperature, and the tetragonal phase exhibits an unusual combination of in-phase ($a^0a^0c^+$) tilts and antiparallel cation displacements along the polar $c$ axis, indicating the AFE feature [243, 244].

For the room temperature FE structure, recent high resolution X-ray diffraction experiment suggests that the structure of BNT is monoclinic ($Cc$, $a^-a^-c^-$) [245, 246]. Further study reveals that the electric field can induce monoclinic ($Cc$) to rhombohedral ($R$3c) phase transformation, and rhombohedral ($R3c$) and monoclinic ($Cc$) phases may coexist in its equilibrium state at room temperature [247, 248]. Concerning the fine structure evolution during temperature increasing, V. Dorcet and Trolliard *et al.* conducted a reinvestigation of phase transition and proposed that BNT has a sequence of several structural phase transitions upon heating [249, 250], as given below:

$$R(R3c,\ \mathrm{FE}) \xrightarrow{200\ ^{\circ}\mathrm{C}} R+O(Pnma+R3c,\ \mathrm{AFE}) \xrightarrow{280\ ^{\circ}\mathrm{C}} O(Pnma,\ \mathrm{AFE}) \xrightarrow{320\ ^{\circ}\mathrm{C}}$$

$$T(P4/mbm+P\tfrac{4}{2}/mnm,\ PE) \xrightarrow{520\ ^{\circ}\mathrm{C}} C(Pm\bar{3}m).$$

In order to shift the AFE transition to room temperature, a second perovskite phase has been extensively incorporated into the BNT matrix, such as BNT-$BaTiO_3$ (BNT-BT) [251], BNT-$NaNbO_3$ [252], BNT-$SrTiO_3$ [37], *etc*. Among them, BNT-BT solid solution is particularly interesting, which not only broadens the temperature region of AFE phase, but also greatly improves the piezoelectric, pyroelectric and energy storage performances due to the inherently rich phase boundaries (Fig. 8d) [239].

**4.2.4. (Bi,Re)$FeO_3$**

$BiFeO_3$ (BFO) is the most studied and important multiferroic material that was first discovered by Royen and Swars in the late of 1950s [253]. The crystal structures and physical properties of BFO are very sensitive to electric field, pressure, and strain, which offers great flexibility to adjust its crystal structure, even the generation of an AFE order [254, 255]. Fig. 9a shows various low-energy phases of BFO, including the ground state FE structure with *R*3*c* symmetry, and two antipolar states with *Pbam* and *Pnma* symmetry. As shown in Fig. 9b, epitaxial strain can modify the relative stability of the AFE phases with respect to the FE *R*3*c* phase, while the AFE phase remains metastable for all lattice constants. Recently, several studies have demonstrated that AFE phases can be stabilized in BFO/$LaFeO_3$ superlattices and $BiFeO_3$/$La_{0.4}Bi_{0.6}FeO_3$ superlattices (Fig. 9c) [188, 256]. The interfacial electrostatic engineering can liberate the otherwise metastable AFE state of BFO (Fig. 9d), and the stability of the AFE phase can be further enhanced by tuning the dielectric constant of the neighboring layers [188].

Besides interfacial engineering, chemical substitution can also stabilize the AFE phase of BFO at room temperature [257], as illustrated by the *P*-*E* hysteresis curves of Nd doped $BiFeO_3$ solid solutions (Fig. 9e). By decreasing the A-site ion size, the introduced chemical pressure enhances tilting of the oxygen octahedra and favors the antipolar structure [258]. Fujino *et al.* discovered a morphotropic phase boundary (MPB) with enhanced piezoelectric response in Sm-doped $BiFeO_3$ thin films, where a FE rhombohedral to an AFE pseudo-orthorhombic phase transition takes place at ~14% Sm doping content [259]. Furthermore, subsequent studies confirmed the universal behavior in structural and electric properties in BFO doped with various rare-earth elements, including $La^{3+}$ [260], $Dy^{3+}$ [261], $Nd^{3+}$ [257, 262]. Fig. 9f shows the phase diagram for rare-earth element substituted $BiFeO_3$, and composition of the MPB can be tuned by systematically adjusting the average A-site ionic radius. This proves that the chemical pressure of the A-site substituted cations can drive the formation of MPB. Although the detailed crystal structure of the new phase is still controversial, the existence of antipolar phase in BFO is beyond doubt, which introduces antiferroelectricity to multiferroics in the doped BFO system.

### 4.3. Non-perovskite antiferroelectrics

#### 4.3.1. Fluorite-structured $Hf_{1-x}Zr_xO_2$

The AFE materials of fluorite structure mainly refer to $HfO_2$-based and $ZrO_2$-based thin films. The ferroelectricity and antiferroelectricity of fluorite structure was first discovered in 2007 and experimentally verified in Si-doped $HfO_2$ thin film in 2011 [263]. Then, the AFE behaviors were obtained in other doped $HfO_2$ and pure $ZrO_2$ thin films [264, 265]. Compared with perovskite-type AFE materials, these fluorite-type AFE materials show more compelling advantages in CMOS compatibility and integration [266, 267]. Therefore, considerable attention has been paid into these fluorite-type materials.

$HfO_2$ is a typical example of fluorite-type AFE materials. As shown in Fig. 10a, $HfO_2$ thin films undergo monoclinic ($M$) → orthorhombic ($O$) → tetragonal ($T$) → cubic ($C$) phase transition with increasing formation energy [268]. $HfO_2$ thin films can be dominated by the metastable state at ambient conditions thermodynamically and kinetically by doping [269], stress [270], defects [271], thickness [272], and electric field [273]. Fig. 10b briefly describes $HfO_2$ samples with different Zr contents under various thicknesses [28]. It can be inferred that the FE phase is more stable in $HfO_2$ while $ZrO_2$ films prefer to be AFE in ultrathin film thickness [274]. To achieve an AFE behavior, extra $ZrO_2$ are usually introduced to stabilize the metastable state, and the electric field induced nonpolar tetragonal → polar orthorhombic phase transition is widely accepted as the origin of AFE behaviors in the fluorite-type material [275]. Recent atomic-resolution Cs-corrected scanning TEM study revealed that antipolar orthorhombic *Pbca* phase with antiparallel oxygen-atom shifts coexist with the FE orthorhombic $Pbc2_1$ phase in $(Hf_{0.5}Zr_{0.5})O_2$ polycrystalline film (Fig. 10c) [276]. The energy between this antipolar phase and the FE phase is slightly different, which makes the reversible phase transition possible [128]. By low voltage cycling, the antipolar phase can be further stabilized and replace the FE $Pbc2_1$ phase, which causes the fatigue in pristine FE thin films [276]. Noting that this kind of fatigue is attributed to the polar-antipolar phase transition, but not to oxygen loss or domain wall pinning, and the fatigued sample can be rejuvenated by applying a slightly higher voltage. Considering the feature of domain walls, this antipolar phase is equivalent to 8-fold 180° domain walls, which can be seen as a typical sub-lattice property of $HfO_2$-based structure [277]. Therefore, the antipolar orthorhombic phase is fundamentally distinguished from the non-polar tetragonal AFE phase. In addition, as schematically shown in Fig. 10d, there is another AFE-like behavior in the doped $HfO_2$ films, manifested by typical double-hysteresis curves. However, this kind of behavior is not stable and only exist for a few cycles, which is also termed as the wake-up process [278]. The reasons that can account for this phenomenon are manifold, *e.g.*, the internal bias voltage caused by oxygen vacancy or local phase inhomogeneity [279, 280], depolarization field, phase transformation and pinned dipoles [281]. Other than this, other open questions, such as the switching kinetics, the origin of high switching fields $E_{AF}$, and so on. Although the origin of $Hf_{1-x}Zr_xO_2$-based antiferroelectricity and its structure details are still obscure, it indeed becomes one of the most promising materials for future utilization in microelectronic system, not only because of its compatibility with the CMOS technique, but also because of its excellent fatigue resistance in nonvolatile random access memory, energy storage, and negative capacitance [62, 282].

#### 4.3.2. Other antiferroelectrics

Apart from the conventional perovskite- and fluorite-type oxides, antiferroelectricity was reported to exist in a wide range of unconventional structures as well, which highly broadened the AFE family and understanding of antiferroelectricity. $(NH_4)H_2PO_4$ (ADP) is a H-bonded compounds reported to have an AFE phase under 148 K [283], and the transition temperature strongly depend on the isotope effect of H. La-doped M-type hexaferrites ($La_{0.2}Sr_{0.7}Fe_{12}O_{19}$) was proved to integrate antiferroelectricity and ferromagnetism as a novel multiferroic candidate in recent years [284]. La substitution can divide the lattice into different pieces, leading to the generation of nano-domains with opposite polarization [285]. In Ruddlesden-Popper structure, a series of $A_3B_2O_7$ type compounds are proposed as a hybrid improper AFE, such as high-temperature antipolar *Pnab* structure $Sr_3Zr_2O_7$ [286]. This greatly promoted the exploration of lead-free AFEs. However, clear AFE double hysteresis loops have not been reported yet probably because of the relatively low bandgap of the material [287].

Specifically, it is worth mentioning that francisite $Cu_3Bi(SeO3)_2O_2Cl$ is a newly reported mineral and undergoes AFE phase transition under 115K with dielectric anomaly [30]. Different from the conventional perovskite-type AFEs which have multiple soft modes [43, 202, 288], the antiferroelectricity in francisite is driven by softening of a single antipolar mode [30], resulting in antiparallel displacement of adjacent Cu and Cl ions. This material is more similar to ideal Kittel's model of antiferroelectricity [25] and offers a good equivalence of antiferromagnets. It opens the door to trace not only the counterpart phenomenon that have been already discovered in antiferromagnets, but also the magnetoelectric coupling for antipolar magnetoelectric excitations in magnetism-rich francisite systems.

Apart from the well-recognized AFEs in inorganic oxides and hydrogen-bonded organic compounds, the exploration of AFEs in organic-inorganic hybrids has made great achievements and greatly enriched the types of AFEs. In 2017, Li *et al.* [78], reported excellent FE and AFE performances in (3-pyrrolinium) $CdBr_3$. In this organic-inorganic hybrid perovskite, the AFE phase can be seen as an intermediate phase between the FE phase (<235.2K) and PE phase(>244.2K). Despite the AFE only exists at a narrow temperature gap (235.2~244.2K) below room temperature, it is the first discovery of the succussive FE−AFE−PE phase transitions in organic-inorganic hybrid perovskites and sheds light on the importance of exploring AFEs based on organic-inorganic hybrid compounds. In 2019, Wu *et al.* [289], reported an above-room-temperature AFE in organic-inorganic Ruddlesden-Popper hybrid perovskite, (i-BA)$_2$CsPb$_2$Br$_7$ (i-BA is isobutylammonium). The antiferroelectricity can be attributed to the synergistic dynamic motion of organic i-BA cations and inorganic Cs atoms. In this compound, the antiferroelectricity exists at a relative wide temperature range (298~353K) and AFE-PE phase transition occurs at 353 K. The discovery of room temperature AFE promotes the understanding of structure-property relationships and future application demonstration. Because of their great chemical diversity and structural tunability, the hybrid organic-inorganic perovskite AFEs are expected to show significant potential in energy storage and electrocaloric cooling applications [289, 290].

## 5. Applications of antiferroelectric thin Films

### 5.1. Energy-storage capacitors

Dielectric capacitors possess faster charge-discharge speed (∼ns), higher power density (up to $10^8$ W/kg), better temperature stability and fatigue resistance, compared with fuel cells, batteries, and supercapacitors, and thus are widely used in electronic devices and electrical power systems [291]. However, capacitors generally have low energy densities (typically less than 10 J/cm$^3$) relative to other energy storage systems. The recoverable energy storage density $J_{reco}$ of an electrostatic dielectric capacitor under an applied electric field $E$ [291]: $J_{reco}=\int_{P_r}^{P_{max}} E dP$, where $P_{max}$ and $P_r$ are the maximum and remnant polarization, respectively. For dielectric capacitors, large breakdown field $E_{BDS}$, small $P_r$, large $P_{max}$, and slim hysteresis (small $J_{loss}$) are needed for high energy storage density and efficiency $\eta$ [292, 293]. Among dielectric materials, including linear dielectrics (LDs), FEs, and relaxor FEs (RFEs), AFEs, AFE materials afford advantages for high $J_{reco}$ by steeply increasing the polarization at a relatively high electric field $E_{AF}$ and fully releasing the stored charge at another high electric field $E_{FA}$, as shown in Fig. 11a.

The charging-discharging speed in AFE thin films is important for energy storage application. The switching dynamics analysis based on the nucleation-limited-switching model has reveal a characteristic switching time as short as 3 ns [91], indicating an ultrafast charging time in AFE films. In addition, the discharging time related to energy release is also important for practical applications. In AFE $Pb_{0.97}La_{0.02}Zr_{0.85}Sn_{0.13}Ti_{0.02}O_3$ films [294], the discharging time was determined to be within 6 ns and a maximum switching current density was achieved as high as 9400 A/cm$^2$, which make AFE thin films the great candidate for pulsed power applications in high-speed multichip modules.

Although AFE thin films exhibit high potential in high energy density and pulse power capacitor applications, the state-of-the-art AFE capacitors currently still face the challenge of low efficiency ($\eta=J_{reco}/J_{total}$) caused by large hysteresis.

To have higher $J_{reco}$, and larger $\eta$, various strategies are proposed to optimize the energy storage characteristics in recent years. Here, we classify these strategies as follows:

① Reducing hysteresis. As shown in Fig. 11a, the green shaded aera corresponds to the energy storage density $J_{reco}$. Obviously, increasing $E_{AF}$ and $E_{FA}$ can increase the green shaded area, that is, increase the energy storage density $J_{reco}$. On the other hand, reducing the difference between $E_{AF}$ and $E_{FA}$, *i.e.*, reducing electric hysteresis, can decrease the hysteresis area, that is, reduce the energy loss density $J_{\rm loss}$ and increase the energy storage efficiency. Previous studies have demonstrated that introducing nanoscale structural heterogeneity into AFEs generally via chemical doping is an effective approach to lower the energy barrier between the AFE and field-driven FE phases and thus reduce the electric hysteresis [158, 295-297].

② Increasing saturation polarization or decreasing remnant polarization. The overall polarization can be modulated by introducing high-polarization phase [298], interface engineering [299], strain engineering [206], and orientation control [20]. Usually, introducing relaxor into AFE results in lower polarization under the same electric field.

Therefore, a synergistic approach with engineered structural heterogeneity and meticulous orientation control can be applied to improve the energy storage density Fig 11b. The application of structure heterogeneity can be seen as the origin of high energy storage efficiency by minimizing the difference between $E_{AF}$ and $E_{FA}$, and the orientation control can effectively enlarge the polarization thus result in a high energy storage density Fig 11c.

For instance, Zhang *et al*. [297] demonstrated that the combination of structure heterogeneity and orientation control in relaxor AFE $(Pb_{0.97}La_{0.02})(Zr_{0.55}Sn_{0.45})O_3$ thin films can simultaneously achieve high energy storage efficiency (~98.5%) and energy storage density (~84.45J/cm$^3$). In that work, introducing $La^{3+}$ and $Sn^{4+}$ dopants into AFE $PbZrO_3$ can engineer the structural heterogeneity and disrupt the long-range ordered AFE domains, which can effectively decrease the polarization hysteresis ($E_{AF}$-$E_{FA}$) [297]. By controlling the orientation, the (111)-oriented films exhibit largest polarization among (111)-, (110)- and (001)-oriented films and gain the highest energy storage density (~84.45J/cm$^3$) [297].

③ Increasing $E_{BDS}$. There are many factors that can affect $E_{BDS}$, such as band gap ($E_g$) [300], porosity/cavity [301], grain size [302], thickness [303], sample geometry [304], and *etc*. The intrinsic $E_{BDS}$ of a semiconductor can be calculated by [305]:

$$E_{BDS}=1.36\times10^{7}\times\left(\frac{E_g}{4}\right)^{3}\ \text{(V/cm)} \qquad (10)$$

Thus, bandgap engineering is an effective method to improve energy storage properties [306]. Besides, $E_{BDS}$ is inversely proportional to the square root of thickness, $d$ [307]:

$$E_{BDS} \propto d^{-1/2} \qquad (11)$$

As a result, thin films [308] show much higher energy density than bulk materials due to the larger $E_{BDS}$, as shown Fig. 11c [172, 234, 237, 252, 296, 302, 309-338] and 11d [21, 90, 100, 145, 182, 235, 339-355].

④ Delaying the saturation of polarization: Imprint is related to the presence of an internal (built-in) electric field within the FE/AFE films, which causes a shift of the P-E loop along the field axis [132, 356]. It has been reported that the formation of defect dipoles by ion bombardment can induce the imprint and delay low-field polarization saturation in RFE $Pb(Mg_{1/3}Nb_{2/3})O_3$-$PbTiO_3$ thin films [292]. Besides, the hysteresis loop shift by imprint effect can effectively enhance the $E_{BDS}$ and thus the $P_{max}$. This has been utilized to improve the energy storage density in RFE thin films [292, 357]. This sheds light on the role of imprint in modifying the AFE hysteresis loops and thus optimization of the energy storage density.

⑤ Improving the stability of AFEs. It is known that the tolerance factor $t$ is a critical indicator for the phase stability of perovskite structure. AFE phase in perovskite oxides can be stabilized by decreasing $t$ within the range of perovskite structural stability [358]. Various doping experiments are conducted to improve the stability of AFE phase, in accompany with the increasing of $E_{AF}$, which shows great values in energy storage applications.

Eventually, by disrupting long-range ordered AFE domains, relaxor antiferroelectrics (RAFEs) can reduce hysteresis upon applying and removing electric fields due to the fast response of nano-domains [296], RAFEs, analogous to RFEs, are a promising solution to overcome this long-standing challenge. That is to say, accompanied with large $E_{BDS}$, small $P_r$ and small hysteresis enable the simultaneous enhancement of large $J_{reco}$ and $\eta$ [296]. This can lead to RAFE one of the most promising dielectric energy storage materials. Although there are many strategies to improve the energy storage characteristics of AFE materials, there is still a big gap between material engineering and practical devices, and further efforts are needed.

Essentially, the following aspects can be further explored: ① From the perspective of environmental protection and human health, lead-based dielectric materials should be replaced in practical application, and more attentions should be paid to developing lead-free materials with good energy storage performance, such as $NaNbO_3$ based, $AgNbO_3$ based, $BiFeO_3$ based compounds. ② From the perspective of practical applications, current studies mainly focus on improving the energy storage density and efficiency [21, 345, 359], while the charging and discharging speed, fatigue performance and thermal stability of thin films are often overlooked. These parameters are also very important for the application in capacitors, so they should be taken into account in future studies. ③ AFE thin films exhibit a huge potential for energy storage due to the great improvement of $E_{BDS}$ (~2-5 MV/cm), which is more than one order of magnitude higher than that in bulks. Exploration and applications of AFE thin films are desired to meet the demand of high performance in microelectronic devices.

### 5.2. Electrocaloric cooling

The electrocaloric effect (*ECE*) is a phenomenon in which a dielectric material shows a temperature change under an external electric field. The effect can be understood as follows. For dielectrics, their electronic bandgap is typically large, and the electronic entropy $S_e$ can be ignored. Thus, the total entropy $S$ has two major components: dipolar or configurational entropy $S_c$ and vibrational entropy $S_p$. The fast phase transition under an external electric field (a few nanoseconds [360, 361]) can be treated as adiabatic and reversible without heat exchange. As a result, the total entropy change is zero, $dS = dS_c + dS_p = dQ/T = 0$, where $T$ is the temperature and $Q$ is the absorbed heat during the switching process. Since applying an electric field typically aligns the diploes and decreases $S_c$, $S_p$ has an increase to keep the total $S$ unchanged. The increase of $S_p$ causes a temperature change $\Delta T$, adiabatic temperature change. Both direct and indirect methods have been employed to measure $\Delta T$ [362-364]. The electric-field-triggered $\Delta T$ is defined as *ECE*, which can be explored to realize the electrocaloric cooling and replace the traditional vapor compression refrigeration. The first *ECE* measurement of AFE materials was reported in 1968 by *Thacher* [365]. That early work has shown positive *ECE* with small magnitude of $\Delta T$. In 2011, the negative ECE in AFEs was observed for the first time [366]. Since then, the ECE of AFEs has attracted broad attention due to the large negative $\Delta T$ [6, 367-369]. As shown in Fig. 12a, the left panel presents the temperature $T$ dependent $P$, $\Delta P$ and $\Delta T$ for AFEs (purple lines) and FEs (red lines), while right panel presents the electric field $E$ dependent $S$, $\Delta S$ and $\Delta T$ for AFE (purple lines) and FE materials (red lines). Noting that entropy in the following refers to total entropy unless otherwise specified. Clearly, for AFEs, with temperature increasing, there is a transition from negative to positive ECE and the negative ECE can be attributed to the phase transition induced instability of AFE dipoles. In contrast, the polarization of FE always decreases with increasing temperature, leading to a positive ECE. From the right panel of Fig. 12a, similar to the temperature induced ECE evolution, the electric field driven ECE also evolves from negative to positive for AFEs. The negative ECE is generally attributed to the non-collinear response of dipoles to an external electric field [6], which increases the entropy and decreases the temperature. While a recent study argued that the negative ECE actually originates from the latent heat related to phonon entropy of the AFE-FE phase transition, rather than the non-collinear response of dipoles [370]. The inconsistency in the origin of the negative ECE in AFEs indicates that there is still a lot of work to be done. On the contrary, for FEs, electric field always decreases the entropy and induces the positive ECE.

For practical applications of the electrocaloric cooling technology, large $\Delta T$ and $\Delta S$ are needed to guarantee the cooling energy density $Q$. The negative and positive ECE in AFEs can be combined to increase the value of $Q$ [371-374]. An efficient routine is raised recently by Li *et al.* [371]. and is presented in Fig. 12b. Specifically, at temperature $T_0$, the PE phase is transformed into FE phase under $E_1$ along with a temperature increase of $\Delta T_1$, that is positive ECE; then contacting the EC material with the external environment, returning the material temperature to $T_0$. Subsequently, removing $E_1$, decreasing the temperature to $T_0$-$\Delta T_1$ and returning the phase to PE. If $\Delta T_1$ is large enough to transform the PE phase to AFE phase, the ECE would become negative; in such case, applying a modest $E_2$ decreases the temperature to $T_0$-$\Delta T_1$-$\Delta T_2$ with AFE phase transformed into FE phase. After that, removing $E_2$, the ordered FE dipoles becomes disordered first, causing an entropy increase and a temperature decrease $\Delta T_3$. Thus, the total $\Delta T$ is $\Delta T_1$+$\Delta T_2$+$\Delta T_3$. Notably, $\Delta T$ for combing negative and positive ECE is $\Delta T_1$+$\Delta T_2$+$\Delta T_3$, which is larger than $\Delta T_1$ for only positive ECE. (noting that $\Delta T_1$, $\Delta T_2$, and $\Delta T_3$ are all positive values) As a result, many research works have been carried out to investigate the (negative) ECE in the AFE materials. Fig. 12c-d summarize the ECE of different AFE material systems, which clearly indicate that with increasing thickness, ECE strength $\Delta T/\Delta E$ decreases significantly [6, 40, 290, 366-369, 371, 373, 375-402]. For example, bulk PLZT system has an ECE strength as large as 0.25 K cm/kV while the PLZT film system only has a value around 0.025 K cm/kV, almost one order of magnitude smaller. From Fig. 12c-d, we can also see that for bulk systems, most studies observed the negative ECE in AFE materials. However, for thin film systems, the ECE phenomena is complex since both negative ECE and positive ECE can be observed [369, 384]. For the unexpected positive ECE in AFE materials, the underlying mechanism could be the AFE-FE or AFE-PE phase transition induced entropy decrease, FE-AFE or FE-PE phase transition, strain effect, substrate effect, and defect, or just the error of indirect measurement. Another reason could be the different nanoregion distributions in different material systems. The underlying physical mechanism about the opposite ECE phenomena in AFE materials is still unclear and needs more investigations [392, 397].

As stated in Section 3.5, the frequency of the external electric field significantly influences the ECE. For PZO films, Wu *et al*. [165], obtained a negative ECE at high frequency (10 kHz) and a positive ECE at low frequency fields (100 Hz). The underlying mechanism is that the domain nucleation and growth is not sufficient at high frequency. In such case, with increasing temperature, the thermal energy promotes the domain nucleation and decreases the pinning force of domain wall movement, leading to the increase of $P$ and negative ECE. While at low frequency, there is enough time for the domain nucleation and growth. As a result, the spontaneous polarization reduces with increasing temperature, and therefore the polarization of the electric-field-driven FE phase reduces, leading to the positive ECE. In addition to frequency, many other methods have been applied to modulate the ECE such as doping [6, 371] and stress [382].

### 5.3. Electrostrain-based applications

Although AFEs lack piezoelectricity due to the presence of the center of symmetry, a sharp deformation can also be induced by phase transitions from AFE to FE phase and AFE to PE phase under electric field or heating, especially in the direction along the field [403]. The volume change accompanied with field-driven phase transition is regarded as the main reason for the electric field induced strain in AFE materials other than domain reorientation, inverse piezoelectric effect, or electrostriction [89]. Recent studies have demonstrated that large strain level of ~0.3-1% can be obtained in AFE $PbZrO_3$ thin films, as shown in Fig. 13a, which is several times larger than traditional FE and electrostrictive materials [404-406]. As early as 1968, *Berlincourt et al.* analyzed the feasibility of using AFE material to make underwater acoustic transducer, and pointed out that transducers based on AFE material own advantages of simple structure, small size, and light weight [407]. Recently, actuators based on AFE thin films have attracted great attention due to its unique structural phase transition. The electric field induced strain can be directly probed via piezoresponse force microscopy as shown in Fig. 13b and 13c [408]. By controlling the crystalline orientation, the giant electrostrain in $PbZrO_3$ thin film is reported, which is strongly anisotropic with respect to the electric field direction as shown in Fig. 13e-f. Apart from the reversible giant electrostrain in conventional AFEs, for some doped AFE compositions with a field-induced metastable FE phase, electric field induced strain can be maintained even after removal of the electric field, which can set back to the initial zero state by heating over the transition temperature. This interesting phenomenon is the so-called shape memory effect [409]. Moreover, AFE exhibited a quicker response time (~100 kHz) than traditional shape memory metallic alloys (~100 Hz), thus, AFE-based devices are promising for practical industrial applications, such as digital displacement transducers.

### 5.4. Negative capacitance

Negative capacitance has been intensely researched in recent few years due to its potential merits to overcome the so-called *Boltzmann* tyranny in conventional transistor and enable the application for ultralow power electronics [410]. For nonlinear dielectric capacitors, capacitance referred to differential capacitance:

$$C=dQ/dV \qquad (12)$$

which is different from the liner dielectric capacitance defined as:

$$C=Q/V \qquad (13)$$

In typical FE, negative capacitance behavior was theoretically predicted by *Landauer* in 1976 [411]. In the free energy landscape, considering the electrical origin of free energy, the applied electric field $E$ can be obtained by:

$$E=dG/dP \qquad (14)$$

where $G$ denotes the free energy of the systems expanded by $P^2$, $P^4$, *etc.* and $P$ denote the polarization. As a result, the differential term can be derived as:

$$C= dQ/dV \sim dP/dE\sim 1/(d^2G/dP^2) \qquad (15)$$

The applied electric field can increase one of the two energy minima with opposite directions in $G(P)$ dependence and drive the polarization to the other one. During the polarization alignment process, the FE state passes through a transient state characterized as:

$$d^2G/dP^2<0 \qquad (16)$$

which indicates the presence of negative differential capacitance. However, it is difficult to directly observe the negative capacitance because of its metastable nature in conventional parallel capacitor devices. Recent results reveal that resistance and dielectrics in electric circuit can slow down the charging process and help to stabilize the negative capacitance state [412-414]. In fact, the negative capacitance was not exclusive in FEs but prevails in any two-state system with an intrinsic energy barrier between states [415].

While S-shaped *P-E* curves have been derived in the Landau–Ginzburg–Devonshire (LGD) theory framework with the double-well free energy landscape $F$ as a function of polarization $P$ [416-418], this phenomenological model assumes the presence of depolarized state and has its limit on the cases of multidomain FEs [419, 420], and polycrystalline FEs where the polarized states are spatially remained [417]. To solve this issue, Park *et al.* proposed an alternative model, *i.e.,* the inhomogeneous stray-field energy (ISE) model to explain the negative capacitance in multidomain FE films [421]. The ISE model suggests that the stray field between neighboring domains contributes to the inhomogeneous electrostatic energy. By assuming that there is no charge injection, the switching of the relative portion of effective upward and downward domains can change the inhomogeneous stray field and its electrostatic energy, which can result in an energy curve with negative curvature near $P \sim 0$. This model succeeds in explaining the observed negative capacitance in several experimental results [418, 421], and in prediction of double S-shaped curve in extended ISE model, which considers the distribution of compensating charges, as verified by experiments [422]. A metal-FE-metal-dielectric-metal structure, where the stray field is cancelled by inserting a metal between dielectric and FE layers, are developed to reinvestigate the possible negative capacitance. The suppression of negative capacitance effect in the structure can be fully explained by the ISE model, while LGD model predict the inconsistent results [423].

As compared with FE, the negative capacitance and its mechanism in AFE received less attention. AFEs can be seen as a well-acknowledge two-state system in a certain degree with respect to the electric field induced AFE-FE transition and the free energy landscape in AFE is shown in Fig. 14a [62]. The ground AFE state can be perturbed by electric field and then a metastable FE state can be induced. Specifically, as the electric field increases, a transient state emerges, which indicates the presence of negative capacitance, shown in Fig. 14b [62]. The negative capacitance in AFE was firstly reported in $PbZrO_3$ thin films (Fig. 14c) [45]. Static negative capacitance and transient capacitance are characterized by the capacitance enhancement and voltage drop during the current charging, respectively. The negative capacitance in $PbZrO_3$ thin films was proposed to be four times than that in FE due to the rich local regions of negative capacitance effect, which makes it more competitive in future application. AFE negative capacitance was also realized in fluorite-structured AFE $ZrO_2$ thin films very recently (see Figs. 14d-e) [62]. As shown in Fig. 14d, by plotting the polarization $P$ versus electric field in the AFE layer $E_a$, two separate negative slope regions were discovered, which confirmed the existence of negative capacitance. By integrating the $P$-$E_a$ curve, the AFE energy landscape of $ZrO_2$ can be further obtained and the normally forbidden regions near the structural phase transition, exhibiting negative capacitance feature, were stabilized in the AFE/dielectric heterostructure (Fig. 14e). The discovery in AFE zirconia films further supports that the negative capacitance can be expected in any two-state system with polarization instability, which broadens the understanding of the negative capacitance.

### 5.5. Memory devices

Although non-volatile FE random access memories (FRAM) have been commercially utilized for years [81], the limited data storage density hinder its further application and market prospect. Recently, the concept of four-state AFE-based random access memories (AFRAM) was proposed with faster write/read circles and enhanced data storage capability [9]. In the prototype AFRAM logic operation shown in Fig. 15a, based on the typical double hysteresis loops, the write voltage can fully saturate the AFE structure at one direction. Correspondingly, the read process is two directional for both positive and negative voltage. Similar process can be realized by writing at an opposite direction and read bidirectionally, leading to 4 optional read-out states. Another AFE-based model for non-volatile memory is proposed in doped-$ZrO_2$, where one branch of the double hysteresis loop can be centered by built-in electric field [282]. In this way, polar and nonpolar states can be effectively obtained with low voltage switching and enhanced endurance. In 2019, Intel patented an AFE based memory cell [424]. In that demonstration, the nonlinear behavior of AFE can boost the charge for different logic states, thus the retention time has an increase. Later, $Hf_xZr_{1-x}O_2$ capacitor based 3D embedded DRAM was experimentally demonstrated [266], and the device displayed promising electrical property for application, including 10 ns polarization switching for write/read operation, small maximum operation voltage (1.8V), more than 1ms retention time and $10^{12}$ endurance cycles. The research on AFE based memories devices are at the early stage and need further exploration and characterization for application. The potential multistate information storage together with the advanced electrical behaviors reported recently make AFE a promising material for future memory device.

For practical applications in memory devices, the read/write time is required to be at the range of 10 ns or less to meet the industry standards [266]. In AFE $Hf_xZr_{1-x}O_2$, the read/write speed is reported to be as fast as 10 ns, and even down to 2 ns within different scenarios from 3-D Embedded-DRAM [266] to FeRAM using AFE capacitors [425]. The high-speed characteristic in AFE $Hf_xZr_{1-x}O_2$-based memories is promising in future high-density, ultrafast memory applications.

### 5.6. Tunnel junctions

Tunnel junction depicts a metal-insulator-metal structure with switchable On-Off states. The insulator layer is required to be very thin. Thus, the quantum tunneling emerges and dominates the conductivity through the two metal electrodes. The FE tunnel junction (FTJ) has been widely explored while the AFE one received much less attention due to its information-volatile nature for memory device. For AFE $PbZrO_3$, recent study revealed that the switchable polar and nonpolar state can also significantly influence the barrier between two electrodes with an ON-OFF ratio up to $10^7$, shown in Fig. 15b [426]. The giant tunneling electroresistance was attributed to the cooperation of macroscopic polarization ordering and Fowler-Nordheim conduction. Although the AFE material retains, normally, no information for memory application, the AFE tunneling junction (AFTJ) highlights the resistive switching capability, which might be possible to make non-volatile using some of the above-mentioned approaches, such as off-centering of the AFE loop by built-in fields in films. Recently, a built-in bias was generated in AFE $Hf_{0.25}Zr_{0.75}O_2$-based AFTJ to form 2 stable nonvolatile states by using asymmetric work function electrodes [427]. This AFTJ exhibits higher endurance property (>$10^9$ cycles) and faster switching speed (<30 ns) than $Hf_{0.5}Zr_{0.5}O_2$-based FTJ. This makes AFEs to be a promising candidate for future cognitive computing.

### 5.7. Thermal switching

With the development of aerospace and advanced semiconductor technology, the integration and power density of electronic devices are increasing. How to solve the high-temperature failure caused by device heating has become a key problem that restricts the efficient and stable operation of devices. Efficient thermal management can solve the impact of extreme environment on device operation and ensure the stable operation of devices within a safe temperature range [364, 428]. The common method is to regulate the thermal conductivity by inducing structural phase transformation, such as solid-liquid phase transition [429] and metal-insulator transition [430]. However, changes in thermal conductivity of these materials can only be realized at the phase transition temperature, limiting their application. To electrically control thermal switching, various material is proposed with different origins [431, 432]. Solid-state thermal conductivity switching has been reported in $PbZrO_3$ thin films recently by Liu *et al*. [10], as shown in Fig. 15c. As the external electric field is loaded, the primitive unit cell of $PbZrO_3$ experiences a substantial change at the AFE to FE phase transition. Consequently, phonon-phonon scattering phase space has a large change across the phonon spectrum before and after the phase transition, leading to a high switching ratio (>2.2). In addition, Liu *et al*. also reported that the $PbZrO_3$-based thermal switching is ultrafast (<150 nanoseconds) and can repeat more than 10 million times. Considering these advantages, AFEs are the key component in solid-state refrigeration and thermal management.

**5.8. Photovoltaics**

In traditional photovoltaics, light irradiation induced electron-hole pairs can be separated by a built-in electric field, leading to the conversion of light energy to electrical energy [61]. In general, the maximum photovoltage is equal to the band gap. For FEs, it was recently confirmed that the intrinsic inversion asymmetry of lattice can induce the bandgap photovoltages. In contrast, for the counterpart of FEs, the AFEs, believed to be centrosymmetric, cannot induce the photovoltaic effect. Although the electric field-induced polar state in AFEs is metastable, the absorbed light energy can help to pin and stabilize the polar state after the external field was removed. As a result, the pined polar state can generate the bandgap photovoltages as shown in Fig. 15d [11]. Among the investigated materials, the largest photoelectric field was discovered recently in $PbZrO_3$ polycrystalline thin films, which is up to the level of million voltage per centimeter [11]. The above bandgap photovoltage in AFEs displays competitive results and offers new prospects on the use of photovoltaic materials in future devices.

**6. Summary and Perspective**

In conclusion, we want to point out that, due to their unique electrical and structural behaviors, research on AFE materials is attracting widespread attention in modern electronic applications. Although the AFE has been discovered and studied for decades in terms of structure, electrical motifs, phase transitions, and potential device applications, the origin of AFE order, polarization modulation, local structure and their influence on properties are still key issues that that further need exploring. Moreover, new AFE thin film materials and phenomena offer plenty of opportunities for further investigations, such as antipolar topological structures, *etc*. All told, the research on AFE thin films and heterostructures, however, is still in its infancy and there remain many open questions, such as:

1. The mechanism and even the definition of AFEs are ambiguous among different reports. On one hand, multiple soft modes during AFE phase transition displays complex results of structure and electrical behavior, which obscure the origin of AFE order. On the other hand, the lack of defect-free AFE material usually results in extrinsic behaviors. In a word, AFE could not be simply seen as an analogue of antiferromagnetic material as initially considered, and the origin of AFE could be due to multiple effects. To solve these problems, advanced theoretical studies should be applied to distinguish the contributions from different soft modes, and advanced synthesis of high quality AFE epitaxial thin films is needed for experimental examination.

2. The structures of AFEs are complex and remain controversial. The observation of weak ferroelectricity especially in thin film form [39], size effect [116, 117], and incommensurate structure [173, 211] makes the intrinsic structure of AFEs deceptive. Besides, to realize practical applications, it is important to understand the phase transition dynamic and fatigue mechanism of AFE thin films. However, these detailed structural evolution under external fields remain elusive, such as the dipole switching kinetics and phase transition pathway. The pursuit of these answers will require advanced characterization and computational supports to explore this complex landscape. In-situ biasing structural characterization techniques such as time-resolved X-ray diffraction and TEM at atomic resolution are all effective and powerful tools to provide in-depth structural information about the phase transitions.

3. In line with requirement of sustainable and eco-friendly society development, new type lead-free AFE materials and their related applications are needed. Machine learning and high-throughput second-principles calculations can potentially assist in discovering new AFE materials with robust and controllable properties. Additionally, the integration of AFE thin films on Si-based production process holds great significance. CMOS-compatible AFE thin film materials, such as $Hf_{1-x}Zr_xO_2$, are needed from theoretical design to experimental test for further device application.

4. Topological structures in FE thin films and heterostructures have become one of the hottest topics in condensed matter physics [151]. As a counterpart of FE, AFE materials do not exhibit net polarization and thus seem to lack a depolarization field, which is considered as one of the origins of forming topological polar structure. However, recent studies have revealed the existence of non-collinear polarization arrangement in AFE thin films [7]. The high sensitivity of AFE order parameter to external stimuli may provide new possibility of polarization rotation and the formation of topological antipolar structures under certain circumstance.

5. Conventional multiferroics are composed of FE and antiferromagnetic/ferromagnetic orders. Whether the AFE order may couple with the ferromagnetic or antiferromagnetic order in a single-phase system remains an open question. The coexistence of antiferroelectric-antiferromagnetic properties in multiferroic $BiFeO_3$-based system could be of significance in scientific interest and great potential in magnetoelectric devices because of its dual antiferroic nature. So far, various methods which have been successfully applied in other ferroic studies can be applied to AFEs as well, such as high entropy [433], anionic substitution [434], composition gradient [435], strain gradient [436], freestanding [437], and *etc*. These methods, which proved to be effective in FEs or magnetic materials for creating novel phenomenon, could also be introduced into AFE materials for unexpected phenomena.

## Acknowledgements

Y.Y. Si, T.F. Zhang, and C.H. Liu contributed equally to this work. This work was funded by National Natural Science Foundation of China (Grant No. 52372105), Shenzhen Science and Technology Innovation project (Grant No. JCYJ20200109112829287), Shenzhen Science and Technology Program (Grant No. KQTD20200820113045083). Z.H.C. has been supported by "the Fundamental Research Funds for the Central Universities" (Grant No. HIT.OCEF.2022038) and "Talent Recruitment Project of Guangdong" (Grant No. 2019QN01C202). S.D. acknowledges Science and Engineering Research Board (SRG/2022/000058) and Indian Institute of Science start up grant for financial support. X.-K.W. was supported by grants of the National High-Level Youth Talents Program (Grant No. 0040/X2450224) and the Xiamen University President's Fund Program (Grant No. 0040/ZK1227). C. Liu was funded by National Natural Science Foundation of China (Grant no. 52206092), Natural Science Foundation of Jiangsu Province (Grant no. BK20210565), Department of Science and Technology of Jiangsu Province (BK20220032), Nanjing Science and Technology Innovation Project for Overseas Students and "Shuangchuang" Doctor program of Jiangsu Province (JSSCBS20210315).

## References

[1] Fernandez A, Acharya M, Lee HG, Schimpf J, Jiang Y, Lou D, et al. Thin-film Ferroelectrics. Adv Mater 2022:2108841. https://doi.org/10.1002/adma.202108841

[2] Ramesh R, Spaldin NA. Multiferroics: progress and prospects in thin films. Nat Mater 2007;6:21-9. https://doi.org/10.1038/nmat1805

[3] Fernandez A, Acharya M, Lee HG, Schimpf J, Jiang Y, Lou D, et al. Thin-Film Ferroelectrics. Adv Mater. 2022;34:e2108841. https://doi.org/10.1002/adma.202108841

[4] Scott JF. Applications of Modern Ferroelectrics. Science 2007;315:954-9. https://doi.org/10.1126/science.1129564

[5] Lidiard AB. Antiferromagnetism. Rep Prog Phys 1954;17:201. https://doi.org/10.1088/0034-4885/17/1/307

[6] Geng WP, Liu Y, Meng XJ, Bellaiche L, Scott JF, Dkhil B, et al. Giant Negative Electrocaloric Effect in Antiferroelectric La-Doped Pb(ZrTi)$O_3$ Thin Films Near Room Temperature. Adv Mater 2015;27:3165-9. https://doi.org/10.1002/adma.201501100

[7] Yao Y, Naden A, Tian M, Lisenkov S, Beller Z, Kumar A, et al. Ferrielectricity in the Archetypal Antiferroelectric, $PbZrO_3$. Adv Mater 2022;35:2206541. https://doi.org/10.1002/adma.202206541

[8] Acharya M, Banyas E, Ramesh M, Jiang Y, Fernandez A, Dasgupta A, et al. Exploring the $Pb_{1-x}Sr_xHfO_3$ System and Potential for High Capacitive Energy Storage Density and Efficiency. Adv Mater 2022;34:2105967. https://doi.org/10.1002/adma.202105967

[9] Vopson MM, Tan X. Four-State Anti-Ferroelectric Random Access Memory. IEDL. 2016;37:1551-4. https://doi.org/10.1109/led.2016.2614841

[10] Liu C, Si Y, Zhang H, Wu C, Deng S, Dong Y, et al. Low voltage-driven high-performance thermal switching in antiferroelectric $PbZrO_3$ thin films. Science 2023;382:1265-9. https://doi.org/science.org/doi/10.1126/science.adj9669

[11] Perez-Tomas A, Lira-Cantu M, Catalan G. Above-Bandgap Photovoltages in Antiferroelectrics. Adv Mater 2016;28:9644-7. https://doi.org/10.1002/adma.201603176

[12] Liu H, Dkhil B. A brief review on the model antiferroelectric $PbZrO_3$ perovskite-like material. Zeitschrift für Kristallographie - Cryst Mater 2011;226:163-70. https://doi.org/10.1524/zkri.2011.1336

[13] Jin L, Li F, Zhang S. Decoding the Fingerprint of Ferroelectric Loops: Comprehension of the Material Properties and Structures. J Am Ceram Soc 2014;97:1-27. https://doi.org/10.1111/jace.12773

[14] Gao R, Reyes-Lillo SE, Xu R, Dasgupta A, Dong Y, Dedon LR, et al. Ferroelectricity in $Pb_{1+\delta}ZrO_3$ Thin Films. Chem Mater 2017;29:6544-51. https://doi.org/10.1021/acs.chemmater.7b02506

[15] Burkovsky RG, Lityagin GA, Ganzha AE, Vakulenko AF, Gao R, Dasgupta A, et al. Field-induced heterophase state in $PbZrO_3$ thin films. Phys Rev B 2022;105:125409. https://doi.org/10.1103/PhysRevB.105.125409

[16] Si Y, Zhang T, Chen Z, Zhang Q, Xu S, Lin T, et al. Phase Competition in High-Quality Epitaxial Antiferroelectric $PbZrO_3$ Thin Films. ACS Appl Mater Inter 2022;14:51096-104. https://doi.org/10.1021/acsami.2c14291

[17] Qiao L, Song C, Wang Q, Zhou Y, Pan F. Polarization Evolution in Nanometer-Thick PbZrO3 Films: Implications for Energy Storage and Pyroelectric Sensors. ACS Applied Nano Mater 2022;5:6083-8. https://doi.org/10.1021/acsanm.2c01132

[18] Kniazeva MA, Ganzha AE, Gao R, Dasgupta A, Filimonov AV, Burkovsky RG. Highly mismatched antiferroelectric films: Transition order and mechanical state. Phys Rev B 2023;107:184113. https://doi.org/10.1103/PhysRevB.107.184113

[19] Roy Chaudhuri A, Arredondo M, Hähnel A, Morelli A, Becker M, Alexe M, et al. Epitaxial strain stabilization of a ferroelectric phase in $PbZrO_3$ thin films. Phys Rev B 2011;84:054112. https://doi.org/10.1103/PhysRevB.84.054112

[20] Ge J, Remiens D, Costecalde J, Chen Y, Dong X, Wang G. Effect of residual stress on energy storage property in $PbZrO_3$ antiferroelectric thin films with different orientations. Appl Phys Lett 2013;103:162903. https://doi.org/10.1063/1.4825336

[21] Nguyen MD, Birkhölzer YA, Houwman EP, Koster G, Rijnders G. Enhancing the Energy-Storage Density and Breakdown Strength in $PbZrO_3/Pb_{0.9}La_{0.1}Zr_{0.52}Ti_{0.48}O_3$-Derived Antiferroelectric/Relaxor-Ferroelectric Multilayers. Adv Energy Mater 2022;12:2200517. https://doi.org/10.1002/aenm.202200517

[22] Liu Z, Lu T, Ye J, Wang G, Dong X, Withers R, et al. Antiferroelectrics for Energy Storage Applications: a Review. Adv Mater Tech 2018;3:1800111. https://doi.org/10.1002/admt.201800111

[23] Hao X, Zhai J, Kong LB, Xu Z. A comprehensive review on the progress of lead zirconate-based antiferroelectric materials. Prog Mater Sci 2014;63:1-57. https://doi.org/10.1016/j.pmatsci.2014.01.002

[24] Randall CA, Fan Z, Reaney I, Chen LQ, Trolier-McKinstry S. Antiferroelectrics: History, fundamentals, crystal chemistry, crystal structures, size effects, and applications. J Am Ceram Soc 2021;104:3775. https://doi.org/10.1111/jace.17834

[25] Kittel C. Theory of Antiferroelectric Crystals. Phys Rev 1951;82:729-32. https://doi.org/10.1103/PhysRev.82.729

[26] Kohli M, Muralt P, Setter N. Removal of 90° domain pinning in (100) $Pb(Zr_{0.15}Ti_{0.85})O_3$ thin films by pulsed operation. Appl Phys Lett 1998;72:3217-9. https://doi.org/10.1063/1.121554

[27] Srivastava N, Weng GJ. A theory of double hysteresis for ferroelectric crystals. J Appl Phys 2006;99:054103. https://doi.org/10.1063/1.2178391

[28] Muller J, Boscke TS, Schroder U, Mueller S, Brauhaus D, Bottger U, et al. Ferroelectricity in Simple Binary $ZrO_2$ and $HfO_2$. Nano Lett 2012;12:4318. https://doi.org/10.1021/nl302049k

[29] Reyes-Lillo SE, Garrity KF, Rabe KM. Antiferroelectricity in thin-film $ZrO_2$ from first principles. Phys Rev B 2014;90:140103. https://doi.org/10.1103/PhysRevB.90.140103

[30] Milesi-Brault C, Toulouse C, Constable E, Aramberri H, Simonet V, de Brion S, et al. Archetypal Soft-Mode-Driven Antipolar Transition in Francisite $Cu_3Bi(SeO_3)_2O_2Cl$. Phys Rev Lett 2020;124:097603. https://doi.org/10.1103/PhysRevLett.124.097603

[31] Tagantsev AK, Vaideeswaran K, Vakhrushev SB, Filimonov AV, Burkovsky RG, Shaganov A, et al. The origin of antiferroelectricity in $PbZrO_3$. Nat Commun 2013;4:2229. https://doi.org/10.1038/ncomms3229

[32] Sawaguchi E, Maniwa H, Hoshino S. Antiferroelectric Structure of Lead Zirconate. Phys Rev 1951;83:1078. https://doi.org/10.1103/PhysRev.83.1078

[33] Shirane G, Newnham R, Pepinsky R. Dielectric Properties and Phase Transitions of NaNbO3 and (Na,K)$NbO_3$. Phys Rev 1954;96:581-8. https://doi.org/10.1103/PhysRev.96.581
[34] Jaffe B. Antiferroelectric ceramics with field-enforced transitions: a new nonlinear circuit element. Proc IRE 1961;49:1264-7. https://doi.org/10.1109/JRPROC.1961.287917
[35] Thacher P. Electrocaloric effects in some ferroelectric and antiferroelectric Pb(Zr,Ti)$O_3$ compounds. J Appl Phys 1968;39:1996-2002. https://doi.org/10.1063/1.1656478
[36] Samara GA. Pressure and Temperature Dependence of the Dielectric Properties and Phase Transitions of the Antiferroelectric Perovskites: $PbZrO_3$ and $PbHfO_3$. Phys Rev B 1970;1:3777-86. https://doi.org/10.1103/PhysRevB.1.3777
[37] Sakata K, Masuda Y. Ferroelectric and antiferroelectric properties of $(Na_{0.5}Bi_{0.5})TiO_3$-$SrTiO_3$ solid solution ceramics. Ferroelectrics. 1974;7:347-9. https://doi.org/10.1080/00150197408238042
[38] Uchino K. Digital displacement transducer using antiferroelectrics. Jpn J Appl Phys 1985;24:460. https://doi.org/10.7567/JJAPS.24S2.460
[39] Dai X, Li JF, Viehland D. Weak ferroelectricity in antiferroelectric lead zirconate. Phys Rev B 1995;51:2651-5. https://doi.org/10.1103/physrevb.51.2651
[40] Mischenko AS, Zhang Q, Scott JF, Whatmore RW, Mathur ND. Giant electrocaloric effect in thin-film $PbZr_{0.95}Ti_{0.05}O_3$. Science. 2006;311:1270-1. https://doi.org/10.1126/science.1123811
[41] Karimi S, Reaney IM, Levin I, Sterianou I. Nd-doped $BiFeO_3$ ceramics with antipolar order. Appl Phys Lett 2009;94:112903. https://doi.org/10.1063/1.3097222
[42] Íñiguez J, Stengel M, Prosandeev S, Bellaiche L. First-principles study of the multimode antiferroelectric transition in $PbZrO_3$. Phys Rev B 2014;90:220103. https://doi.org/10.1103/PhysRevB.90.220103
[43] Hlinka J, Ostapchuk T, Buixaderas E, Kadlec C, Kuzel P, Gregora I, et al. Multiple soft-mode vibrations of lead zirconate. Phys Rev Lett 2014;112:197601. https://doi.org/10.1103/PhysRevLett.112.197601
[44] Park MH, Lee YH, Kim HJ, Kim YJ, Moon T, Kim KD, et al. Ferroelectricity and antiferroelectricity of doped thin $HfO_2$-based films. Adv Mater 2015;27:1811. https://doi.org/10.1002/adma.201404531
[45] Qiao L, Song C, Sun Y, Fayaz MU, Lu T, Yin S, et al. Observation of negative capacitance in antiferroelectric $PbZrO_3$ Films. Nat Commun 2021;12:4215. https://doi.org/10.1038/s41467-021-24530-w
[46] Aryana K, Tomko JA, Gao R, Hoglund ER, Mimura T, Makarem S, et al. Observation of solid-state bidirectional thermal conductivity switching in antiferroelectric lead zirconate ($PbZrO_3$). Nat Commun 2022;13:1573. https://doi.org/10.1038/s41467-022-29023-y
[47] Sawaguchi E, Shirane G, Takagi Y. Phase transition in lead zirconate. J Phys Soc Jpn 1951;6:333-9. https://doi.org/10.1143/JPSJ.6.333
[48] Roberts S. Dielectric Properties of Lead Zirconate and Barium-Lead Zirconate. J Am Ceram Soc 1950;33:63-6. https://doi.org/10.1111/j.1151-2916.1950.tb14168.x
[49] Shirane G, Sawaguchi E, Takeda A. On the Phase Transition in Lead Zirconate. Phys Rev. 1950;80:485. https://doi.org/10.1103/PhysRev.80.485

[50] Shirane G, Sawaguchi E, Takagi Y. Dielectric Properties of Lead Zirconate. Phys Rev 1951;84:476-81. https://doi.org/10.1103/PhysRev.84.476

[51] Shirane G, Pepinsky R. Phase Transitions in Antiferroelectric $PbHfO_3$. Phys Rev 1953;91:812-5. https://doi.org/10.1103/PhysRev.91.812

[52] Francombe MH, Lewis B. Structural and Electrical Properties of Silver Niobate and Silver Tantalate. Acta Cryst 1958;11:175. https://doi.org/10.1107/S0365110X58000463

[53] Iida S, Terauchi H. Dipole-Glass Phase in Random Mixture of Ferroelectric and Antiferroelectric: $Rb_{1-x}(NH_4)_xH_2PO_4$. J Phys Soc Jpn 1983;52:4044-7. https://doi.org/10.1143/JPSJ.52.4044

[54] Chandani ADL, Gorecka E, Ouchi Y, Takezoe H, Fukuda A. Antiferroelectric Chiral Smectic Phases Responsible for the Trislable Switching in MHPOBC. Jpn J Appl Phys 1989;28:L1265. https://doi.org/10.1143/JJAP.28.L1265

[55] Pan WY, Dam CQ, Zhang QM, Cross LE. Large displacement transducers based on electric field forced phase transitions in the tetragonal $(Pb_{0.97}La_{0.02})$ $(Ti,Zr,Sn)O_3$ family of ceramics. J Appl Phys 1989;66:6014-23. https://doi.org/10.1063/1.343578

[56] Fesenko OE, Kolesova RV, Sindeyev YG. The structural phase transitions in lead zirconate in super-high electric fields. Ferroelectrics 1978;20:177-8. https://doi.org/10.1080/00150197808237203

[57] Fesenko OE, Balyunis LE. The temperature-electric field phase diagram of lead hafnate. Ferroelectrics 1980;29:95-8. https://doi.org/10.1080/00150198008009017

[58] Ayyub P, Chattopadhyay S, Pinto R, Multani M. Ferroelectric behavior in thin films of antiferroelectric materials. Phys Rev B 1998;57:R5559. https://doi.org/10.1103/PhysRevB.57.R5559

[59] Pintilie L, Boldyreva K, Alexe M, Hesse D. Coexistence of ferroelectricity and antiferroelectricity in epitaxial $PbZrO_3$ films with different orientations. J Appl Phys 2008;103:024101. https://doi.org/10.1063/1.2831023

[60] Lai Y-H, Zheng J-D, Lu S-C, Wang Y-K, Duan C-G, Yu P, et al. Antiferroelectric $PbSnO_3$ Epitaxial Thin Films. Adv Sci 2022;9:2203863. https://doi.org/10.1002/advs.202203863

[61] Yang SY, Seidel J, Byrnes SJ, Shafer P, Yang CH, Rossell MD, et al. Above-bandgap voltages from ferroelectric photovoltaic devices. Nat Nanotechnol 2010;5:143. https://doi.org/10.1038/nnano.2009.451

[62] Hoffmann M, Wang Z, Tasneem N, Zubair A, Ravindran PV, Tian M, et al. Antiferroelectric negative capacitance from a structural phase transition in zirconia. Nat Commun 2022;13:1228. https://doi.org/10.1038/s41467-022-28860-1

[63] Zhang Y, Bellaiche L, Xu B. Ultrahigh energy storage density in lead-free antiferroelectric rare-earth-substituted bismuth ferrite. Phys. Rev. Materials. 2022;6:051401. https://doi.org/10.1103/PhysRevMaterials.6.L05140

[64] Xu B, Hellman O, Bellaiche L. Order-disorder transition in the prototypical antiferroelectric $PbZrO_3$. Phys Rev B. 2019;100:020102. https://doi.org/10.1103/PhysRevB.100.020102

[65] Wei XK, Tagantsev AK, Kvasov A, Roleder K, Jia CL, Setter N. Ferroelectric translational antiphase boundaries in nonpolar materials. Nat Commun. 2014;5:3031. https://doi.org/10.1038/ncomms4031

[66] Haun MJ, Harvin TJ, Lanagan MT, Zhuang ZQ, Jang SJ, Cross LE. Thermodynamic theory of $PbZrO_3$. J Appl Phys 1989;65:3173-80. https://doi.org/10.1063/1.342668

[67] Roleder K, Dee J. The defect-induced ferroelectric phase in thin $PbZrO_3$ single crystals. J Phys: Condens Matter 1989;1:1503. https://doi.org/10.1088/0953-8984/1/8/013

[68] Ujma Z. Dielectric properties and phase transitions in $PbZrO_3$ with oxygen vacancies. Phase Transitions 1984;4:169-81. https://doi.org/10.1080/01411598408218593

[69] Vales-Castro P, Roleder K, Zhao L, Li J-F, Kajewski D, Catalan G. Flexoelectricity in antiferroelectrics. Appl Phys Lett 2018;113:132903. https://doi.org/10.1063/1.5044724

[70] Mani BK, Lisenkov S, Ponomareva I. Finite-temperature properties of antiferroelectric $PbZrO_3$ from atomistic simulations. Phys Rev B 2015;91:134112. https://doi.org/10.1103/PhysRevB.91.134112

[71] Patel K, Prosandeev S, Yang Y, Xu B, Íñiguez J, Bellaiche L. Atomistic mechanism leading to complex antiferroelectric and incommensurate perovskites. Phys Rev B 2016;94:054107. https://doi.org/10.1103/PhysRevB.94.054107

[72] Burkovsky RG. Dipole-dipole interactions and incommensurate order in perovskite structures. Phys Rev B 2018;97:184109. https://doi.org/10.1103/PhysRevB.97.184109

[73] Lu T, Studer AJ, Yu D, Withers RL, Feng Y, Chen H, et al. Critical role of the coupling between the octahedral rotation and A-site ionic displacements in $PbZrO_3$-based antiferroelectric materials investigated by in situ neutron diffraction. Phys Rev B 2017;96:214108. https://doi.org/10.1103/PhysRevB.96.214108

[74] Shapovalov K, Stengel M. Tilt-driven antiferroelectricity in $PbZrO_3$. arXiv preprint arXiv:12167. 2021. https://doi.org/10.48550/arXiv.2112.12167

[75] Amisi S. Ab initio investigation in $PbZrO_3$ antiferroelectric: structural and vibrational properties. Eur Phys J Plus 2021;136:653. https://doi.org/10.1140/epjp/s13360-021-01639-x

[76] Zhang TF, Tang XG, Liu QX, Jiang YP, Huang XX, Zhou QF. Energy-storage properties and high-temperature dielectric relaxation behaviors of relaxor ferroelectric $Pb(Mg_{1/3}Nb_{2/3})O_3$–$PbTiO_3$ ceramics. J Phys D: Appl Phys 2016;49:095302. https://doi.org/10.1088/0022-3727/49/9/095302

[77] Lanfredi S, Lente M, Eiras J. Phase transition at low temperature in $NaNbO_3$ ceramic. Appl Phys Lett 2002;80:2731-3. https://doi.org/10.1063/1.1470260

[78] Li PF, Liao WQ, Tang YY, Ye HY, Zhang Y, Xiong RG. Unprecedented ferroelectric–antiferroelectric–paraelectric phase transitions discovered in an organic–inorganic hybrid perovskite. J Am Chem Soc 2017;139:8752-7. https://doi.org/10.1021/jacs.7b04693

[79] Wei XK, Prokhorenko S, Wang BX, Liu Z, Xie YJ, Nahas Y, et al. Ferroelectric phase-transition frustration near a tricritical composition point. Nat Commun 2021;12:5322. https://doi.org/10.1038/s41467-021-25543-1

[80] Dufour P, Maroutian T, Vallet M, Patel K, Chanthbouala A, Jacquemont C, et al. Ferroelectric phase transitions in epitaxial antiferroelectric $PbZrO_3$ thin films. Appl Phys Rev 2023;10:021405. https://doi.org/10.1063/5.0143892

[81] Kato Y, Kaneko Y, Tanaka H, Kaibara K, Koyama S, Isogai K, et al. Overview and future challenge of ferroelectric random access memory technologies. J Jap Appl Phys 2007;46:2157. https://doi.org/10.1143/JJAP.46.2157

[82] Zhang M-H, Fulanović L, Egert S, Ding H, Groszewicz PB, Kleebe H-J, et al. Electric-field-induced antiferroelectric to ferroelectric phase transition in polycrystalline $NaNbO_3$. Acta Mater 2020;200:127-35. https://doi.org/10.1016/j.actamat.2020.09.002

[83] Tan X, Frederick J, Ma C, Jo W, Rodel J. Can an electric field induce an antiferroelectric phase out of a ferroelectric phase? Phys Rev Lett 2010;105:255702. https://doi.org/10.1103/PhysRevLett.105.255702

[84] Feng M, Feng Y, Zhang T, Li J, Chen Q, Chi Q, et al. Recent advances in multilayer-structure dielectrics for energy storage application. Adv Sci 2021;8:2102221. https://doi.org/10.1002/advs.202102221

[85] Feng Y, Wei X, Wang D, Xu Z, Yao X. Dielectric behaviors of antiferroelectric–ferroelectric transition under electric field. Ceram Int 2004;30:1389-92. https://doi.org/10.1016/j.ceramint.2003.12.090

[86] Wang J, Yang T, Chen S, Yao X, Pelaiz-Barranco A. DC electric field dependence for the dielectric permittivity in antiferroelectric and ferroelectric states. J Alloy Compd 2014;587:827-9. https://doi.org/10.1016/j.jallcom.2013.10.251

[87] Sigman J, Norton D, Christen H, Fleming P, Boatner L. Antiferroelectric behavior in symmetric $KNbO_3/KTaO_3$ superlattices. Phys Rev Lett 2002;88:097601. https://doi.org/10.1103/PhysRevLett.88.097601

[88] Christen HM, Specht ED, Silliman SS, Harshavardhan KS. Ferroelectric and antiferroelectric coupling in superlattices of paraelectric perovskites at room temperature. Phys Rev B 2003;68:020101. https://doi.org/10.1103/PhysRevB.68.020101

[89] Shebanov L, Kusnetsov M, Sternberg A. Electric field-induced antiferroelectric-to-ferroelectric phase transition in lead zirconate titanate stannate ceramics modified with lanthanum. J Appl Phys 1994;76:4301-4. https://doi.org/10.1063/1.357315

[90] Tang Z, Hu S, Yao D, Li Z, Liu Z, Guo X, et al. Enhanced energy-storage density and temperature stability of $Pb_{0.89}La_{0.06}Sr_{0.05}(Zr_{0.95}Ti_{0.05})O_3$ anti-ferroelectric thin film capacitor. J Materiomics 2022;8:239-46. https://doi.org/10.1016/j.jmat.2020.12.012

[91] Liu C, Lin SX, Qin MH, Lu XB, Gao XS, Zeng M, et al. Energy storage and polarization switching kinetics of (001)-oriented $Pb_{0.97}La_{0.02}(Zr_{0.95}Ti_{0.05})O_3$ antiferroelectric thick films. Appl Phys Lett 2016;108:112903. https://doi.org/10.1063/1.4944645

[92] Bharadwaja SSN, Krupanidhi SB. Backward switching phenomenon from field forced ferroelectric to antiferroelectric phases in antiferroelectric $PbZrO_3$ thin films. J Appl Phys 2001;89:4541-7. https://doi.org/10.1063/1.1331659

[93] Si M, Lyu X, Shrestha PR, Sun X, Wang H, Cheung KP, et al. Ultrafast measurements of polarization switching dynamics on ferroelectric and anti-ferroelectric hafnium zirconium oxide. Appl Phys Lett 2019;115:072107. https://doi.org/10.1063/1.5098786

[94] Parsonnet E, Huang YL, Gosavi T, Qualls A, Nikonov D, Lin CC, et al. Toward Intrinsic Ferroelectric Switching in Multiferroic $BiFeO_3$. Phys Rev Lett 2020;125:067601. https://doi.org/10.1103/PhysRevLett.125.067601

[95] Janovec V. On the theory of the coercive field of single-domain crystals of $BaTiO_3$. Cechoslovackij Fiziceskij Zurnal 1958;8:3-15. https://doi.org/10.1007/BF01688741

[96] Kay H, Dunn JW. Thickness dependence of the nucleation field of triglycine sulphate. Philos Mag 1962;7:2027-34. https://doi.org/10.1080/14786436208214471

[97] Tasneem N, Yousry YM, Tian M, Dopita M, Reyes-Lillo SE, Kacher J, et al. A Janovec-Kay-Dunn-Like Behavior at Thickness Scaling in Ultra-Thin Antiferroelectric $ZrO_2$ Films. Adv Elec Mater 2021;7:2100485. https://doi.org/10.1002/aelm.202100485
[98] Zhai J, Chen H. Direct current field and temperature dependent behaviors of antiferroelectric to ferroelectric switching in highly (100)-oriented $PbZrO_3$ thin films. Appl Phys Lett 2003;82:2673-5. https://doi.org/10.1063/1.1569420
[99] Tan X, Ma C, Frederick J, Beckman S, Webber KG, Green DJ. The Antiferroelectric ↔ Ferroelectric Phase Transition in Lead-Containing and Lead-Free Perovskite Ceramics. J Am Ceram Soc 2011;94:4091-107. https://doi.org/10.1111/j.1551-2916.2011.04917.x
[100] Ahn CW, Amarsanaa G, Won SS, Chae SA, Lee DS, Kim IW. Antiferroelectric Thin-Film Capacitors with High Energy-Storage Densities, Low Energy Losses, and Fast Discharge Times. ACS Appl Mater Interfaces. 2015;7:26381. https://doi.org/10.1021/acsami.5b08786
[101] Lu T, Studer AJ, Noren L, Hu W, Yu D, McBride B, et al. Electric-field-induced AFE-FE transitions and associated strain/preferred orientation in antiferroelectric PLZST. Sci Rep 2016;6:1-8. https://doi.org/10.1038/srep23659
[102] Liu H, Fan L, Sun S, Lin K, Ren Y, Tan X, et al. Electric-field-induced structure and domain texture evolution in $PbZrO_3$-based antiferroelectric by in-situ high-energy synchrotron X-ray diffraction. Acta Mater 2020;184:41-9. https://doi.org/10.1016/j.actamat.2019.11.050
[103] Wei XK, Jia CL, Roleder K, Dunin-Borkowski RE, Mayer J. In Situ Observation of Point-Defect-Induced Unit-Cell-Wise Energy Storage Pathway in Antiferroelectric $PbZrO_3$. Adv Funct Mater 2021;31:2008609. https://doi.org/10.1002/adfm.202008609
[104] Wei XK, Jia CL, Du HC, Roleder K, Mayer J, Dunin-Borkowski RE. An Unconventional Transient Phase with Cycloidal Order of Polarization in Energy-Storage Antiferroelectric $PbZrO_3$. Adv Mater 2020;32:1907208. https://doi.org/10.1002/adma.201907208
[105] Grigoriev A, Sichel R, Lee HN, Landahl EC, Adams B, Dufresne EM, et al. Nonlinear piezoelectricity in epitaxial ferroelectrics at high electric fields. Phys Rev Lett 2008;100:027604. https://doi.org/10.1103/PhysRevLett.100.027604
[106] Li Q, Stoica VA, Paściak M, Zhu Y, Yuan Y, Yang T, et al. Subterahertz collective dynamics of polar vortices. Nature 2021;592:376-80. https://doi.org/10.1038/s41586-021-03342-4
[107] Tagantsev AK, Stolichnov I, Colla EL, Setter N. Polarization fatigue in ferroelectric films: Basic experimental findings, phenomenological scenarios, and microscopic features. J Appl Phys 2001;90:1387-402. https://doi.org/10.1063/1.1381542
[108] Zhou L, Zimmermann A, Zeng YP, Aldinger F. Fatigue of Field-Induced Strain in Antiferroelectric $Pb_{0.97}La_{0.02}(Zr_{0.77}Sn_{0.14}Ti_{0.09})O_3$ Ceramics. J Am Ceram Soc 2004;87:1591-3. https://doi.org/10.1111/j.1551-2916.2004.01591.x
[109] Lou X, Wang J. Unipolar and bipolar fatigue in antiferroelectric lead zirconate thin films and evidences for switching-induced charge injection inducing fatigue. Appl Phys Lett 2010;96:102906. https://doi.org/10.1063/1.3358138
[110] Geng W, Lou X, Xu J, Zhang F, Liu Y, Dkhil B, et al. Effective driving voltage on polarization fatigue in $(Pb,La)(Zr,Ti)O_3$ antiferroelectric thin films. Ceram Int 2015;41:109-14. https://doi.org/10.1016/j.ceramint.2014.08.041

[111] Zhai J, Chen H. Electric fatigue in Pb (Nb,Zr,Sn,Ti)$O_3$ thin films grown by a sol–gel process. Appl Phys Lett 2003;83:978-80. https://doi.org/10.1063/1.1594843

[112] Hao X, Zhai J, Yao X. Improved energy storage performance and fatigue endurance of Sr-doped $PbZrO_3$ antiferroelectric thin films. J Am Ceram Soc 2009;92:1133-5. https://doi.org/10.1111/j.1551-2916.2009.03015.x

[113] Tagantsev AK, Fousek J, Cross LE. Domains in Ferroic Crystals and Thin Films: Springer New York, NY; 2010. https://link.springer.com/book/10.1007/978-1-4419-1417-0

[114] Junquera J, Ghosez P. Critical thickness for ferroelectricity in perovskite ultrathin films. Nature 2003;422:506-9. https://doi.org/10.1038/nature01501

[115] Fong DD, Stephenson GB, Streiffer SK, Eastman JA, Auciello O, Fuoss PH, et al. Ferroelectricity in Ultrathin Perovskite Films. Science 2004;304:1650-3. https://doi.org/10.1126/science.1098252

[116] Mani BK, Chang CM, Lisenkov S, Ponomareva I. Critical Thickness for Antiferroelectricity in $PbZrO_3$. Phys Rev Lett 2015;115:097601. https://doi.org/10.1103/PhysRevLett.115.097601

[117] Chattopadhyay S, Ayyub P, Palkar VR, Multani MS, Pai SP, Purandare SC, et al. Dielectric properties of oriented thin films of $PbZrO_3$ on Si produced by pulsed laser ablation. J Appl Phys. 1998;83:7808-12. https://doi.org/10.1063/1.367955

[118] Boldyreva K, Pintilie L, Lotnyk A, Misirlioglu IB, Alexe M, Hesse D. Thickness-driven antiferroelectric-to-ferroelectric phase transition of thin $PbZrO_3$ layers in epitaxial $PbZrO_3/Pb(Zr_{0.8}Ti_{0.2})O_3$ multilayers. Appl Phys Lett. 2007;91:122915. https://doi.org/10.1063/1.2789401

[119] Bratkovsky AM, Levanyuk AP. Smearing of Phase Transition due to a Surface Effect or a Bulk Inhomogeneity in Ferroelectric Nanostructures. Phys Rev Lett 2005;94:107601. https://doi.org/10.1103/PhysRevLett.94.107601

[120] Eliseev EA, Glinchuk MD. Size-induced appearance of ferroelectricity in thin antiferroelectric films. Phys B: Conden Matter 2007;400:106-13. https://doi.org/10.1016/j.physb.2007.06.034

[121] Xu R, Crust KJ, Harbola V, Arras R, Patel KY, Prosandeev S, et al. Size-Induced Ferroelectricity in Antiferroelectric Oxide Membranes. Adv Mater 2023;35:2210562. https://doi.org/10.1002/adma.202210562

[122] Jiang R-J, Cao Y, Geng W-R, Zhu M-X, Tang Y-L, Zhu Y-L, et al. Atomic Insight into the Successive Antiferroelectric–Ferroelectric Phase Transition in Antiferroelectric Oxides. Nano Lett 2023;23:1522-9. https://doi.org/10.1021/acs.nanolett.2c04972

[123] Aramberri H, Cazorla C, Stengel M, Íñiguez J. On the possibility that $PbZrO_3$ not be antiferroelectric. npj Comput Mater 2021;7:1-10. https://doi.org/10.1038/s41524-021-00671-w

[124] Chaudhuri AR, Arredondo M, Hähnel A, Morelli A, Becker M, Alexe M, et al. Epitaxial strain stabilization of a ferroelectric phase in $PbZrO_3$ thin films. Phys Rev B 2011;84:054112. https://doi.org/10.1103/PhysRevB.84.054112

[125] Hyuk Park M, Joon Kim H, Jin Kim Y, Lee W, Moon T, Seong Hwang C. Evolution of phases and ferroelectric properties of thin $Hf_{0.5}Zr_{0.5}O_2$ films according to the thickness and annealing temperature. Appl Phys Lett 2013;102:242905. https://doi.org/10.1063/1.4811483

[126] Park MH, Kim HJ, Kim YJ, Lee YH, Moon T, Kim KD, et al. Study on the size effect in $Hf_{0.5}Zr_{0.5}O_2$ films thinner than 8 nm before and after wake-up field cycling. Appl Phys Lett 2015;107. https://doi.org/10.1063/1.4935588
[127] Richter C, Schenk T, Park MH, Tscharntke FA, Grimley ED, LeBeau JM, et al. Si Doped Hafnium Oxide—A “Fragile” Ferroelectric System. Adv Elec Mater 2017;3:1700131. https://doi.org/10.1002/aelm.201700131
[128] Materlik R, Künneth C, Kersch A. The origin of ferroelectricity in $Hf_{1-x}Zr_xO_2$: A computational investigation and a surface energy model. J Appl Phys 2015;117. https://doi.org/10.1063/1.4916707
[129] Cheema SS, Shanker N, Hsu S-L, Rho Y, Hsu C-H, Stoica VA, et al. Emergent ferroelectricity in subnanometer binary oxide films on silicon. Science 2022;376:648-52. https://doi.org/doi:10.1126/science.abm8642
[130] Lomenzo PD, Materano M, Mittmann T, Buragohain P, Gruverman A, Kiguchi T, et al. Harnessing Phase Transitions in Antiferroelectric $ZrO_2$ Using the Size Effect. Adv. Electron. Mater. 2022;8:2100556. https://doi.org/10.1002/aelm.202100556
[131] Matthews J, Blakeslee A. Defects in epitaxial multilayers: I. Misfit dislocations. J Cryst Growth 1974;27:118-25. https://doi.org/10.1016/S0022-0248(74)80055-2
[132] Damodaran AR, Breckenfeld E, Chen Z, Lee S, Martin LW. Enhancement of ferroelectric Curie temperature in $BaTiO_3$ films via strain-induced defect dipole alignment. Adv Mater 2014;26:6341. https://doi.org/10.1002/adma.201400254
[133] Pertsev NA, Koukhar VG. Polarization Instability in Polydomain Ferroelectric Epitaxial Thin Films and the Formation of Heterophase Structures. Phys Rev Lett 2000;84:3722-5. https://doi.org/10.1103/PhysRevLett.84.3722
[134] Reyes-Lillo SE, Rabe KM. Antiferroelectricity and ferroelectricity in epitaxially strained $PbZrO_3$ from first principles. Phys Rev B 2013;88:180102. https://doi.org/10.1103/PhysRevB.88.180102
[135] Zhu L, Wang X, Lou X. Effects of Epitaxial Strain on Antiferrodistortion of $AgNbO_3$ from First-Principle Calculations. Phys Status Solidi - Rapid Res Lett 2018;12:1800007. https://doi.org/10.1002/pssr.201800007
[136] Patel K, Prosandeev S, Xu B, Xu C, Bellaiche L. Properties of (001) $NaNbO_3$ films under epitaxial strain: A first-principles study. Phys Rev B 2021;103:094103. https://doi.org/10.1103/PhysRevB.103.094103
[137] Ge J, Remiens D, Dong X, Chen Y, Costecalde J, Gao F, et al. Enhancement of energy storage in epitaxial $PbZrO_3$ antiferroelectric films using strain engineering. Appl Phys Lett 2014;105:112908. https://doi.org/10.1063/1.4896156
[138] Janolin P-E. Strain on ferroelectric thin films. J Mater Sci 2009;44:5025-48. https://doi.org/10.1007/s10853-009-3553-1
[139] Taylor TR, Hansen PJ, Acikel B, Pervez N, York RA, Streiffer SK, et al. Impact of thermal strain on the dielectric constant of sputtered barium strontium titanate thin films. Appl Phys Lett 2002;80:1978-80. https://doi.org/10.1063/1.1459482
[140] Schroeder U, Park MH, Mikolajick T, Hwang CS. The fundamentals and applications of ferroelectric $HfO_2$. Nat Rev Mater 2022;7:653-69. https://doi.org/10.1038/s41578-022-00431-2

[141] Zhang H, Kalantari K, Marincel D, Trolier-McKinstry S, MacLaren I, Ramasse Q, et al. The effect of substrate clamping on the paraelectric to antiferroelectric phase transition in Nd-doped $BiFeO_3$ thin films. Thin Solid Films 2016;616:767-72. https://doi.org/10.1016/j.tsf.2016.10.004

[142] Hyuk Park M, Joon Kim H, Jin Kim Y, Moon T, Seong Hwang C. The effects of crystallographic orientation and strain of thin $Hf_{0.5}Zr_{0.5}O_2$ film on its ferroelectricity. Appl Phys Lett 2014;104:072901. https://doi.org/10.1063/1.4866008

[143] Ramesh R, Schlom D. Orienting ferroelectric films. Science 2002;296:1975-6. https://doi.org/10.1126/science.1072855

[144] Xu R, Gao R, Reyes-Lillo SE, Saremi S, Dong Y, Lu H, et al. Reducing Coercive-Field Scaling in Ferroelectric Thin Films via Orientation Control. ACS Nano 2018;12:4736-43. https://doi.org/10.1021/acsnano.8b01399

[145] Tsai MF, Zheng YZ, Lu SC, Zheng JD, Pan H, Duan CG, et al. Antiferroelectric Anisotropy of Epitaxial $PbHfO_3$ Films for Flexible Energy Storage. Adv Funct Mater 2021;31:2105060. https://doi.org/10.1002/adfm.202105060

[146] Lisenkov S, Yao Y, Bassiri-Gharb N, Ponomareva I. Prediction of high-strain polar phases in antiferroelectric $PbZrO_3$ from a multiscale approach. Phys Rev B 2020;102:104101. https://doi.org/10.1103/PhysRevB.102.104101

[147] Streiffer SK, Eastman JA, Fong DD, Thompson C, Munkholm A, Ramana Murty MV, et al. Observation of nanoscale 180 degrees stripe domains in ferroelectric $PbTiO_3$ thin films. Phys Rev Lett 2002;89:067601. https://doi.org/10.1103/PhysRevLett.89.067601

[148] Fan Z, Ma T, Wei J, Yang T, Zhou L, Tan X. TEM investigation of the domain structure in $PbHfO_3$ and $PbZrO_3$ antiferroelectric perovskites. J Mater Sci 2020;55:4953-61. https://doi.org/10.1007/s10853-020-04361-8

[149] Lummen TTA, Gu Y, Wang J, Lei S, Xue F, Kumar A, et al. Thermotropic phase boundaries in classic ferroelectrics. Nat Commun 2014;5:3172. https://doi.org/10.1038/ncomms4172

[150] Wei X-K, Jia C-L, Sluka T, Wang B-X, Ye Z-G, Setter N. Néel-like domain walls in ferroelectric $Pb(Zr,Ti)O_3$ single crystals. Nat Commun 2016;7:12385. https://doi.org/10.1038/ncomms12385

[151] Nataf GF, Guennou M, Gregg JM, Meier D, Hlinka J, Salje EKH, et al. Domain-wall engineering and topological defects in ferroelectric and ferroelastic materials. Nat Rev Phys 2020;2:634-48. https://doi.org/10.1038/s42254-020-0235-z

[152] Das S, Tang YL, Hong Z, Gonçalves MAP, McCarter MR, Klewe C, et al. Observation of room-temperature polar skyrmions. Nature 2019;568:368-72. https://doi.org/10.1038/s41586-019-1092-8

[153] Seidel J, Martin LW, He Q, Zhan Q, Chu YH, Rother A, et al. Conduction at domain walls in oxide multiferroics. Nat Mater 2009;8:229. https://doi.org/10.1038/nmat2373

[154] Tanaka M, Saito R, Tsuzuki K. Electron microscopic studies on domain structure of PbZrO3. J Jap Appl Phys 1982;21:291. https://doi.org/10.1143/JJAP.21.291

[155] Wei X-K, Vaideeswaran K, Sandu CS, Jia C-L, Setter N. Preferential Creation of Polar Translational Boundaries by Interface Engineering in Antiferroelectric $PbZrO_3$ Thin Films. Adv Mater Interfaces 2015;2:1500349. https://doi.org/10.1002/admi.201500349

[156] Vakhrushev SB, Andronikova D, Bronwald I, Koroleva EY, Chernyshov D, Filimonov AV, et al. Electric field control of antiferroelectric domain pattern. Phys Rev B 2021;103:214108. https://doi.org/10.1103/PhysRevB.103.214108

[157] Uchino K. Antiferroelectric Shape Memory Ceramics. Actuators 2016;5:11. https://doi.org/10.3390/act5020011

[158] Zhu L-F, Deng S, Zhao L, Li G, Wang Q, Li L, et al. Heterovalent-doping-enabled atom-displacement fluctuation leads to ultrahigh energy-storage density in $AgNbO_3$-based multilayer capacitors. Nat Commun 2023;14:1166. https://doi.org/10.1038/s41467-023-36919-w

[159] Flores Gonzales JE, Ganzha A, Kniazeva M, Andronikova D, Vakulenko A, Filimonov A, et al. Thickness independence of antiferroelectric domain characteristic sizes in epitaxial $PbZrO_3/SrRuO_3/SrTiO_3$ films. J Appl Crystallogr 2023;56:697-706. https://doi.org/10.1107/S1600576723002868

[160] Yuan S, Chen Z, Prokhorenko S, Nahas Y, Bellaiche L, Liu C, et al. Hexagonal Close-Packed Polar-Skyrmion Lattice in Ultrathin Ferroelectric $PbTiO_3$ Films. Phys Rev Lett 2023;130:226801. https://doi.org/10.1103/PhysRevLett.130.226801

[161] Prosandeev S, Lisenkov S, Bellaiche L. Kittel Law in $BiFeO_3$ Ultrathin Films: A First-Principles-Based Study. Phys Rev Lett 2010;105:147603. https://doi.org/10.1103/PhysRevLett.105.147603

[162] Sun Y, Yang J, Li S, Wang D. Defect engineering in perovskite oxide thin films. Chem Commun 2021;57:8402-20. https://doi.org/10.1039/D1CC02276H

[163] Pan H, Tian Z, Acharya M, Huang X, Kavle P, Zhang H, et al. Defect-Induced, Ferroelectric-Like Switching and Adjustable Dielectric Tunability in Antiferroelectrics. Adv Mater 2023:2300257. https://doi.org/10.1002/adma.202300257

[164] Ning S, Kumar A, Klyukin K, Cho E, Kim JH, Su T, et al. An antisite defect mechanism for room temperature ferroelectricity in orthoferrites. Nat Commun 2021;12:4298. https://doi.org/10.1038/s41467-021-24592-w

[165] Wu M, Song D, Vats G, Ning S, Guo M, Zhang D, et al. Defect-controlled electrocaloric effect in $PbZrO_3$ thin films. Journal of Materials Chemistry C 2018;6:10332-40. https://doi.org/10.1039/c8tc03965h

[166] Garcia-Munoz JL, Suaaidi M, Martinez-Lope MJ, Alonso JA. Influence of carrier injection on the metal-insulator transition in electron- and hole-doped R1-xAxNiO3 perovskites. Phys Rev B 1995;52:13563-9. https://doi.org/10.1103/physrevb.52.13563

[167] Takagi H, Ido T, Ishibashi S, Uota M, Uchida S, Tokura Y. Superconductor-to-nonsuperconductor transition in $(La_{1-x}Sr_x)_2CuO_4$ as investigated by transport and magnetic measurements. Phys Rev B 1989;40:2254-61. https://doi.org/10.1103/physrevb.40.2254

[168] Cheema SS, Shanker N, Wang L-C, Hsu C-H, Hsu S-L, Liao Y-H, et al. Ultrathin ferroic HfO2–ZrO2 superlattice gate stack for advanced transistors. Nature 2022;604:65-71. https://doi.org/10.1038/s41586-022-04425-6

[169] Zhang N, Yokota H, Glazer AM, Ren Z, Keen DA, Keeble DS, et al. The missing boundary in the phase diagram of $PbZr_{1-x}Ti_xO_3$. Nat Commun 2014;5:5231. https://doi.org/10.1038/ncomms6231

[170] Gonnard P, Troccaz M. Dopant distribution between A and B sites in the PZT ceramics of type $ABO_3$. J Solid State Chem 1978;23:321-6. https://doi.org/10.1016/0022-4596(78)90080-4
[171] Lee HJ, Won SS, Cho KH, Han CK, Mostovych N, Kingon AI, et al. Flexible high energy density capacitors using La-doped $PbZrO_3$ antiferroelectric thin films. Appl Phys Lett 2018;112:092901. https://doi.org/10.1063/1.5018003
[172] Wang H, Liu Y, Yang T, Zhang S. Ultrahigh Energy-Storage Density in Antiferroelectric Ceramics with Field-Induced Multiphase Transitions. Adv Funct Mater 2019;29:1807321. https://doi.org/10.1002/adfm.201807321
[173] Ma T, Fan Z, Xu B, Kim TH, Lu P, Bellaiche L, et al. Uncompensated Polarization in Incommensurate Modulations of Perovskite Antiferroelectrics. Phys Rev Lett 2019;123:217602. https://doi.org/10.1103/PhysRevLett.123.217602
[174] Li B, Liu QX, Tang XG, Zhang TF, Jiang YP, Li WH, et al. Antiferroelectric to relaxor ferroelectric phase transition in PbO modified $(Pb_{0.97}La_{0.02})(Zr_{0.95}Ti_{0.05})O_3$ ceramics with a large energy-density for dielectric energy storage. RSC Adv 2017;7:43327-33. https://doi.org/10.1039/c7ra08621k
[175] Halliyal A, Gururaja T, Kumar U, Safari A. Stability of Perovskite Phase in $P(Zn_{1/3}Nb_{2/3})O_3$ and Other A(b'b")$O_3$ Perovskites. Sixth IEEE International Symposium on Applications of Ferroelectrics: IEEE; 1986. 437-41. https://doi.org/10.1109/ISAF.1986.201177
[176] Shimizu H, Guo H, Reyes-Lillo SE, Mizuno Y, Rabe KM, Randall CA. Lead-free antiferroelectric: $xCaZrO_3$-(1-x)$NaNbO_3$ system ($0 \leq x \leq 0.10$). Dalton Trans 2015;44:10763-72. https://doi.org/10.1039/C4DT03919J
[177] Xu Y, Guo Y, Liu Q, Bai J, Huang J, Lin L, et al. Antiferroelectricity in new silver niobate lead-free antiferroelectric ceramics (1-x)$AgNbO_3$-x$CaZrO_3$ (x= 0.00–0.01). J Alloy Compd 2019;782:469-76. https://doi.org/10.1016/j.jallcom.2018.12.206
[178] Guo H, Shimizu H, Mizuno Y, Randall CA. Strategy for stabilization of the antiferroelectric phase (Pbma) over the metastable ferroelectric phase (P21ma) to establish double loop hysteresis in lead-free (1-x)$NaNbO_3$-x$SrZrO_3$ solid solution. J Appl Phys 2015;117:214103. https://doi.org/10.1063/1.4921876
[179] Park MH, Kim HJ, Kim YJ, Moon T, Kim KD, Hwang CS. Thin $Hf_xZr_{1-x}O_2$ Films: A New Lead-Free System for Electrostatic Supercapacitors with Large Energy Storage Density and Robust Thermal Stability. Adv. Energy Mater. 2014;4:1400610. https://doi.org/10.1002/aenm.201400610
[180] Mueller S, Mueller J, Singh A, Riedel S, Sundqvist J, Schroeder U, et al. Incipient Ferroelectricity in Al-Doped $HfO_2$ Thin Films. Adv Funct Mater 2012;22:2412-7. https://doi.org/10.1002/adfm.201103119
[181] Lee CK, Cho E, Lee HS, Hwang CS, Han S. First-principles study on doping and phase stability of $HfO_2$. Phys Rev B 2008;78:012102. https://doi.org/10.1103/PhysRevB.78.012102
[182] Lomenzo PD, Chung CC, Zhou C, Jones JL, Nishida T. Doped $Hf_{0.5}Zr_{0.5}O_2$ for high efficiency integrated supercapacitors. Appl Phys Lett 2017;110:232904. https://doi.org/10.1063/1.4985297
[183] Ohtomo A, Hwang H. A high-mobility electron gas at the $LaAlO_3/SrTiO_3$ heterointerface. Nature 2004;427:423-6. https://doi.org/10.1038/nature02308

[184] Chen S, Yuan S, Hou Z, Tang Y, Zhang J, Wang T, et al. Recent Progress on Topological Structures in Ferroic Thin Films and Heterostructures. Adv Mater 2021;33:2000857. https://doi.org/10.1002/adma.202000857
[185] Bousquet E, Dawber M, Stucki N, Lichtensteiger C, Hermet P, Gariglio S, et al. Improper ferroelectricity in perovskite oxide artificial superlattices. Nature 2008;452:732-6. https://doi.org/10.1038/nature06817
[186] Aramberri H, Fedorova NS, Íñiguez J. Ferroelectric/paraelectric superlattices for energy storage. Sci Adv 2022;8:4880. https://doi.org/10.1126/sciadv.abn4880
[187] Dong W, Peters JJP, Rusu D, Staniforth M, Brunier AE, Lloyd-Hughes J, et al. Emergent Antipolar Phase in $BiFeO_3$-$La_{0.7}Sr_{0.3}MnO_3$ Superlattice. Nano Lett 2020;20:6045-50. https://doi.org/10.1021/acs.nanolett.0c02063
[188] Mundy JA, Grosso BF, Heikes CA, Ferenc Segedin D, Wang Z, Shao YT, et al. Liberating a hidden antiferroelectric phase with interfacial electrostatic engineering. Sci Adv 2022;8:eabg5860. https://doi.org/10.1126/sciadv.abg5860
[189] Caretta L, Shao Y-T, Yu J, Mei AB, Grosso BF, Dai C, et al. Non-volatile electric-field control of inversion symmetry. Nat Mater 2023;22:207-15. https://doi.org/10.1038/s41563-022-01412-0
[190] Liu SZ, Geng WR, Tang YL, Zhu YL, Wang YJ, Cao Y, et al. Engineering antiferroelectric nucleation in ferroelectric films with enhanced piezoelectricity. Acta Mater 2023;250:118885. https://doi.org/10.1016/j.actamat.2023.118885
[191] Park JY, Lee DH, Yang K, Kim SH, Yu GT, Park GH, et al. Engineering Strategies in Emerging Fluorite-Structured Ferroelectrics. ACS Appl Elec Mater 2022;4:1369-80. https://doi.org/10.1021/acsaelm.1c00792
[192] Park MH, Kim HJ, Lee G, Park J, Lee YH, Kim YJ, et al. A comprehensive study on the mechanism of ferroelectric phase formation in hafnia-zirconia nanolaminates and superlattices. Appl Phys Rev 2019;6:041403. https://doi.org/10.1063/1.5118737
[193] Wang CY, Wang CI, Yi SH, Chang TJ, Chou CY, Yin YT, et al. Paraelectric/antiferroelectric/ferroelectric phase transformation in As-deposited $ZrO_2$ thin films by the TiN capping engineering. Mater Des 2020;195:109020. https://doi.org/10.1016/j.matdes.2020.109020
[194] Yi SH, Lin BT, Hsu TY, Shieh J, Chen MJ. Modulation of ferroelectricity and antiferroelectricity of nanoscale $ZrO_2$ thin films using ultrathin interfacial layers. J Eur Ceram Soc 2019;39:4038-45. https://doi.org/10.1016/j.jeurceramsoc.2019.05.065
[195] He J, Borisevich A, Kalinin SV, Pennycook SJ, Pantelides ST. Control of Octahedral Tilts and Magnetic Properties of Perovskite Oxide Heterostructures by Substrate Symmetry. Phys Rev Lett 2010;105:227203. https://doi.org/10.1103/PhysRevLett.105.227203
[196] Kim TH, Puggioni D, Yuan Y, Xie L, Zhou H, Campbell N, et al. Polar metals by geometric design. Nat 2016;533:68-72. https://doi.org/10.1038/nature17628
[197] Rondinelli JM, Spaldin NA. Substrate coherency driven octahedral rotations in perovskite oxide films. Phys Rev B 2010;82:113402. https://doi.org/10.1103/PhysRevB.82.113402
[198] Jeong SG, Han G, Song S, Min T, Mohamed AY, Park S, et al. Propagation Control of Octahedral Tilt in $SrRuO_3$ via Artificial Heterostructuring. Adv Sci 2020;7:2001643. https://doi.org/10.1002/advs.202001643

[199] Gao R, Dong Y, Xu H, Zhou H, Yuan Y, Gopalan V, et al. Interfacial Octahedral Rotation Mismatch Control of the Symmetry and Properties of $SrRuO_3$. ACS Appl Mater Interface 2016;8:14871-8. https://doi.org/10.1021/acsami.6b02864

[200] Bellaiche L, Íñiguez J. Universal collaborative couplings between oxygen-octahedral rotations and antiferroelectric distortions in perovskites. Phys Rev B 2013;88:014104. https://doi.org/10.1103/PhysRevB.88.014104

[201] Zhao HJ, Chen P, Prosandeev S, Artyukhin S, Bellaiche L. Dzyaloshinskii–Moriya-like interaction in ferroelectrics and antiferroelectrics. Nat Mater 2021;20:341-5. https://doi.org/10.1038/s41563-020-00821-3

[202] Burkovsky RG, Bronwald I, Andronikova D, Lityagin G, Piecha J, Souliou SM, et al. Triggered incommensurate transition in $PbHfO_3$. Phys Rev B 2019;100:014107. https://doi.org/10.1103/PhysRevB.100.014107

[203] Kniazeva MA, Ganzha AE, Jankowska-Sumara I, Paściak M, Majchrowski A, Filimonov AV, et al. Ferroelectric to incommensurate fluctuations crossover in $PbHfO_3$-$PbSnO_3$. Phys Rev B 2022;105:014101. https://doi.org/10.1103/PhysRevB.105.014101

[204] Uecker R, Bertram R, Brützam M, Galazka Z, Gesing TM, Guguschev C, et al. Large-lattice-parameter perovskite single-crystal substrates. J Cryst Growth 2017;457:137-42. https://doi.org/10.1016/j.jcrysgro.2016.03.014

[205] James KK, Krishnaprasad PS, Hasna K, Jayaraj MK. Structural and optical properties of La-doped $BaSnO_3$ thin films grown by PLD. J Phys Chem Solids 2015;76:64-9. https://doi.org/10.1016/j.jpcs.2014.07.024

[206] Choi KJ, Biegalski M, Li Y, Sharan A, Schubert J, Uecker R, et al. Enhancement of ferroelectricity in strained $BaTiO_3$ thin films. Science 2004;306:1005-9. https://doi.org/10.1126/science.1103218

[207] Wei XK, Jia CL, Roleder K, Setter N. Polarity of translation boundaries in antiferroelectric $PbZrO_3$. Mater Res Bull 2015;62:101-5. https://doi.org/10.1016/j.materresbull.2014.11.024

[208] Mtebwa M, Feigl L, Yudin P, McGilly LJ, Shapovalov K, Tagantsev AK, et al. Room temperature concurrent formation of ultra-dense arrays of ferroelectric domain walls. Appl Phys Lett 2015;107:142903. https://doi.org/10.1063/1.4932524

[209] Sawaguchi E. Ferroelectricity versus antiferroelectricity in the solid solutions of $PbZrO_3$ and $PbTiO_3$. J Phys Soc Jpn 1953;8:615-29. https://doi.org/10.1143/JPSJ.8.615

[210] Li Z, Fu Z, Cai H, Hu T, Yu Z, Luo Y, et al. Discovery of electric devil's staircase in perovskite antiferroelectric. Sci Adv 2022;8:9088. https://doi.org/10.1126/sciadv.abl9088

[211] Gao M, Tang X, Dai S, Li J, Viehland D. Depth dependent ferroelectric to incommensurate/commensurate antiferroelectric phase transition in epitaxial lanthanum modified lead zirconate titanate thin films. Appl Phys Lett 2019;115:072901. https://doi.org/10.1063/1.5113720

[212] Bak P, von Boehm J. Ising model with solitons, phasons, and "the devil's staircase". Phys Rev B 1980;21:5297-308. https://doi.org/10.1103/PhysRevB.21.5297

[213] Wei XK, Jia CL, Du HC, Roleder K, Mayer J, Dunin-Borkowski RE. An unconventional transient phase with cycloidal order of polarization in energy-storage antiferroelectric $PbZrO_3$. Adv Mater 2020;32:1907208. https://doi.org/10.1002/adma.201907208

[214] Fthenakis ZG, Ponomareva I. Intrinsic dynamics of the electric-field-induced phase switching in antiferroelectric $PbZrO_3$ ultrathin films. Phys Rev B 2018;98:054107. https://doi.org/10.1103/PhysRevB.98.054107
[215] Knjazeva M, Bronwald Y, Andronikova D, Lityagin G, Bosak A, Paraskevas P, et al. Modulated Structures in $PbHfO_3$ Crystals at High-Pressure-High-Temperature Conditions. Def Diff Forum 2018;386:149-55. https://doi.org/10.4028/www.scientific.net/DDF.386.149
[216] Bosak A, Svitlyk V, Arakcheeva A, Burkovsky R, Diadkin V, Roleder K, et al. Incommensurate crystal structure of $PbHfO_3$. Acta Crystallogr B Struct Sci Cryst Eng Mater 2020;76:7-12. https://doi.org/10.1107/S205252061901494X
[217] Matthias B, Remeika J. Dielectric properties of sodium and potassium niobates. Phys Rev. 1951;82:727. https://doi.org/10.1103/PhysRev.82.727
[218] Yang D, Gao J, Shu L, Liu YX, Yu J, Zhang Y, et al. Lead-free antiferroelectric niobates $AgNbO_3$ and $NaNbO_3$ for energy storage applications. J Mater Chem A 2020;8:23724-37. https://doi.org/10.1039/d0ta08345c
[219] Guo H, Shimizu H, Randall CA. Direct evidence of an incommensurate phase in $NaNbO_3$ and its implication in $NaNbO_3$-based lead-free antiferroelectrics. Appl Phys Lett 2015;107:112904. https://doi.org/10.1063/1.4930067
[220] Mishra SK, Mittal R, Pomjakushin VY, Chaplot SL. Phase stability and structural temperature dependence in sodium niobate: A high-resolution powder neutron diffraction study. Phys Rev B 2011;83:134105. https://doi.org/10.1103/PhysRevB.83.134105
[221] Qi H, Xie A, Fu J, Zuo R. Emerging antiferroelectric phases with fascinating dielectric, polarization and strain response in $NaNbO_3$-$(Bi_{0.5}Na_{0.5})TiO_3$ lead-free binary system. Acta Mater 2021;208:116710. https://doi.org/10.1016/j.actamat.2021.116710
[222] Darlington C, Megaw HD. The low-temperature phase transition of sodium niobate and the structure of the low-temperature phase. Acta Crystallogr B Struc Crystallogr Cryst Chem 1973;29:2171-85. https://doi.org/10.1107/S0567740873006308
[223] A K, J K. $Ag_{1-x}Na_xNbO_3$ (ANN) solid solutions: from disordered antiferroelectric $AgNbO_3$ to normal antiferroelectric $NaNbO_3$. J Phys Condens Matter 1999;11:8933. https://doi.org/10.1088/0953-8984/11/45/316
[224] Yang Y, Xu B, Xu C, Ren W, Bellaiche L. Understanding and revisiting the most complex perovskite system via atomistic simulations. Phys Rev B 2018;97:174106. https://doi.org/10.1103/PhysRevB.97.174106
[225] Guo H, Shimizu H, Mizuno Y, Randall CA. Domain configuration changes under electric field-induced antiferroelectric-ferroelectric phase transitions in $NaNbO_3$-based ceramics. J Appl Phys 2015;118:054102. https://doi.org/10.1063/1.4928153
[226] Zhang M-H, Hadaeghi N, Egert S, Ding H, Zhang H, Groszewicz PB, et al. Design of Lead-Free Antiferroelectric (1–x)$NaNbO_3$–x$SrSnO_3$ Compositions Guided by First-Principles Calculations. Chem Mater 2021;33:266-74. https://doi.org/10.1021/acs.chemmater.0c03685
[227] Yuzyuk YI, Shakhovoy R, Raevskaya S, Raevski I, El Marssi M, Karkut MG, et al. Ferroelectric Q-phase in a $NaNbO_3$ epitaxial thin film. Appl Phys Lett 2010;96:222904. https://doi.org/10.1063/1.3437090
[228] Reisman A, Holtzberg F. Heterogeneous equilibria in the systems $Li_2O$-, $Ag_2O$-$Nb_2O_5$ and oxide-models. J Am Chem Soc 1958;80:6503-7. https://doi.org/10.1021/ja01557a010

[229] Kania A, Roleder K, Łukaszewski M. The ferroelectric phase in $AgNbO_3$. Ferroelectrics 1984;52:265-9. https://doi.org/10.1080/00150198408209394
[230] Gao J, Li Q, Zhang S, Li J-F. Lead-free antiferroelectric $AgNbO_3$: Phase transitions and structure engineering for dielectric energy storage applications. J Appl Phys 2020;128:070903. https://doi.org/10.1063/5.0018373
[231] Fu D, Endo M, Taniguchi H, Taniyama T, Itoh M. $AgNbO_3$: A lead-free material with large polarization and electromechanical response. Appl Phys Lett 2007;90:252907. https://doi.org/10.1063/1.2751136
[232] Levin I, Krayzman V, Woicik JC, Karapetrova J, Proffen T, Tucker MG, et al. Structural changes underlying the diffuse dielectric response in $AgNbO_3$. Phys Rev B 2009;79:104113. https://doi.org/10.1103/PhysRevB.79.104113
[233] Yashima M, Matsuyama S, Sano R, Itoh M, Tsuda K, Fu D. Structure of Ferroelectric Silver Niobate $AgNbO_3$. Chem Mater 2011;23:1643-5. https://doi.org/10.1021/cm103389q
[234] Luo N, Han K, Cabral MJ, Liao X, Zhang S, Liao C, et al. Constructing phase boundary in AgNbO3 antiferroelectrics: pathway simultaneously achieving high energy density and efficiency. Nat Commun 2020;11:4824. https://doi.org/10.1038/s41467-020-18665-5
[235] Yanle Zhang, Xiaobo Li, Jianmin Song, Suwei Zhang, Jing Wang, Xiuhong Dai, et al. AgNbO3 antiferroelectric film with high energy storage performance. J Materiomics 2021;7:1294-300. https://doi.org/10.1016/j.jmat.2021.02.018
[236] Sakurai H, Yamazoe S, Wada T. Ferroelectric and antiferroelectric properties of $AgNbO_3$ films fabricated on (001), (110), and (111)$SrTiO_3$ substrates by pulsed laser deposition. Appl Phys Lett 2010;97:042901. https://doi.org/10.1063/1.3467137
[237] Ren P, Ren D, Sun L, Yan F, Yang S, Zhao G. Grain size tailoring and enhanced energy storage properties of two-step sintered $Nd^{3+}$-doped $AgNbO_3$. J Eur Ceram Soc 2020;40:4495-502. https://doi.org/10.1016/j.jeurceramsoc.2020.05.076
[238] Smolenskii G, Isupov V, Agranovskaya A, Krainik N. New materials of AIIBIVOVI type. Trans Sov Phys Solid State 1961;2:2651-4.
[239] Ma C, Tan X. In situ Transmission Electron Microscopy Study on the Phase Transitionsin Lead-Free (1−x)$(Bi_{1/2}Na_{1/2})TiO_3$–x$BaTiO_3$ Ceramics. J Am Ceram Soc 2011;94:4040-4. https://doi.org/10.1111/j.1551-2916.2011.04670.x
[240] Carter J, Aksel E, Iamsasri T, Forrester JS, Chen J, Jones JL. Structure and ferroelectricity of nonstoichiometric $(Na_{0.5}Bi_{0.5})TiO_3$. Appl Phys Lett 2014;104:112904. https://doi.org/10.1063/1.4868109
[241] Luo H, Qi H, Sun S, Wang L, Ren Y, Liu H, et al. Structural origin for the high piezoelectric performance of $(Na_{0.5}Bi_{0.5})TiO_3$-$BaTiO_3$-$BiAlO_3$ lead-free ceramics. Acta Mater 2021;218:117202. https://doi.org/10.1016/j.actamat.2021.117202
[242] Li L, Fan P, Wang M, Takesue N, Salamon D, Vtyurin AN, et al. Review of lead-free Bi-based dielectric ceramics for energy-storage applications. J Phys D: Appl Phys 2021;54:293001. https://doi.org/10.1088/1361-6463/abf860
[243] Zvirgzds JA, Kapostin PP, Zvirgzde JV, Kruzina TV. X-ray study of phase transitions in efrroelectric $Na_{0.5}Bi_{0.5}TiO_3$. Ferroelectrics. 1982;40:75-7. https://doi.org/10.1080/00150198208210600

[244] Jones G, Thomas P. Investigation of the structure and phase transitions in the novel A-site substituted distorted perovskite compound $Na_{0.5}Bi_{0.5}TiO_3$. Acta Crystallogr Sect B: Struct Sci. 2002;58:168-78. https://doi.org/10.1107/s0108768101020845
[245] Gorfman S, Thomas P. Evidence for a non-rhombohedral average structure in the lead-free piezoelectric material $Na_{0.5}Bi_{0.5}TiO_3$. J Appl Crystallogr. 2010;43:1409-14. https://doi.org/10.1107/S002188981003342X
[246] Aksel E, Forrester JS, Jones JL, Thomas PA, Page K, Suchomel MR. Monoclinic crystal structure of polycrystalline $Na_{0.5}Bi_{0.5}TiO_3$. Appl Phys Lett. 2011;98:152901. https://doi.org/10.1063/1.3573826
[247] Rao BN, Ranjan R. Electric-field-driven monoclinic-to-rhombohedral transformation in $Na_{1/2}Bi_{1/2}TiO_3$. Phys Rev B 2012;86:134103. https://doi.org/10.1103/PhysRevB.86.134103
[248] Rao BN, Fitch AN, Ranjan R. Ferroelectric-ferroelectric phase coexistence in $Na_{1/2}Bi_{1/2}TiO_3$. Phys Rev B 2013;87:060102. https://doi.org/10.1103/PhysRevB.87.060102
[249] Trolliard G, Dorcet V. Reinvestigation of phase transitions in $Na_{0.5}Bi_{0.5}TiO_3$ by TEM. Part II: Second order orthorhombic to tetragonal phase transition. Chem Mater 2008;20:5074-82. https://doi.org/10.1021/cm800464d
[250] Dorcet V, Trolliard G, Boullay P. Reinvestigation of phase transitions in $Na_{0.5}Bi_{0.5}TiO_3$ by TEM. Part I: First order rhombohedral to orthorhombic phase transition. Chem Mater 2008;20:5061-73. https://doi.org/10.1021/cm8004634
[251] Zhang ST, Kounga AB, Aulbach E, Deng Y. Temperature-Dependent Electrical Properties of $0.94Bi_{0.5}Na_{0.5}TiO_3$-$0.06BaTiO_3$ Ceramics. J Am Ceram Soc 2008;91:3950-4. https://doi.org/10.1111/j.1551-2916.2008.02778.x
[252] Qi H, Zuo R. Linear-like lead-free relaxor antiferroelectric $(Bi_{0.5}Na_{0.5})TiO_3$–$NaNbO_3$ with giant energy-storage density/efficiency and super stability against temperature and frequency. J Mater Chem A 2019;7:3971-8. https://doi.org/10.1039/c8ta12232f
[253] Royen P, Swars K. Das System Wismutoxyd-Eisenoxyd im Bereich von 0 bis 55 mol% Eisenoxyd. Angew Chem 1957;69:779-. https://doi.org/10.1002/ange.19570692407
[254] Prosandeev S, Wang D, Ren W, Íñiguez J, Bellaiche L. Novel nanoscale twinned phases in perovskite oxides. Adv Funct Mater 2013;23:234-40. https://doi.org/10.1002/adfm.201201467
[255] Ravindran P, Vidya R, Kjekshus A, Fjellvåg H, Eriksson O. Theoretical investigation of magnetoelectric behavior in $BiFeO_3$. Phys Rev B 2006;74:224412. https://doi.org/10.1103/PhysRevB.74.224412
[256] Carcan B, Bouyanfif H, El Marssi M, Le Marrec F, Dupont L, Davoisne C, et al. Phase Diagram of $BiFeO_3$/$LaFeO_3$ Superlattices: Antiferroelectric-Like State Stability Arising from Strain Effects and Symmetry Mismatch at Heterointerfaces. Adv Mater Interfaces 2017;4:1601036. https://doi.org/10.1002/admi.201601036
[257] Xu B, Iniguez J, Bellaiche L. Designing lead-free antiferroelectrics for energy storage. Nat Commun 2017;8:1-8. https://doi.org/10.1038/ncomms15682
[258] Kan D, Long CJ, Steinmetz C, Lofland SE, Takeuchi I. Combinatorial search of structural transitions: Systematic investigation of morphotropic phase boundaries in chemically substituted $BiFeO_3$. J Mater Res 2012;27:2691-704. https://doi.org/10.1557/jmr.2012.314

[259] Fujino S, Murakami M, Anbusathaiah V, Lim SH, Nagarajan V, Fennie CJ, et al. Combinatorial discovery of a lead-free morphotropic phase boundary in a thin-film piezoelectric perovskite. Appl Phys Lett 2008;92. https://doi.org/10.1063/1.2931706

[260] Chen D, Nelson CT, Zhu X, Serrao CR, Clarkson JD, Wang Z, et al. A Strain-Driven Antiferroelectric-to-Ferroelectric Phase Transition in La-Doped $BiFeO_3$ Thin Films on Si. Nano Lett 2017;17:5823-9. https://doi.org/10.1021/acs.nanolett.7b03030

[261] Lennox RC, Price MC, Jamieson W, Jura M, Daoud-Aladine A, Murray CA, et al. Strain driven structural phase transformations in dysprosium doped $BiFeO_3$ ceramics. J Mater Chem C 2014;2:3345-60. https://doi.org/10.1039/c3tc32345e

[262] Levin I, Tucker MG, Wu H, Provenzano V, Dennis CL, Karimi S, et al. Displacive Phase Transitions and Magnetic Structures in Nd-Substituted $BiFeO_3$. Chem Mater 2011;23:2166-75. https://doi.org/10.1021/cm1036925

[263] Böscke T, Müller J, Bräuhaus D, Schröder U, Böttger U. Ferroelectricity in hafnium oxide thin films. Appl Phys Lett 2011;99:102903. https://doi.org/10.1063/1.3634052

[264] Lomenzo PD, Materano M, Richter C, Alcala R, Mikolajick T, Schroeder U. A Gibbs energy view of double hysteresis in $ZrO_2$ and Si-doped $HfO_2$. Appl Phys Lett 2020;117:142904. https://doi.org/10.1063/5.0018199

[265] Collins L, Celano U. Revealing antiferroelectric switching and ferroelectric wakeup in hafnia by advanced piezoresponse force microscopy. ACS Appl Mater Interfaces 2020;12:41659-65. https://doi.org/10.1021/acsami.0c07809

[266] Chang S-C, Haratipour N, Shivaraman S, Brown-Heft TL, Peck J, Lin C-C, et al. Anti-ferroelectric $Hf_xZr_{1-x}O_2$ capacitors for high-density 3-D embedded-DRAM. 2020 IEEE International Electron Devices Meeting (IEDM); 2020.28.1.1-.1.4. https://doi.org/10.1109/IEDM13553.2020.9372011

[267] Pešić M, Schroeder U. Antiferroelectric One Transistor/One Capacitor Memory Cell. Ferroelectricity in Doped Hafnium Oxide: Materials, Properties and Devices: Elsevier; 2019. p. 425-35. https://doi.org/10.1016/B978-0-08-102430-0.00020-6

[268] Johnson B, Jones JL. Structures, Phase Equilibria, and Properties of $HfO_2$. Ferroelectricity in Doped Hafnium Oxide: Materials, Properties and Devices: Elsevier; 2019. p. 25-45. https://doi.org/10.1016/C2017-0-01145-X

[269] Park MH, Schenk T, Schroeder U. Dopants in atomic layer deposited $HfO_2$ thin films. Ferroelectricity in Doped Hafnium Oxide: Materials, Properties and Devices: Elsevier; 2019. p. 49-74. https://doi.org/10.1016/B978-0-08-102430-0.00005-X

[270] Qi Y, Singh S, Lau C, Huang F-T, Xu X, Walker FJ, et al. Stabilization of Competing Ferroelectric Phases of $HfO_2$ under Epitaxial Strain. Phys Rev Lett 2020;125:257603. https://doi.org/10.1103/PhysRevLett.125.257603

[271] Goh Y, Cho SH, Park S-HK, Jeon S. Oxygen vacancy control as a strategy to achieve highly reliable hafnia ferroelectrics using oxide electrode. Nanoscale 2020;12:9024-31. https://doi.org/10.1039/d0nr00933d

[272] Song T, Bachelet R, Saint-Girons G, Dix N, Fina I, Sánchez F. Thickness effect on the ferroelectric properties of La-doped $HfO_2$ epitaxial films down to 4.5 nm. J Mater Chem C 2021;9:12224-30. https://doi.org/10.1039/D1TC02512K

[273] Nukala P, Ahmadi M, Wei Y, De Graaf S, Stylianidis E, Chakrabortty T, et al. Reversible oxygen migration and phase transitions in hafnia-based ferroelectric devices. Science 2021;372:630-5. https://doi.org/10.1126/science.abf3789
[274] Park MH, Lee YH, Kim HJ, Schenk T, Lee W, Do Kim K, et al. Surface and grain boundary energy as the key enabler of ferroelectricity in nanoscale hafnia-zirconia: A comparison of model and experiment. Nanoscale 2017;9:9973-86. https://doi.org/10.1039/C7NR02121F
[275] Tashiro Y, Shimizu T, Mimura T, Funakubo H. Comprehensive study on the kinetic formation of the orthorhombic ferroelectric phase in epitaxial Y-doped ferroelectric $HfO_2$ thin films. ACS Appl Elec Mater 2021;3:3123-30. https://doi.org/10.1021/acsaelm.1c00342
[276] Cheng Y, Gao Z, Ye KH, Park HW, Zheng Y, Zheng Y, et al. Reversible transition between the polar and antipolar phases and its implications for wake-up and fatigue in $HfO_2$-based ferroelectric thin film. Nat Commun 2022;13:1-8. https://doi.org/10.1038/s41467-022-28236-5
[277] Ding W, Zhang Y, Tao L, Yang Q, Zhou Y. The atomic-scale domain wall structure and motion in $HfO_2$-based ferroelectrics: A first-principle study. Acta Mater 2020;196:556-64. https://doi.org/10.1016/j.actamat.2020.07.012
[278] Zhou D, Xu J, Li Q, Guan Y, Cao F, Dong X, et al. Wake-up effects in Si-doped hafnium oxide ferroelectric thin films. Appl Phys Lett 2013;103:192904. https://doi.org/10.1063/1.4829064
[279] Schenk T, Hoffmann M, Ocker J, Pešić M, Mikolajick T, Schroeder U. Complex internal bias fields in ferroelectric hafnium oxide. ACS Appl Mater Interfaces 2015;7:20224-33. https://doi.org/10.1021/acsami.5b05773
[280] Chouprik A, Spiridonov M, Zarubin S, Kirtaev R, Mikheev V, Lebedinskii Y, et al. Wake-up in a $Hf_{0.5}Zr_{0.5}O_2$ film: A cycle-by-cycle emergence of the remnant polarization via the domain depinning and the vanishing of the anomalous polarization switching. ACS Appl Elec Mater 2019;1:275-87. https://doi.org/10.1021/acsaelm.8b00046
[281] Pešić M, Fengler FPG, Larcher L, Padovani A, Schenk T, Grimley ED, et al. Physical mechanisms behind the field-cycling behavior of $HfO_2$-based ferroelectric capacitors. Adv Funct Mater 2016;26:4601-12. https://doi.org/10.1002/adfm.201600590
[282] Pešić M, Hoffmann M, Richter C, Mikolajick T, Schroeder U. Nonvolatile random access memory and energy storage based on antiferroelectric like hysteresis in $ZrO_2$. Adv Funct Mater. 2016;26:7486-94. https://doi.org/10.1002/adfm.201603182
[283] Courtens E, Rosenbaum TF, Nagler SE, Horn PM. Short-range ordering and freezing in a randomly mixed ferroelectric-antiferroelectric crystal. Phys Rev B 1984;29:515-8. https://doi.org/10.1103/PhysRevB.29.515
[284] Duan CC, Tan GL. Tuning ferroelectrics to antiferroelectrics in multiferroic $La_xSr_{1-x}Fe_{12}O_{19}$ ceramics. J Mater Res 2022;37:1651-63. https://doi.org/10.1557/s43578-022-00568-4
[285] Tan G, Nan N, Sharma P, Kumar A. Antiferroelectric and magnetic performance in $La_{0.2}Sr_{0.7}Fe_{12}O_{19}$ system. J Mater Sci: Mater Elec 2021;32:21697-708. https://doi.org/10.1007/s10854-021-06689-6
[286] Yoshida S, Fujita K, Akamatsu H, Hernandez O, Sen Gupta A, Brown FG, et al. Ferroelectric $Sr_3Zr_2O_7$: Competition between Hybrid Improper Ferroelectric and

Antiferroelectric Mechanisms. Adv Funct Mater 2018;28:1801856. https://doi.org/10.1002/adfm.201801856

[287] Randall CA, Fan Z, Reaney I, Chen LQ, Trolier-McKinstry S. Antiferroelectrics: History, fundamentals, crystal chemistry, crystal structures, size effects, and applications. J Am Ceram Soc 2021;104:3775-810. https://doi.org/10.1111/jace.17834

[288] Burkovsky RG, Bronwald I, Andronikova D, Wehinger B, Krisch M, Jacobs J, et al. Critical scattering and incommensurate phase transition in antiferroelectric $PbZrO_3$ under pressure. Sci Rep 2017;7:41512. https://doi.org/10.1038/srep41512

[289] Wu Z, Liu X, Ji C, Li L, Wang S, Peng Y, et al. Discovery of an above-room-temperature antiferroelectric in two-dimensional hybrid perovskite. J Am Chem Soc 2019;141:3812-6. https://doi.org/10.1021/jacs.8b13827

[290] Li MF, Han SG, Liu Y, Luo JH, Hong MC, Sun ZH. Soft Perovskite-Type Antiferroelectric with Giant Electrocaloric Strength near Room Temperature. J Am Chem Soc 2020;142:20744-51. https://doi.org/10.1021/jacs.0c09601

[291] Wei XK, Domingo N, Sun Y, Balke N, Dunin-Borkowski RE, Mayer J. Progress on Emerging Ferroelectric Materials for Energy Harvesting, Storage and Conversion. Adv Energy Mater 2022;12:2201199. https://doi.org/10.1002/aenm.202201199

[292] Jieun Kim, Sahar Saremi, Megha Acharya, Gabriel Velarde, Eric Parsonnet, Patrick Donahue, et al. Ultrahigh capacitive energy density in ion-bombarded relaxor ferroelectric films. Science 2020;369:81-4. https://doi.org/10.1126/science.abb0631

[293] Chen L, Deng S, Liu H, Wu J, Qi H, Chen J. Giant energy-storage density with ultrahigh efficiency in lead-free relaxors via high-entropy design. Nat Commun 2022;13:3089. https://doi.org/10.1038/s41467-022-30821-7

[294] Xu B, Moses P, Pai NG, Cross LE. Charge release of lanthanum-doped lead zirconate titanate stannate antiferroelectric thin films. Appl Phys Lett 1998;72:593-5. https://doi.org/10.1063/1.120817

[295] Pan H, Li F, Liu Y, Zhang Q, Wang M, Lan S, et al. Ultrahigh-energy density lead-free dielectric films via polymorphic nanodomain design. Science. 2019;365:578-82. https://doi.org/10.1126/science.aaw8109

[296] Qi H, Zuo R, Xie A, Tian A, Fu J, Zhang Y, et al. Ultrahigh Energy-Storage Density in $NaNbO_3$-Based Lead-Free Relaxor Antiferroelectric Ceramics with Nanoscale Domains. Adv Funct Mater 2019;29:1903877. https://doi.org/10.1002/adfm.201903877

[297] Zhang T, Si Y, Deng S, Wang H, Wang T, Shao J, et al. Superior Energy Storage Performance in Antiferroelectric Epitaxial Thin Films via Structural Heterogeneity and Orientation Control. Adv Funct Mater 2023:2311160. https://doi.org/10.1002/adfm.202311160

[298] Zhu C, Cai Z, Luo B, Cheng X, Guo L, Jiang Y, et al. Multiphase engineered BNT-based ceramics with simultaneous high polarization and superior breakdown strength for energy storage applications. ACS Appl Mater Interfaces 2021;13:28484-92. https://doi.org/10.1021/acsami.1c06075

[299] Lu H, Liu X, Burton JD, Bark CW, Wang Y, Zhang Y, et al. Enhancement of ferroelectric polarization stability by interface engineering. Adv Mater 2012;24:1209. https://doi.org/10.1002/adma.201104398

[300] Sun Y, Bealing C, Boggs S, Ramprasad R. 50+ years of intrinsic breakdown. IEEE Elec Insul Mag 2013;29:8-15. https://doi: 10.1109/MEI.2013.6457595

[301] Owate IO, Freer R. AC breakdown characteristics of ceramic materials. J Appl Phys. 1992;72:2418-22. https://doi.org/10.1063/1.351586

[302] Xie A, Qi H, Zuo R. Achieving Remarkable Amplification of Energy-Storage Density in Two-Step Sintered $NaNbO_3$-$SrTiO_3$ Antiferroelectric Capacitors through Dual Adjustment of Local Heterogeneity and Grain Scale. ACS Appl Mater Interfaces 2020;12:19467-75. https://doi.org/10.1021/acsami.0c00831

[303] Tong S. Size and temperature effects on dielectric breakdown of ferroelectric films. J Adv Ceram 2021;10:181-6. https://doi.org/10.1007/s40145-020-0426-1

[304] Wang PJ, Zhou D, Guo HH, Liu WF, Su JZ, Fu MS, et al. Ultrahigh enhancement rate of the energy density of flexible polymer nanocomposites using core–shell $BaTiO_3$@MgO structures as the filler. J Mater Chem A 2020;8:11124-32. https://doi.org/10.1039/d0ta03304a

[305] Wang G, Lu Z, Li Y, Li L, Ji H, Feteira A, et al. Electroceramics for High-Energy Density Capacitors: Current Status and Future Perspectives. Chem Rev 2021;121:6124–72. https://doi.org/10.1021/acs.chemrev.0c01264

[306] Chen L, Long F, Qi H, Liu H, Deng S, Chen J. Outstanding Energy Storage Performance in High-Hardness $(Bi_{0.5}K_{0.5})TiO_3$-Based Lead-Free Relaxors via Multi-Scale Synergistic Design. Adv Funct Mater 2021;32:2110478. https://doi.org/10.1002/adfm.202110478

[307] Neusel C, Jelitto H, Schmidt D, Janssen R, Felten F, Schneider GA. Thickness-dependence of the breakdown strength: Analysis of the dielectric and mechanical failure. J Eur Ceram Soc 2015;35:113-23. https://doi.org/10.1016/j.jeurceramsoc.2014.08.028

[308] Li L, Bai Y, Wang S, Zhang T. A Superhydrophobic Smart Coating for Flexible and Wearable Sensing Electronics. Adv Mater 2017;29:1702517. https://doi.org/10.1002/adma.201702517

[309] Zhao P, Wang S, Tang H, Jian X, Zhao X, Yao Y, et al. Superior energy storage density and giant negative electrocaloric effects in $(Pb_{0.98}La_{0.02})(Zr, Sn)O_3$ antiferroelectric ceramics. Scripta Mater 2021;200:113920. https://doi.org/10.1016/j.scriptamat.2021.113920

[310] Zhang Y, Liu P, Kandula KR, Li W, Meng S, Qin Y, et al. Achieving excellent energy storage density of $Pb_{0.97}La_{0.02}(ZrSn_{0.05}Ti_{0.95})O_3$ ceramics by the B-site modification. J Eur Ceram Soc 2021;41:360-7. https://doi.org/10.1016/j.jeurceramsoc.2020.08.039

[311] Meng X, Zhao Y, Li Y, Hao X. Systematical investigation on energy-storage behavior of PLZST antiferroelectric ceramics by composition optimizing. J Am Ceram Soc 2021;104:2170-80. https://doi.org/10.1111/jace.17669

[312] Liu X, Li Y, Li Y, Hao X. Giant Energy-Storage Density and Thermally Activated Phase Transition in $(Pb_{0.96}La_{0.04})(Zr_{0.99}Ti_{0.01})O_3$ Antiferroelectric Ceramics. ACS Appl Energy Mater 2021;4:4897-902. https://doi.org/10.1021/acsaem.1c00474

[313] Liu X, Li Y, Sun N, Hao X. High energy-storage performance of PLZS antiferroelectric multilayer ceramic capacitors. Inorg Chem Front 2020;7:756-64. https://doi.org/10.1039/c9qi01416k

[314] Xie J, Yao M, Gao W, Su Z, Yao X. Ultrahigh breakdown strength and energy density in PLZST@ PBSAZM antiferroelectric ceramics based on core-shell structure. J Eur Ceram Soc 2019;39:1050-6. https://doi.org/10.1016/j.jeurceramsoc.2018.12.044

[315] Liu X, Zhao Y, Sun N, Li Y, Hao X. Ultra-high energy density induced by diversified enhancement effects in $(Pb_{0.98-x}La_{0.02}Ca_x)(Zr_{0.7}Sn_{0.3})_{0.995}O_3$ antiferroelectric multilayer ceramic capacitors. Chem Eng J 2020;417:128032. https://doi.org/10.1016/j.cej.2020.128032

[316] Liu X, Li Y, Hao X. Ultra-high energy-storage density and fast discharge speed of $(Pb_{0.98-x}La_{0.02}Sr_x)(Zr_{0.9}Sn_{0.1})_{0.995}O_3$ antiferroelectric ceramics prepared via the tape-casting method. J Mater Chem A 2019;7:11858-66. https://doi.org/10.1039/c9ta02149c

[317] Ge G, Shi C, Chen C, Shi Y, Yan F, Bai H, et al. Tunable Domain Switching Features of Incommensurate Antiferroelectric Ceramics Realizing Excellent Energy Storage Properties. Adv Mater 2022:e2201333. https://doi.org/10.1002/adma.202201333

[318] Chao W, Tian L, Yan T, Li Y, Liu Z. Excellent energy storage performance achieved in novel PbHfO3-based antiferroelectric ceramics via grain size engineering. Chem Eng J 2022;433:133814. https://doi.org/10.1016/j.cej.2021.133814

[319] Chao W, Yang T, Li Y. Achieving high energy efficiency and energy density in PbHfO3-based antiferroelectric ceramics. J Mater Chem C 2020;8:17016-24. https://doi.org/10.1039/d0tc04617e

[320] Guo J, Yang T. Giant energy storage density in Ba, La co-doped $PbHfO_3$-based antiferroelectric ceramics by a rolling process. J Alloys Compd 2021;888:161539. https://doi.org/10.1016/j.jallcom.2021.161539

[321] Ge PZ, Tang XG, Meng K, Huang XX, Li SF, Liu QX, et al. Energy storage density and charge–discharge properties of $PbHf_{1-x}Sn_xO_3$ antiferroelectric ceramics. Chem Eng J 2022;429:132540. https://doi.org/10.1016/j.cej.2021.132540

[322] Xie A, Fu J, Zuo R, Zhou C, Qiao Z, Li T, et al. NaNbO3-CaTiO3 lead-free relaxor antiferroelectric ceramics featuring giant energy density, high energy efficiency and power density. Chem Eng J 2022;429:132534. https://doi.org/10.1016/j.cej.2021.132534

[323] Tian A, Zuo R, Qi H, Shi M. Large energy-storage density in transition-metal oxide modified $NaNbO_3$–$Bi(Mg_{0.5}Ti_{0.5})O_3$ lead-free ceramics through regulating the antiferroelectric phase structure. J Mater Chem A 2020;8:8352-9. https://doi.org/10.1039/d0ta02285c

[324] Ye J, Wang G, Zhou M, Liu N, Chen X, Li S, et al. Excellent comprehensive energy storage properties of novel lead-free $NaNbO_3$-based ceramics for dielectric capacitor applications. J Mater Chem C 2019;7:5639-45. https://doi.org/10.1039/c9tc01414d

[325] Wang X, Wang X, Huan Y, Li C, Ouyang J, Wei T. A Combined Optimization Strategy for Improvement of Comprehensive Energy Storage Performance in Sodium Niobate-Based Antiferroelectric Ceramics. ACS Appl Mater Interfaces 2022;14:9330-9. https://doi.org/10.1021/acsami.1c23914

[326] Chen J, Qi H, Zuo R. Realizing Stable Relaxor Antiferroelectric and Superior Energy Storage Properties in $(Na_{1-x/2}La_{x/2})(Nb_{1-x}Ti_x)O_3$ Lead-Free Ceramics through A/B-Site Complex Substitution. ACS Appl Mater Interfaces 2020;12:32871-9. https://doi.org/10.1021/acsami.0c09876

[327] Shi R, Pu Y, Wang W, Guo X, Li J, Yang M, et al. A novel lead-free $NaNbO_3$–$Bi(Zn_{0.5}Ti_{0.5})O_3$ ceramics system for energy storage application with excellent stability. J Alloys Compd 2020;815:152356. https://doi.org/10.1016/j.jallcom.2019.152356

[328] Zhu LF, Zhao L, Yan Y, Leng H, Li X, Cheng LQ, et al. Composition and strain engineered $AgNbO_3$-based multilayer capacitors for ultra-high energy storage capacity. J Mater Chem A 2021;9:9655-64. https://doi.org/10.1039/d1ta00973g
[329] Shi P, Wang X, Lou X, Zhou C, Liu Q, He L, et al. Significantly enhanced energy storage properties of $Nd^{3+}$ doped $AgNbO_3$ lead-free antiferroelectric ceramics. J Alloys Compd 2021;877:160162. https://doi.org/10.1016/j.jallcom.2021.160162
[330] Li J, Jin L, Tian Y, Chen C, Lan Y, Hu Q, et al. Enhanced energy storage performance under low electric field in $Sm^{3+}$ doped $AgNbO_3$ ceramics. J Materiomics. 2022;8:266-73. https://doi.org/10.1016/j.jmat.2021.10.005
[331] Xu Y, Guo Y, Liu Q, Wang G, Bai J, Tian J, et al. High energy storage properties of lead-free Mn-doped (1-x)$AgNbO_3$-x$Bi_{0.5}Na_{0.5}TiO_3$ antiferroelectric ceramics. J Eur Ceram Soc 2020;40:56-62. https://doi.org/10.1016/j.jeurceramsoc.2019.09.022
[332] Gao J, Liu Q, Dong J, Wang X, Zhang S, Li JF. Local Structure Heterogeneity in Sm-Doped $AgNbO_3$ for Improved Energy-Storage Performance. ACS Appl Mater Interfaces 2020;12:6097-104. https://doi.org/10.1021/acsami.9b20803
[333] Lu Z, Bao W, Wang G, Sun S-K, Li L, Li J, et al. Mechanism of enhanced energy storage density in $AgNbO_3$-based lead-free antiferroelectrics. Nano Energy 2021;79:105423. https://doi.org/10.1016/j.nanoen.2020.105423
[334] Wang Z, Kang R, Zhang L, Mao P, Sun Q, Kang F, et al. Remarkably enhanced energy-storage density and excellent thermal stability under low electric fields of $(Na_{0.5}Bi_{0.5})TiO_3$-based ceramics via composition optimization strategy. J Eur Ceram Soc 2021;41:1917-24. https://doi.org/10.1016/j.jeurceramsoc.2020.10.047
[335] Huang J, Qi H, Gao Y, Xie A, Zhang Y, Li Y, et al. Expanded linear polarization response and excellent energy-storage properties in $(Bi_{0.5}Na_{0.5})TiO_3$-$KNbO_3$ relaxor antiferroelectrics with medium permittivity. Chem Eng J 2020;398:125639. https://doi.org/10.1016/j.cej.2020.125639
[336] Zhang L, Jing R, Huang Y, Hu Q, Alikin DO, Shur VY, et al. Enhanced antiferroelectric-like relaxor ferroelectric characteristic boosting energy storage performance of $(Bi_{0.5}Na_{0.5})TiO_3$-based ceramics via defect engineering. J Materiomics 2022;8:527-36. https://doi.org/10.1016/j.jmat.2022.01.007
[337] Li T, Jiang X, Li J, Xie A, Fu J, Zuo R. Ultrahigh Energy-Storage Performances in Lead-free $Na_{0.5}Bi_{0.5}TiO_3$-Based Relaxor Antiferroelectric Ceramics through a Synergistic Design Strategy. ACS Appl Mater Interfaces 2022. https://doi.org/10.1021/acsami.2c01287
[338] Shi W, Zhang L, Jing R, Hu Q, Zeng X, Alikin D, et al. Relaxor antiferroelectric-like characteristic boosting enhanced energy storage performance in eco-friendly $(Bi_{0.5}Na_{0.5})TiO_3$-based ceramics. J Eur Ceram Soc 2022;42:4528-38. https://doi.org/10.1016/j.jeurceramsoc.2022.04.057
[339] Wu Q, Zhao Y, Zhou Y, Chen X, Wu X, Zhao S. Energy storage properties of composite films with relaxor antiferroelectric behaviors. J Alloys Compd 2021;881:160576. https://doi.org/10.1016/j.jallcom.2021.160576
[340] Zhang AH, Wang W, Li QJ, Zhu JY, Wang DD, Lu XB, et al. Internal-strain release and remarkably enhanced energy storage performance in PLZT–$SrTiO_3$ multilayered films. Appl Phys Lett 2020;117:252901. https://doi.org/10.1063/5.0030279

[341] Li YZ, Lin JL, Bai Y, Li Y, Zhang ZD, Wang ZJ. Ultrahigh-Energy Storage Properties of (PbCa)$ZrO_3$ Antiferroelectric Thin Films via Constructing a Pyrochlore Nanocrystalline Structure. ACS Nano 2020;14:6857-65. https://doi.org/10.1021/acsnano.0c00791
[342] Cai H, Yan S, Dong X, Cao F, Wang G. Significantly enhanced energy storage performance by constructing $TiO_2$ nanowire arrays in $PbZrO_3$-based antiferroelectric films. Ceram Int 2020;46:6436-42. https://doi.org/10.1016/j.ceramint.2019.11.123
[343] Lin Z, Chen Y, Liu Z, Wang G, Remiens D, Dong X. Large energy storage density, low energy loss and highly stable $(Pb_{0.97}La_{0.02})(Zr_{0.66}Sn_{0.23}Ti_{0.11})O_3$ antiferroelectric thin-film capacitors. J Eur Ceram Soc 2018;38:3177-81. https://doi.org/10.1016/j.jeurceramsoc.2018.03.004
[344] Li YZ, Wang ZJ, Bai Y, Zhang ZD. High energy storage performance in Ca-doped $PbZrO_3$ antiferroelectric films. J Eur Ceram Soc 2020;40:1285-92. https://doi.org/10.1016/j.jeurceramsoc.2019.11.063
[345] Ko DL, Hsin T, Lai YH, Ho SZ, Zheng Y, Huang R, et al. High-stability transparent flexible energy storage based on $PbZrO_3$/muscovite heterostructure. Nano Energy 2021;87:106149. https://doi.org/10.1016/j.nanoen.2021.106149
[346] Acharya M, Banyas E, Ramesh M, Jiang Y, Fernandez A, Dasgupta A, et al. Exploring the $Pb_{1-x}Sr_xHfO_3$ System and Potential for High Capacitive Energy Storage Density and Efficiency. Adv Mater 2022;334:2105967. https://doi.org/10.1002/adma.202105967
[347] Huang XX, Zhang TF, Wang W, Ge PZ, Tang XG. Tailoring energy-storage performance in antiferroelectric $PbHfO_3$ thin films. Mater Des 2021;204:109666. https://doi.org/10.1016/j.matdes.2021.109666
[348] Hanrahan B, Milesi-Brault C, Leff A, Payne A, Liu S, Guennou M, et al. The other model antiferroelectric: $PbHfO_3$ thin films from ALD precursors. APL Mater 2021;9. https://doi.org/10.1063/5.0035730
[349] Beppu K, Shimasaki T, Fujii I, Imai T, Adachi H, Wada T. Energy storage properties of antiferroelectric $0.92NaNbO_3$-$0.08SrZrO_3$ film on (001)$SrTiO_3$ substrate. Phys Lett A 2020;384:126690. https://doi.org/10.1016/j.physleta.2020.126690
[350] Zhang WL, Mao YH, Cui L, Tang MH, Su PY, Long XJ, et al. Impact of the radiation effect on the energy storage density and wake-up behaviors of antiferroelectric-like Al-doped $HfO_2$ thin films. Phys Chem Chem Phys 2020;22:21893-9. https://doi.org/10.1039/d0cp04196c
[351] Payne A, Brewer O, Leff A, Strnad NA, Jones JL, Hanrahan B. Dielectric, energy storage, and loss study of antiferroelectric-like Al-doped $HfO_2$ thin films. Appl Phys Lett. 2020;117:221104. https://doi.org/10.1063/5.0029706
[352] Do Kim K, Lee YH, Gwon T, Kim YJ, Kim HJ, Moon T, et al. Scale-up and optimization of $HfO_2$-$ZrO_2$ solid solution thin films for the electrostatic supercapacitors. Nano Energy 2017;39:390-9. https://doi.org/10.1016/j.nanoen.2017.07.017
[353] Kozodaev MG, Chernikova AG, Khakimov RR, Park MH, Markeev AM, Hwang CS. La-doped $Hf_{0.5}Zr_{0.5}O_2$ thin films for high-efficiency electrostatic supercapacitors. Appl Phys Lett 2018;113:123902. https://doi.org/10.1063/1.5045288
[354] Ali F, Liu X, Zhou D, Yang X, Xu J, Schenk T, et al. Silicon-doped hafnium oxide anti-ferroelectric thin films for energy storage. J Appl Phys 2017;122:144105. https://doi.org/10.1063/1.4989908

[355] Serralta-Macías JdJ, Rodriguez-Davila RA, Quevedo-Lopez M, Olguín D, Castillo SJ, Young CD, et al. Energy storage performance in lead-free antiferroelectric 0.92($Bi_{0.54}Na_{0.46}$)$TiO_3$-0.08$BaTiO_3$ ultrathin films by pulsed laser deposition. J Vacuum Sci Tech 2022;40:033417. https://doi.org/10.1116/6.0001755
[356] Zhou Y, Chan HK, Lam CH, Shin FG. Mechanisms of imprint effect on ferroelectric thin films. J Appl Phys 2005;98. https://doi.org/10.1063/1.1984075
[357] Nguyen MD, Houwman EP, Do MT, Rijnders G. Relaxor-ferroelectric thin film heterostructure with large imprint for high energy-storage performance at low operating voltage. Energy Storage Mater 2020;25:193-201. https://doi.org/10.1016/j.ensm.2019.10.015
[358] Luo N, Han K, Zhuo F, Xu C, Zhang G, Liu L, et al. Aliovalent A-site engineered $AgNbO_3$ lead-free antiferroelectric ceramics toward superior energy storage density. J Mater Chem A 2019;7:14118-28. https://doi.org/10.1039/c9ta02053e
[359] Sun Z, Ma C, Liu M, Cui J, Lu L, Lu J, et al. Ultrahigh Energy Storage Performance of Lead-Free Oxide Multilayer Film Capacitors via Interface Engineering. Adv Mater 2017;29:1604427. https://doi.org/10.1002/adma.201604427
[360] Scott JF, Araujo CAPd. Ferroelectric Memories. Science. 1989;246:1400-5. https://doi.org/doi:10.1126/science.246.4936.1400
[361] Jiang Y, Parsonnet E, Qualls A, Zhao W, Susarla S, Pesquera D, et al. Enabling ultra-low-voltage switching in $BaTiO_3$. Nat Mater 2022;21:779-85. https://doi.org/10.1038/s41563-022-01266-6
[362] Liu Y, Scott JF, Dkhil B. Direct and indirect measurements on electrocaloric effect: Recent developments and perspectives. Appl Phys Rev 2016;3:031102. https://doi.org/10.1063/1.4958327
[363] Scott JF. Electrocaloric Materials. Annual Rev Mater Res 2011;41:229-40. https://doi.org/10.1146/annurev-matsci-062910-100341
[364] Peláiz-Barranco A, Wang J, Yang T. Direct and indirect analysis of the electrocaloric effect for lanthanum-modified lead zirconate titanate antiferroelectric ceramics. Ceram Int. 2016;42:229-33. https://doi.org/10.1016/j.ceramint.2015.08.097
[365] Thacher PD. Electrocaloric Effects in Some Ferroelectric and Antiferroelectric Pb(Zr,Ti)$O_3$ Compounds. J Appl Phys 1968;39:1996-2002. https://doi.org/10.1063/1.1656478
[366] Bai Y, Zheng GP, Shi SQ. Abnormal electrocaloric effect of $Na_{0.5}Bi_{0.5}TiO_3$-$BaTiO_3$ lead-free ferroelectric ceramics above room temperature. Mater Res Bull 2011;46:1866-9. https://doi.org/10.1016/j.materresbull.2011.07.038
[367] Park MH, Kim HJ, Kim YJ, Moon T, Do Kim K, Lee YH, et al. Giant Negative Electrocaloric Effects of $Hf_{0.5}Zr_{0.5}O_2$ Thin Films. Adv Mater 2016;28:7956-61. https://doi.org/10.1002/adma.201602787
[368] Zhuo FP, Li Q, Gao JH, Ji YJ, Yan QF, Zhang YL, et al. Giant Negative Electrocaloric Effect in (Pb,La)(Zr,Sn,Ti)$O_3$ Antiferroelectrics Near Room Temperature. ACS Appl Mater Inter 2018;10:11747-55. https://doi.org/10.1021/acsami.8b00744
[369] Wu M, Song DS, Guo MY, Bian JH, Li JN, Yang YD, et al. Remarkably Enhanced Negative Electrocaloric Effect in $PbZrO_3$ Thin Film by Interface Engineering. ACS Appl Mater Interface 2019;11:36863-70. https://doi.org/10.1021/acsami.9b13143

[370] Liu C, Si Y, Hao M, Tao Y, Deng S, Lu P, et al. Phonon entropy engineering for caloric cooling. Appl Phys Rev 2023;10:031411. https://doi.org/10.1063/5.0152301

[371] Li JJ, Li JT, Wu HH, Qin SQ, Su XP, Wang Y, et al. Giant Electrocaloric Effect and Ultrahigh Refrigeration Efficiency in Antiferroelectric Ceramics by Morphotropic Phase Boundary Design. ACS Appl Mater Interface 2020;12:45005-14. https://doi.org/10.1021/acsami.0c13734

[372] Ponomareva I, Lisenkov S. Bridging the Macroscopic and Atomistic Descriptions of the Electrocaloric Effect. Phys Rev Lett 2012;108:167604. https://doi.org/10.1103/PhysRevLett.108.167604

[373] Vats G, Kumar A, Ortega N, Bowen CR, Katiyar RS. Giant pyroelectric energy harvesting and a negative electrocaloric effect in multilayered nanostructures. Energ Environ Sci 2016;9:1335-45. https://doi.org/10.1039/C5EE03641K

[374] Liu C, Si W, Wu C, Yang J, Chen Y, Dames C. The ignored effects of vibrational entropy and electrocaloric effect in $PbTiO_3$ and $PbZr_{0.5}Ti_{0.5}O_3$ as studied through first-principles calculation. Acta Mater 2020;191:221-9. https://doi.org/10.1016/j.actamat.2020.03.059

[375] Huang XX, Zhang TF, Gao RZ, Huang HB, Ge PZ, Tang H, et al. Large Room Temperature Negative Electrocaloric Effect in Novel Antiferroelectric $PbHfO_3$ Films. ACS Appl Mater Interface 2021;13:21331-7. https://doi.org/10.1021/acsami.1c03079

[376] Peng BL, Zhang MM, Tang SL, Jiang JT, Zhao WG, Zou BS, et al. Frequency dependent electrocaloric effect in Nb-doped PZST relaxor thin film with the coexistence of tetragonal antiferroelectric and rhombohedral ferroelectric phases. Ceram Int 2020;46:4300-6. https://doi.org/10.1016/j.ceramint.2019.10.151

[377] Wang WH, Chen XQ, Sun Q, Xin TZ, Ye M. Tailoring the negative electrocaloric effect of PbZrO3 antiferroelectric thin films by Yb doping. J Alloys Compd 2020;830:154581. https://doi.org/10.1016/j.jallcom.2020.154581

[378] Ye M, Li T, Sun Q, Liu ZK, Peng BL, Huang CW, et al. A giant negative electrocaloric effect in Eu-doped $PbZrO_3$ thin films. J Mater Chem C 2016;4:3375-8. https://doi.org/10.1039/C6TC00218H

[379] Guo MY, Wu M, Gao WW, Sun BW, Lou XJ. Giant negative electrocaloric effect in antiferroelectric $PbZrO_3$ thin films in an ultra-low temperature range. J Mater Chem C. 2019;7:617-21. https://doi.org/10.1039/C8TC05108A

[380] Allouche B, Hwang HJ, Yoo TJ, Lee BH. A negative electrocaloric effect in an antiferroelectric zirconium dioxide thin film. Nanoscale 2020;12:3894-901. https://doi.org/10.1039/c9nr07293d

[381] Parui J, Krupanidhi SB. Electrocaloric effect in antiferroelectric $PbZrO_3$ thin films. Phys Status Solidi-R 2008;2:230-2. https://doi.org/10.1002/pssr.200802128

[382] Peng BL, Zhang Q, Lyu YN, Liu LJ, Lou XJ, Shaw C, et al. Thermal strain induced large electrocaloric effect of relaxor thin film on $LaNiO_3$/Pt composite electrode with the coexistence of nanoscale antiferroelectric and ferroelectric phases in a broad temperature range. Nano Energy 2018;47:285-93. https://doi.org/10.1016/j.nanoen.2018.03.003

[383] Park MH, Kim HJ, Kim YJ, Moon T, Kim KD, Hwang CS. Toward a multifunctional monolithic device based on pyroelectricity and the electrocaloric effect of thin antiferroelectric $Hf_xZr_{1-x}O_2$ films. Nano Energy 2015;12:131-40. https://doi.org/10.1016/j.nanoen.2014.09.025

[384] Peng BL, Fan HQ, Zhang Q. A Giant Electrocaloric Effect in Nanoscale Antiferroelectric and Ferroelectric Phases Coexisting in a Relaxor $Pb_{0.8}Ba_{0.2}ZrO_3$ Thin Film at Room Temperature. Adv Funct Mater 2013;23:2987-92. https://doi.org/10.1002/adfm.201202525

[385] Maiwa H, Tsutsui N. Electrocaloric Effect in Antiferroelectric Lead Zirconate Thin Films Negative electrocaloric effects in $PbZrO_3$ films prepared by chemical solution deposition. 2019 IEEE International Symposium on Applications of Ferroelectrics (ISAF) (2019) 1-3. https://doi.org/10.1109/ISAF43169.2019.9034946

[386] Hao XH, Zhai JW. Electric-field tunable electrocaloric effects from phase transition between antiferroelectric and ferroelectric phase. Appl Phys Lett 2014;104:022902. https://doi.org/10.1063/1.4862171

[387] Hao XH, Zhao Y, Zhang Q. Phase Structure Tuned Electrocaloric Effect and Pyroelectric Energy Harvesting Performance of $(Pb_{0.97}La_{0.02})(Zr,Sn,Ti)O_3$ Antiferroelectric Thick Films. J Phys Chem C 2015;119:18877-85. https://doi.org/10.1021/acs.jpcc.5b04178

[388] Zhao Y, Gao HC, Hao XH, Zhang Q. Orientation-dependent energy-storage performance and electrocaloric effect in PLZST antiferroelectric thick films. Mater Res Bull 2016;84:177-84. https://doi.org/10.1016/j.materresbull.2016.08.005

[389] Zhao Y, Hao XH, Zhang Q. A giant electrocaloric effect of a $Pb_{0.97}La_{0.02}(Zr_{0.75}Sn_{0.18Ti}i_{0.07})O_3$ antiferroelectric thick film at room temperature. J Mater Chem C 2015;3:1694-9. https://doi.org/10.1039/C4TC02381A

[390] Zhao Y, Hao XH, Zhang Q. Energy-Storage Properties and Electrocaloric Effect of $Pb_{(1-3x/2)}La_xZr_{0.85}Ti_{0.15}O_3$ Antiferroelectric Thick Films. ACS Appl Mater Interface 2014;6:11633-9. https://doi.org/10.1021/am502415z

[391] Huang D, Wang JB, Zhong XL, Li B, Zhang Y, Jin C, et al. Giant negative electrocaloric effect in $PbZrO_3$/0.88$BaTiO_3$-0.12$Bi(Mg_{1/2}Ti_{1/2})O_3$ multilayered composite ferroelectric thin film for solid-state refrigeration. J Appl Phys 2017;122:194103. https://doi.org/10.1063/1.4991994

[392] Vales-Castro P, Faye R, Vellvehi M, Nouchokgwe Y, Perpina X, Caicedo JM, et al. Origin of large negative electrocaloric effect in antiferroelectric $PbZrO_3$. Phys Rev B 2021;103:054112. https://doi.org/10.1103/PhysRevB.103.054112

[393] Kimmel AV, Gindele OT, Duffy DM, Cohen RE. Giant electrocaloric effect at the antiferroelectric-to-ferroelectric phase boundary in $Pb(Zr_xTi_{1-x})O_3$. Appl Phys Lett 2019;115:023902. https://doi.org/10.1063/1.5096592

[394] Jiang XJ, Luo LH, Wang BY, Li WP, Chen HB. Electrocaloric effect based on the depolarization transition in (1-x)$Bi_{0.5}Na_{0.5}TiO_3$-x$KNbO_3$ lead-free ceramics. Ceram Int 2014;40:2627-34. https://doi.org/10.1016/j.ceramint.2013.10.066

[395] Zheng XC, Zheng GP, Lin Z, Jiang ZY. Electro-caloric behaviors of lead-free $Bi_{0.5}Na_{0.5}TiO_3$-$BaTiO_3$ ceramics. J Electroceram 2012;28:20-6. https://doi.org/10.1007/s10832-011-9673-4

[396] Zannen M, Lahmar A, Asbani B, Khemakhem H, El Marssi M, Kutnjak Z, et al. Electrocaloric effect and luminescence properties of lanthanide doped $(Na_{1/2}Bi_{1/2})TiO_3$ lead free materials. Appl Phys Lett 2015;107:032905. https://doi.org/10.1063/1.4927280

[397] Li JJ, Wu HH, Li JT, Su XP, Yin RW, Qin SQ, et al. Room-Temperature Symmetric Giant Positive and Negative Electrocaloric Effect in $PbMg_{0.5}W_{0.5}O_3$ Antiferroelectric Ceramic. Adv Funct Mater 2021;31:2101176. https://doi.org/10.1002/adfm.202101176

[398] Liu NT, Liang RH, Zhang GZ, Zhou ZY, Yan SG, Li XB, et al. Colossal negative electrocaloric effects in lead-free bismuth ferrite-based bulk ferroelectric perovskite for solid-state refrigeration. J Mater Chem C 2018;6:10415-21. https://doi.org/10.1039/C8TC04125C

[399] Niu ZH, Jiang YP, Tang XG, Liu QX, Li WH, Lin XW, et al. Giant negative electrocaloric effect in B-site non-stoichiometric $(Pb_{0.97}La_{0.02})(Zr_{0.95}Ti_{0.05})_{1+y}O_{-3}$ anti-ferroelectric ceramics. Mater Res Lett 2018;6:384-9. https://doi.org/10.1080/21663831.2018.1466737

[400] Zhao YC, Liu QX, Tang XG, Jiang YP, Li B, Li WH, et al. Giant Negative Electrocaloric Effect in Anti-Ferroelectric $(Pb_{0.97}La_{0.02})(Zr_{0.95}Ti_{0.05})O_3$ Ceramics. Acs Omega 2019;4:14650-4. https://doi.org/10.1021/acsomega.9b02149

[401] Xu HJ, Guo WQ, Wang JQ, Ma Y, Han SG, Liu Y, et al. A Metal-Free Molecular Antiferroelectric Material Showing High Phase Transition Temperatures and Large Electrocaloric Effects. J Am Chem Soc 2021;143:14379-85. https://doi.org/10.1021/jacs.1c07521

[402] Zhuo FP, Li Q, Gao JH, Wang YJ, Yan QF, Zhang YL, et al. Coexistence of multiple positive and negative electrocaloric responses in $(Pb,La)(Zr,Sn,Ti)O_3$ single crystal. Appl Phys Lett 2016;108:082904. https://doi.org/10.1063/1.4941816

[403] Hao J, Li W, Zhai J, Chen H. Progress in high-strain perovskite piezoelectric ceramics. Mater Sci Eng R: Rep 2019;135:1-57. https://doi.org/10.1016/j.mser.2018.08.001

[404] Nguyen MD, Rijnders G. Comparative study of piezoelectric response and energy-storage performance in normal ferroelectric, antiferroelectric and relaxor-ferroelectric thin films. Thin Solid Films 2020;697:137843. https://doi.org/10.1016/j.tsf.2020.137843

[405] Zhuo F, Li Q, Zhou Y, Ji Y, Yan Q, Zhang Y, et al. Large field-induced strain, giant strain memory effect, and high thermal stability energy storage in $(Pb,La)(Zr,Sn,Ti)O_3$ antiferroelectric single crystal. Acta Mater 2018;148:28-37. https://doi.org/10.1016/j.actamat.2018.01.021

[406] Guo Y, Liu Y, Withers RL, Brink F, Chen H. Large electric field-induced strain and antiferroelectric behavior in $(1-x)(Na_{0.5}Bi_{0.5})TiO_3-xBaTiO_3$ ceramics. Chem Mater 2011;23:219-28. https://doi.org/10.1021/cm102719k

[407] Berlincourt D. Piezoelectric and ferroelectric energy conversion. IEEE Trans Sonics Ultrason 1968;15:89-96. https://doi.org/10.1109/T-SU.1968.29453

[408] Lu H, Glinsek S, Buragohain P, Defay E, Iñiguez J, Gruverman A. Probing Antiferroelectric-Ferroelectric Phase Transitions in $PbZrO_3$ Capacitors by Piezoresponse Force Microscopy. Adv Funct Mater 2020;30. https://doi.org/10.1002/adfm.202003622

[409] Zhuo F, Damjanovic D, Li Q, Zhou Y, Ji Y, Yan Q, et al. Giant shape memory and domain memory effects in antiferroelectric single crystals. Mater Horiz 2019;6:1699-706. https://doi.org/10.1039/c9mh00352e

[410] Íñiguez J, Zubko P, Luk'yanchuk I, Cano A. Ferroelectric negative capacitance. Nat Rev Mater 2019;4:243-56. https://doi.org/10.1038/s41578-019-0089-0

[411] Landauer R. Can capacitance be negative. Collect Phenom 1976;2:167-70. https://scholar.google.com/scholar_lookup?title=Can%20capacitance%20be%20negative%3F

&journal=Collect.%20Phenom.&volume=2&pages=167-170&publication_year=1976&author=Landauer%2CR

[412] Karda K, Jain A, Mouli C, Alam MA. An anti-ferroelectric gated Landau transistor to achieve sub-60 mV/dec switching at low voltage and high speed. Appl Phys Lett 2015;106:163501. https:// doi.org/10.1063/1.4918649

[413] Appleby DJ, Ponon NK, Kwa KS, Zou B, Petrov PK, Wang T, et al. Experimental observation of negative capacitance in ferroelectrics at room temperature. Nano Lett 2014;14:3864. https://doi.org/10.1021/nl5017255

[414] Yadav AK, Nguyen KX, Hong Z, Garcia-Fernandez P, Aguado-Puente P, Nelson CT, et al. Spatially resolved steady-state negative capacitance. Nat 2019;565:468-71. https://doi.org/10.1038/s41586-018-0855-y

[415] Khan AI, Chatterjee K, Wang B, Drapcho S, You L, Serrao C, et al. Negative capacitance in a ferroelectric capacitor. Nat Mater 2015;14:182. https://doi.org/10.1038/nmat4148

[416] Gao W, Khan A, Marti X, Nelson C, Serrao C, Ravichandran J, et al. Room-temperature negative capacitance in a ferroelectric–dielectric superlattice heterostructure. Nano Lett 2014;14:5814-9. https://doi.org/10.1021/nl502691u

[417] Hoffmann M, Fengler FPG, Herzig M, Mittmann T, Max B, Schroeder U, et al. Unveiling the double-well energy landscape in a ferroelectric layer. Nature 2019;565:464-7. https://doi.org/10.1038/s41586-018-0854-z

[418] Kim KD, Kim YJ, Park MH, Park HW, Kwon YJ, Lee YB, et al. Transient Negative Capacitance Effect in Atomic-Layer-Deposited $Al_2O_3/Hf_{0.3}Zr_{0.7}O_2$ Bilayer Thin Film. Adv Funct Mater 2019;29:1808228. https://doi.org/10.1002/adfm.201808228

[419] Kim YJ, Yamada H, Moon T, Kwon YJ, An CH, Kim HJ, et al. Time-Dependent Negative Capacitance Effects in $Al_2O_3/BaTiO_3$ Bilayers. Nano Lett 2016;16:4375-81. https://doi.org/10.1021/acs.nanolett.6b01480

[420] Park HW, Roh J, Lee YB, Hwang CS. Modeling of Negative Capacitance in Ferroelectric Thin Films. Adv Mater 2019;31:1805266. https://doi.org/10.1002/adma.201805266

[421] Park HW, Oh M, Hwang CS. Negative capacitance from the inhomogenous stray field in a ferroelectric–dielectric structure. Adv Funct Mater 2022;32:2200389. https://doi.org/10.1002/adfm.202200389

[422] Park HW, Oh M, Lee IS, Byun S, Jang YH, Lee YB, et al. Double S-Shaped Polarization–Voltage Curve and Negative Capacitance from $Al_2O_3$-$Hf_{0.5}Zr_{0.5}O_2$-$Al_2O_3$ Triple-Layer Structure. Adv Funct Mater 2023;33:2206637. https://doi.org/10.1002/adfm.202206637

[423] Park HW, Byun S, Kim KD, Ryoo SK, Lee IS, Lee YB, et al. Exploring the Physical Origin of the Negative Capacitance Effect in a Metal–Ferroelectric–Metal–Dielectric Structure. Adv Funct Mater 2023:2304754. https://doi.org/10.1002/adfm.202304754

[424] Morris DH, Avci UE, Young IA. Anti-ferroelectric capacitor memory cell. European Patent EP3576092. https://data.epo.org/publication-server/document?iDocId=6101648&iFormat=0

[425] Chang S-C, Haratipour N, Shivaraman S, Neumann C, Atanasov S, Peck J, et al. FeRAM using anti-ferroelectric capacitors for high-speed and high-density embedded memory. 2021 IEEE International Electron Devices Meeting 2021. p. 33.2. 1-.2. 4. https://doi.org/10.1109/IEDM19574.2021.9720510

[426] Apachitei G, Peters JJ, Sanchez AM, Kim DJ, Alexe M. Antiferroelectric tunnel junctions. Adv Elec Mater 2017;3:1700126. https://doi.org/10.1002/aelm.201700126

[427] Goh Y, Hwang J, Jeon S. Excellent reliability and high-speed antiferroelectric HfZrO2 tunnel junction by a high-pressure annealing process and built-in bias engineering. ACS Appl Mater Interfaces 2020;12:57539-46. https://doi.org/10.1021/acsami.0c15091

[428] Liu C, Wu C, Song T, Zhao Y, Yang J, Lu P, et al. An Efficient Strategy for Searching High Lattice Thermal Conductivity Materials. ACS Appl Energy Mater 2022;5:15356–64. https://doi.org/10.1021/acsaem.2c02970

[429] Guo M, Qian Y, Qi H, Bi K, Chen Y. Experimental measurements on the thermal conductivity of strained monolayer graphene. Carbon 2020;157:185-90. https://doi.org/10.1016/j.carbon.2019.10.027

[430] Lee S, Hippalgaonkar K, Yang F, Hong J, Ko C, Suh J, et al. Anomalously low electronic thermal conductivity in metallic vanadium dioxide. Science. 2017;355:371-4. https://doi.org/10.1126/science.aag0410

[431] Liu C, Chen Z, Wu C, Qi J, Hao M, Lu P, et al. Large Thermal Conductivity Switching in Ferroelectrics by Electric Field-Triggered Crystal Symmetry Engineering. ACS Appl Mater Interface 2022;14:46716-25. https://doi.org/10.1021/acsami.2c11530

[432] Liu C, Chen Y, Dames C. Electric-field-controlled thermal switch in ferroelectric materials using first-principles calculations and domain-wall engineering. Phys Rev Appl 2019;11:044002. https://doi.org/10.1103/PhysRevApplied.11.044002

[433] Yang B, Zhang Y, Pan H, Si W, Zhang Q, Shen Z, et al. High-entropy enhanced capacitive energy storage. Nat Mater 2022;21:1074-80. https://doi.org/10.1038/s41563-022-01274-6

[434] Garcia-Castro AC, Ma Y, Romestan Z, Bousquet E, Cen C, Romero AH. Engineering of Ferroic Orders in Thin Films by Anionic Substitution. Adv Funct Mater 2021;32:2107135. https://doi.org/10.1002/adfm.202107135

[435] Phuoc NN, Ong C. Anomalous Temperature Dependence of Magnetic Anisotropy in Gradient-Composition Sputterred Thin Films. Adv Mater 2013;25:980-4. https://doi.org/10.1002/adma.201203995

[436] Catalan G, Noheda B, McAneney J, Sinnamon L, Gregg J. Strain gradients in epitaxial ferroelectrics. Phys Rev B 2005;72:020102. https://doi.org/10.1103/PhysRevB.72.020102

[437] Wang J, Wylie-van Eerd B, Sluka T, Sandu C, Cantoni M, Wei XK, et al. Negative-pressure-induced enhancement in a freestanding ferroelectric. Nat Mat 2015;14:985-90. https://doi.org/10.1038/nmat4365

[438] Lefkowitz I, Lukaszewicz K, Megaw HD. The high-temperature phases of sodium niobate and the nature of transitions in pseudosymmetric structures. Acta Crystallogr. 1966;20:670-83. https://doi.org/10.1107/S0365110X66001592

[439] Mishra S, Choudhury N, Chaplot S, Krishna P, Mittal R. Competing antiferroelectric and ferroelectric interactions in $NaNbO_3$: Neutron diffraction and theoretical studies. Phys Rev B 2007;76:024110. https://doi.org/10.1103/PhysRevB.76.024110

[440] Kania, Kwapulinski. $Ag_{1-x}Na_xNbO_3$ (ANN) solid solutions: from disordered antiferroelectric $AgNbO_3$ to normal antiferroelectric $NaNbO_3$. J Phys Condens Matter 1999;11:8933. https://doi.org/10.1088/0953-8984/11/45/316

[441] He X, Chen C, Li C, Zeng H, Yi Z. Ferroelectric, Photoelectric, and Photovoltaic Performance of Silver Niobate Ceramics. Adv Funct Mater 2019;29. https://doi.org/10.1002/adfm.201900918
[442] Sciau P, Kania A, Dkhil B, Suard E, Ratuszna A. Structural investigation of $AgNbO_3$ phases using x-ray and neutron diffraction. J Phys Condens Matter 2004;16:2795-810. https://doi.org/doi.org/10.1088/0953-8984/16/16/004
[443] Levin I, Krayzman V, Woicik JC, Karapetrova J, Proffen T, Tucker MG, et al. Structural changes underlying the diffuse dielectric response in $AgNbO_3$. Phys Rev B 2009;79. https://doi.org/10.1103/PhysRevB.79.104113
[444] Jones G, Thomas P. The tetragonal phase of $Na_{0.5}Bi_{0.5}TiO_3$–a new variant of the perovskite structure. Acta Crystallogr Sect B: Struct Sci. 2000;56:426-30. https://doi.org/10.1107/S0108768100001166
[445] Jones G, Thomas P. Investigation of the structure and phase transitions in the novel A-site substituted distorted perovskite compound $Na_{0.5}Bi_{0.5}TiO_3$. Acta Crystallogr Sect B: Struct Sci. 2002;58:168-78. https://doi.org/10.1107/S0108768101020845
[446] Parija B, Badapanda, T., Senthil, V., Rout, S. K., Panigrahi, S. Diffuse phase transition, piezoelectric and optical study of $Bi_{0.5}Na_{0.5}TiO_3$ ceramic. Bull Mater Sci 2012; 35:197-202. https://doi.org/10.1007/s12034-012-0276-8
[447] Hagiyev MS, Ismailzade IH, Abiyev AK. Pyroelectric properties of (Na½Bi½)TiO3 ceramics. Ferroelectrics. 2011;56:215-7. https:// doi.org/10.1080/00150198408221371
[448] Yamasaki K, Soejima Y, Fischer K. Superstructure determination of PbZrO3. Acta Crystallogr Sect B: Struct Sci. 1998;54:524-30. https://doi.org/10.1107/S0108768197018466
[449] Nokhrin S, Laguta V, Glinchuk M, Bykov I, Lyaschenko A, Jastrabik L, et al. Influence of La impurities on $PbZrO_3$ dielectric permittivity. J Korean Phys Soc 1998;32:S308-S11. https://elibrary.ru/item.asp?id=27754766
[450] Lanagan MT, Kim J, JANG SJ, Newnham RE. Microwave dielectric properties of antiferroelectric lead zirconate. J Am Ceram Soc. 1988;71:311-6. https://doi.org/10.1111/j.1151-2916.1988.tb05864.x
[451] Kupriyanov M, Petrovich E, Dutova E, Kabirov YV. Sequence of phase transitions in PbHfO3. CryRp. 2012;57:205-7. https://doi.org/10.1134/S1063774512020125
[452] Corker D, Glazer A, Kaminsky W, Whatmore R, Dec J, Roleder K. Investigation into the crystal structure of the perovskite lead hafnate, PbHfO3. Acta Crystallogr Sect B: Struct Sci. 1998;54:18-28. https://doi.org/10.1107/S0108768197009208
[453] Madigout V, Baudour JL, Bouree F, Favotto C, Roubin M, Nihoul G. Crystallographic structure of lead hafnate (PbHfO3) from neutron powder diffraction and electron microscopy. Philos Mag A. 1999;79:847-58. https://doi.org/10.1080/01418619908210335
[454] Bussmann-Holder A, Kim TH, Lee BW, Ko JH, Majchrowski A, Soszynski A, et al. Phase transitions and interrelated instabilities in PbHfO3 single crystals. J Phys Condens Matter. 2015;27:105901. https://doi.org/10.1088/0953-8984/27/10/105901
[455] Dedon LR, Chen Z, Gao R, Qi Y, Arenholz E, Martin LW. Strain-Driven Nanoscale Phase Competition near the Antipolar-Nonpolar Phase Boundary in $Bi_{0.7}La_{0.3}FeO_3$ Thin Films. ACS Appl Mater Interfaces. 2018;10:14914-21. https://doi.org/10.1021/acsami.8b02597

**Figure Captions**:

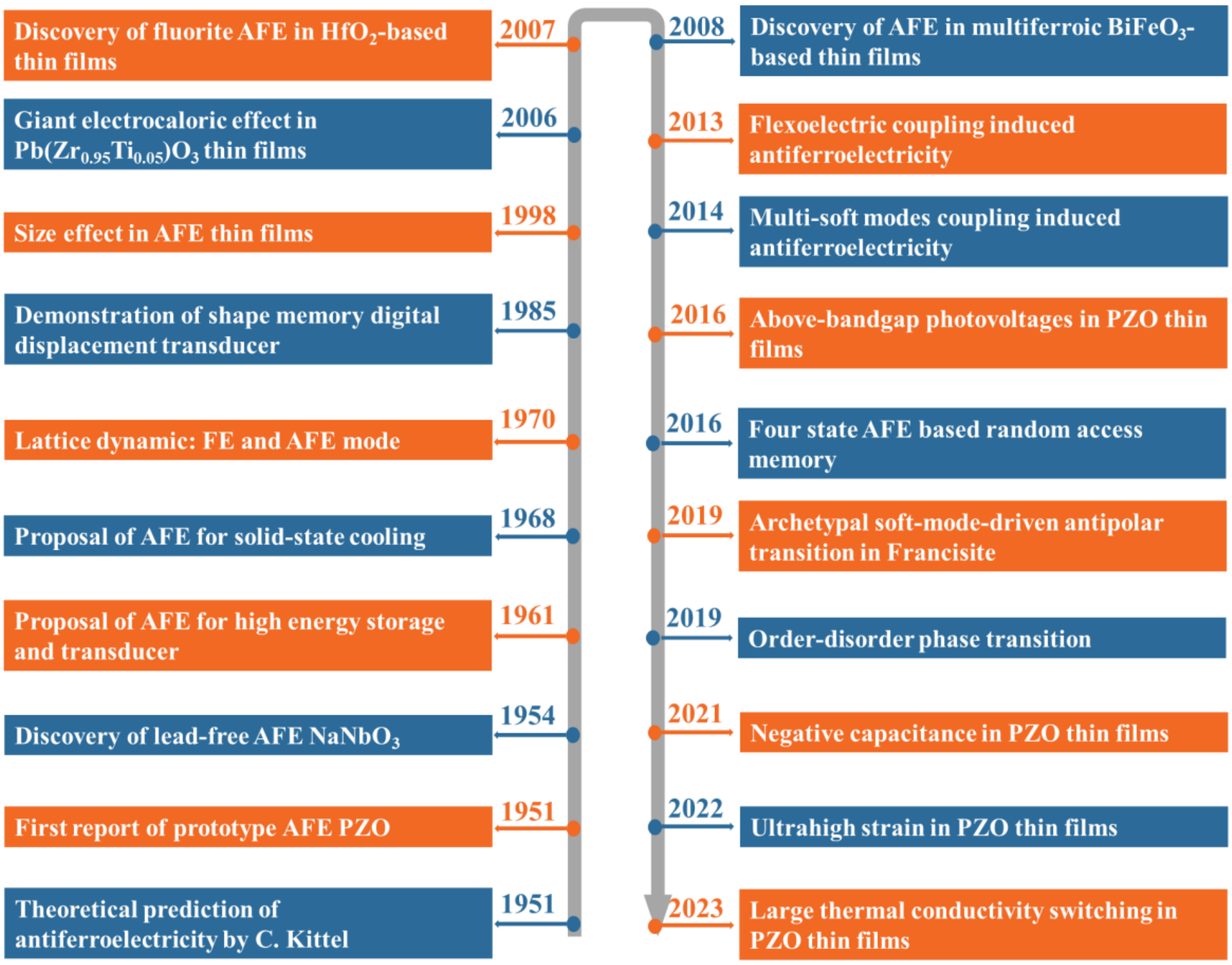

**Fig. 1.** Key progress in the development history of AFEs [7, 9-11, 25, 30-46].

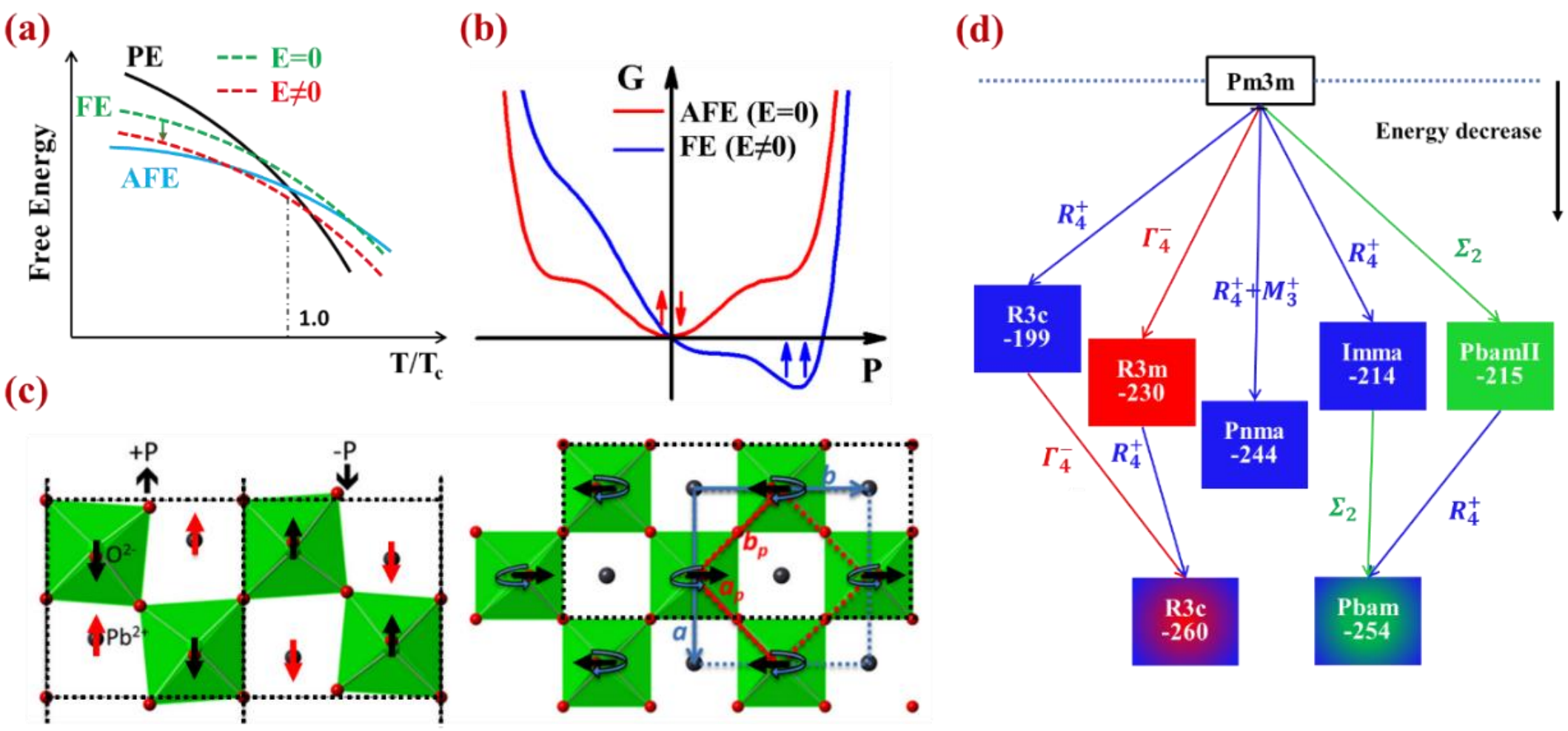

**Fig. 2.** Antiferroelectrics from the energy and structure perspective. (a,b) Electric field lowers the energy of the FE phase with respect to the AFE and paraelectric phases. Reproduced with permission from [50]. (c) The atomic scale picture for AFE ($\Sigma_2$) (left panel) and AFD ($R_4^+$) distortion modes (right panel). Reproduced with permission from [73]. One can distinguish two bands of polarization in the orthorhombic unit cell on the left directed up and down, respectively. (d) Relative energies of variously distorted virtual structures of $PbZrO_3$ crystal. Reproduced with permission from [75].

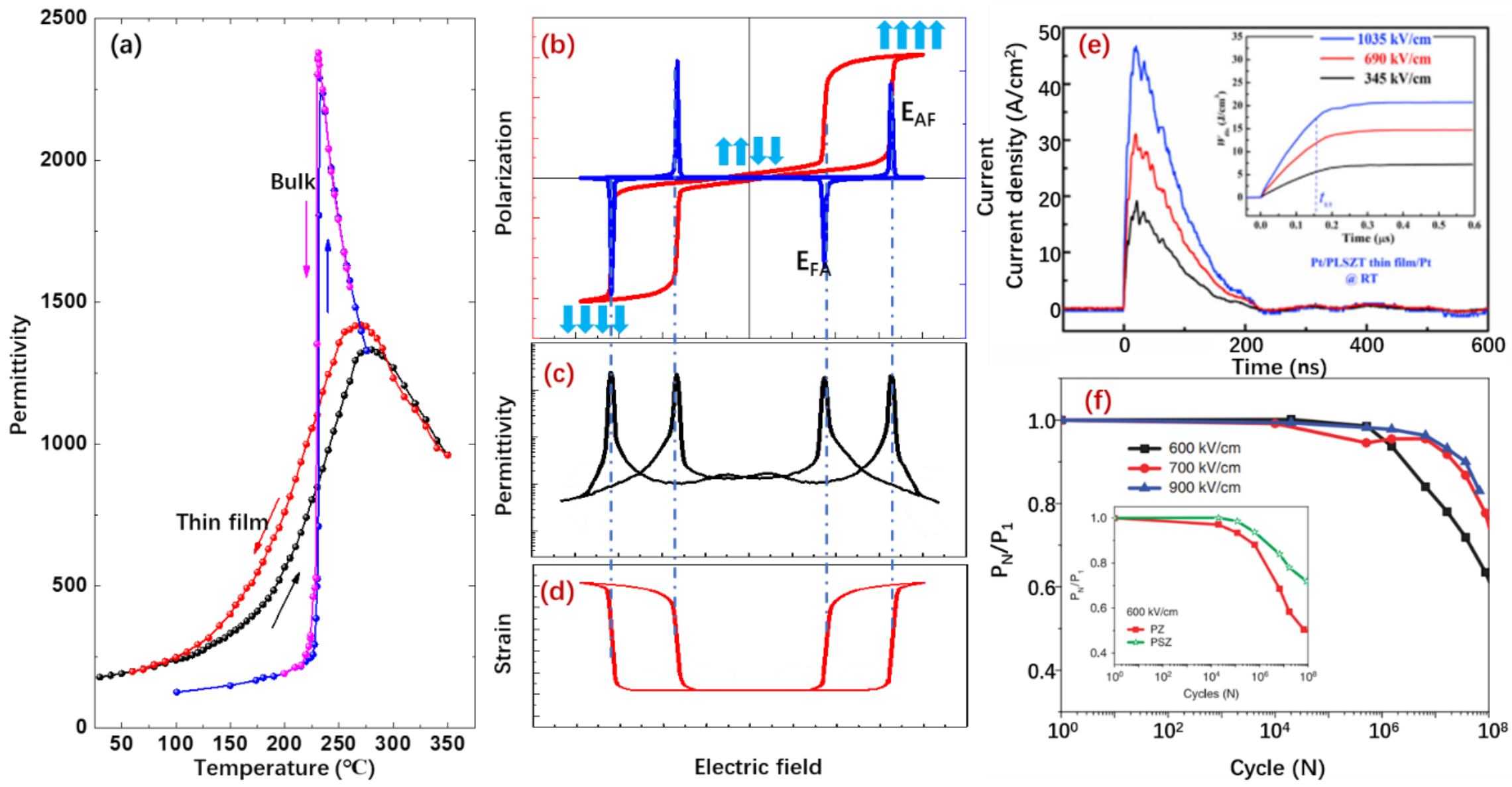

Fig. 3. Featured antiferroelectric behavior in different scenarios. (a) Temperature dependent dielectric permittivity spectrum of $PbZrO_3$. Reproduced with permission from [16, 50]. (b-d) Schematic view of polarization (b), dielectric permittivity (c), and strain (d) versus electric field. (e) Temporal evolution of voltage, current, and polarization curves during the switching in $(Pb_{0.89}La_{0.06}Sr_{0.05})(Zr_{0.95}Ti_{0.05})O_3$ AFE thin film. Reproduced with permission from [90]. (f) Modified fatigue performance of Sr-doped $PbZrO_3$-based antiferroelectric thin films. Reproduced with permission from [112].

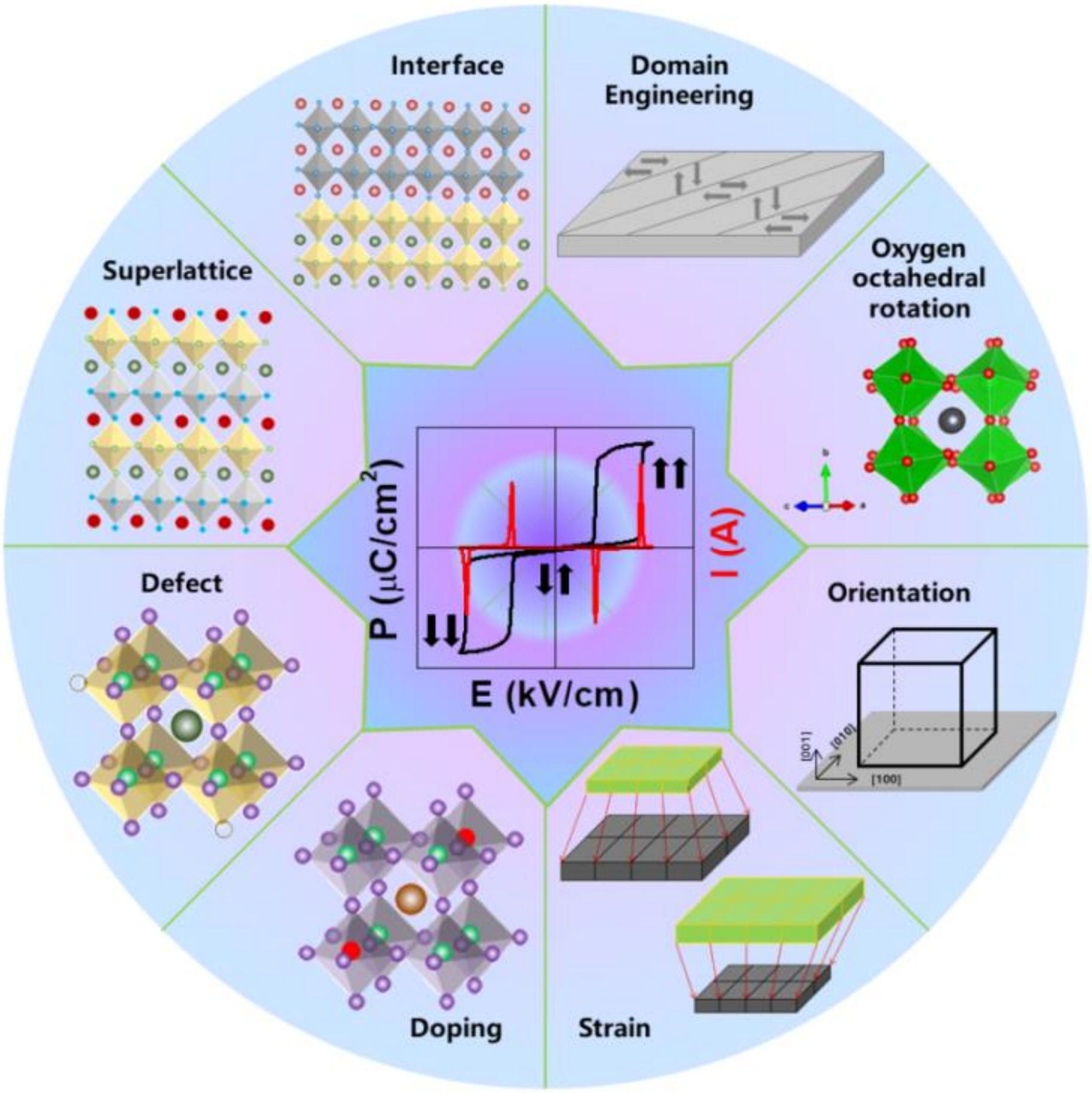

**Fig. 4.** Schematic illustration showing engineering strategies in thin films.

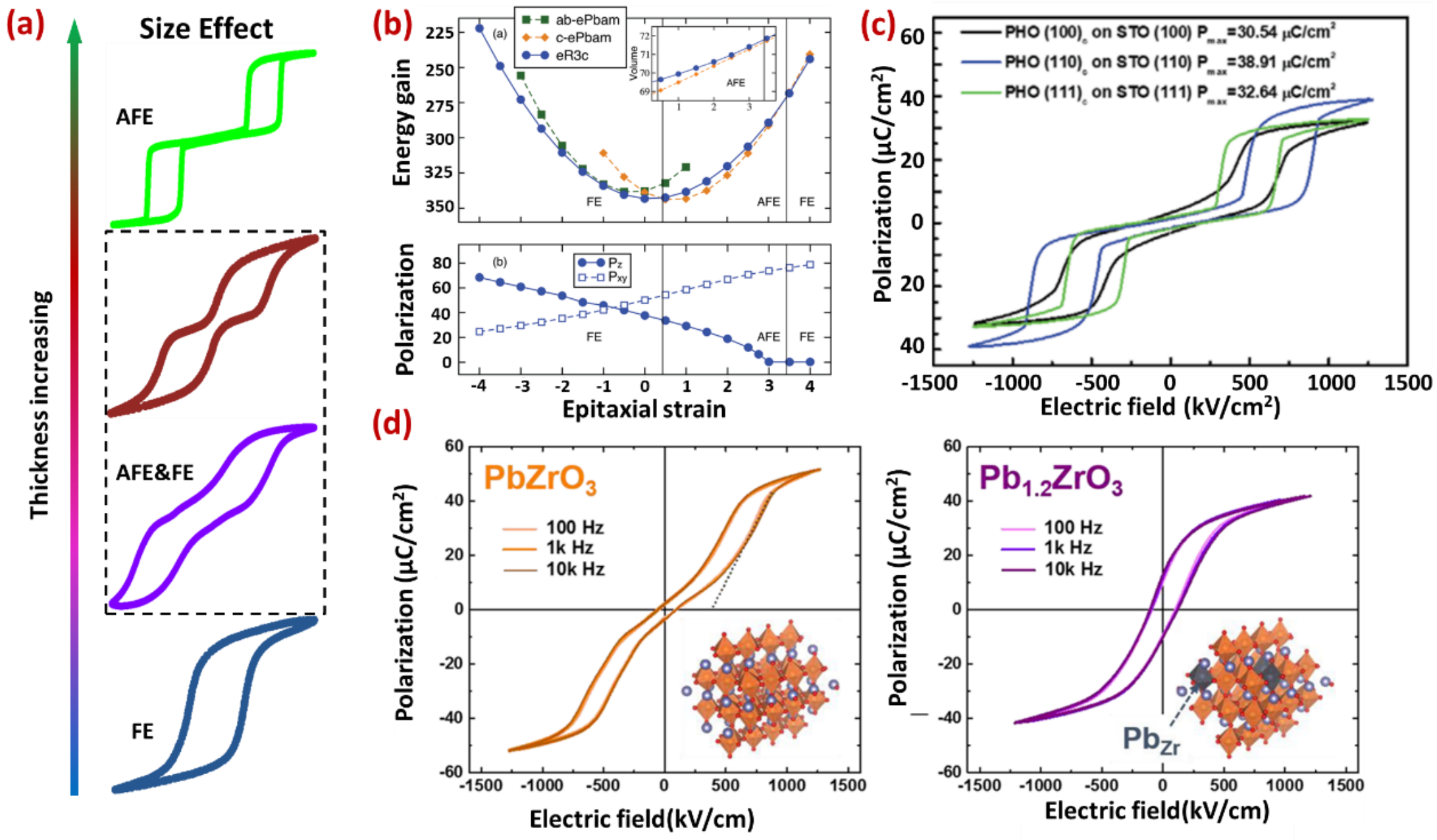

**Fig. 5.** Typical methods for manipulating AFE thin films (a) schematic view of Size effect in AFE thin film. (b) Energy *vs* epitaxial strain diagram of $PbZrO_3$ at 0 K. Reproduced with permission from [134]. (c) The maximum polarization values of three oriented $PbHfO_3$ films. Reproduced with permission from [145]. (d) Polarization-electric field loops of $Pb_{1.2}ZrO_3$ and $PbZrO_3$ thin film. Reproduced with permission from [14].

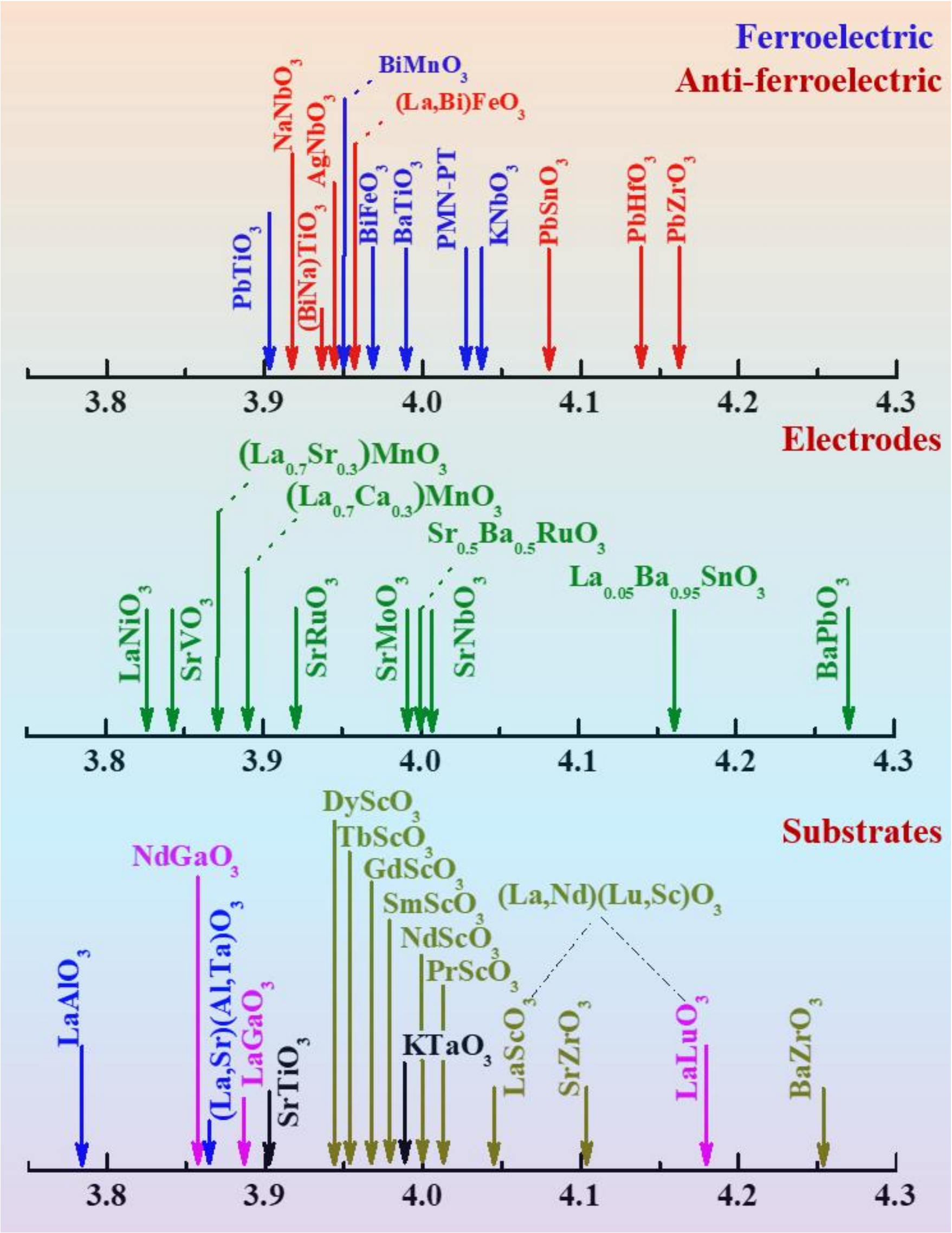

**Fig. 6.** Survey of pseudo-cubic lattice parameters (Å) of commercially available single crystal substrates, perovskite electrodes and ferroelectric/antiferroelectric materials at room temperature.

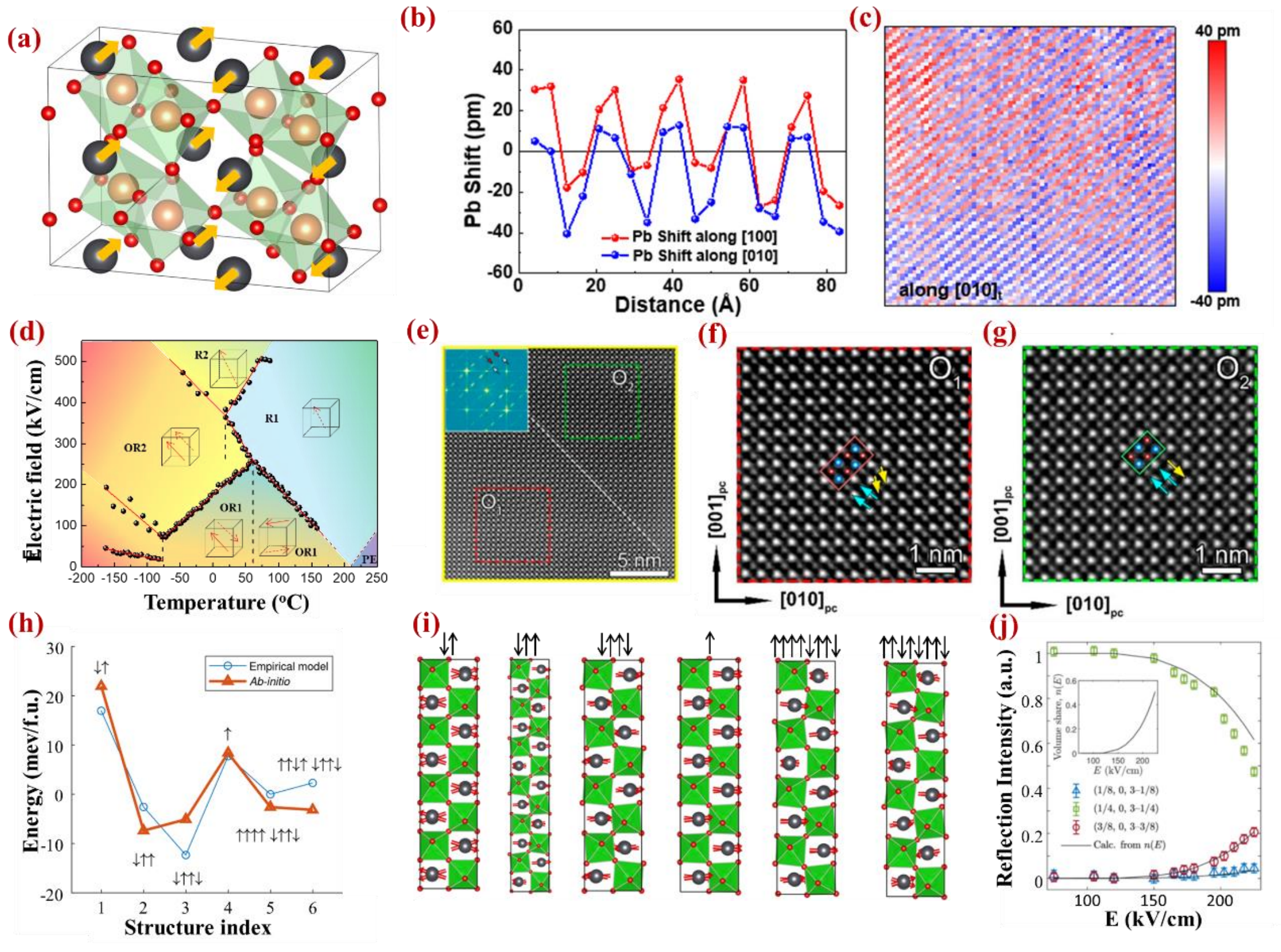

**Fig. 7.** Structure variation in antiferroelectric $PbZrO_3$ thin films. (a) standard bulk structure of AFE $PbZrO_3$, (b) Uncompensated polarization in $PbZrO_3$ thin films, and (c) Fluctuation of $Pb^{2+}$ shift in $PbZrO_3$ thin films extracted from TEM results. Reproduced with permission from [16]. (d) temperature-electric field phase diagram of $PbZrO_3$ bulks. Reproduced with permission from [56]. (e-g) Local ferrielectric structure in $PbZrO_3$ thin films. Reproduced with permission from [122]. (h, i) Energy comparison for different virtual transversely modulated structures of PbZrO3, as obtained by 0 K DFT calculations. Circles in (h) show a parametrization of energy by one of the widely used Hamiltonians in magnetic models. Reproduced with permission from [15]. (j) Evolution of superstructure reflections with electric field in $PbZrO_3$ thin film, and also the recalculated fraction of the ferrielectric-like phase in the inset. Reproduced with permission from [15].

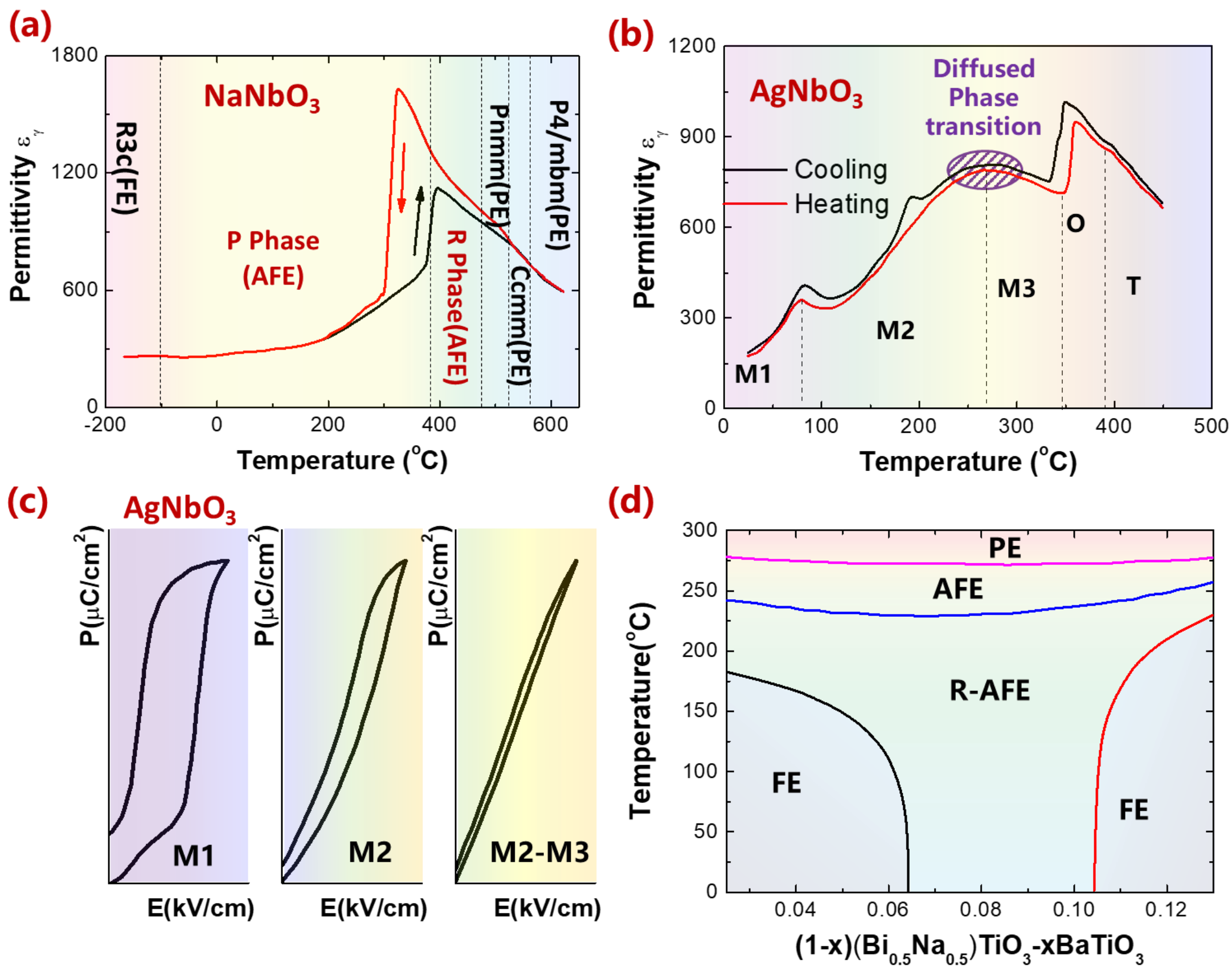

**Fig. 8.** Phase diagrams of conventional lead-free AFEs. (a) Temperature dependences of dielectric permittivity for $NaNbO_3$ ceramics. Reproduced with permission from [223]. (b) dielectric spectrum of $AgNbO_3$. Reproduced with permission from [230]. (c) P-E loops schematic diagram of $AgNbO_3$ in various AFE phases. Reproduced with permission from [234]. (d) the phase diagram for unpoled ceramics in the (1-$x$)BNT-$x$BT binary system. Reproduced with permission from [239].

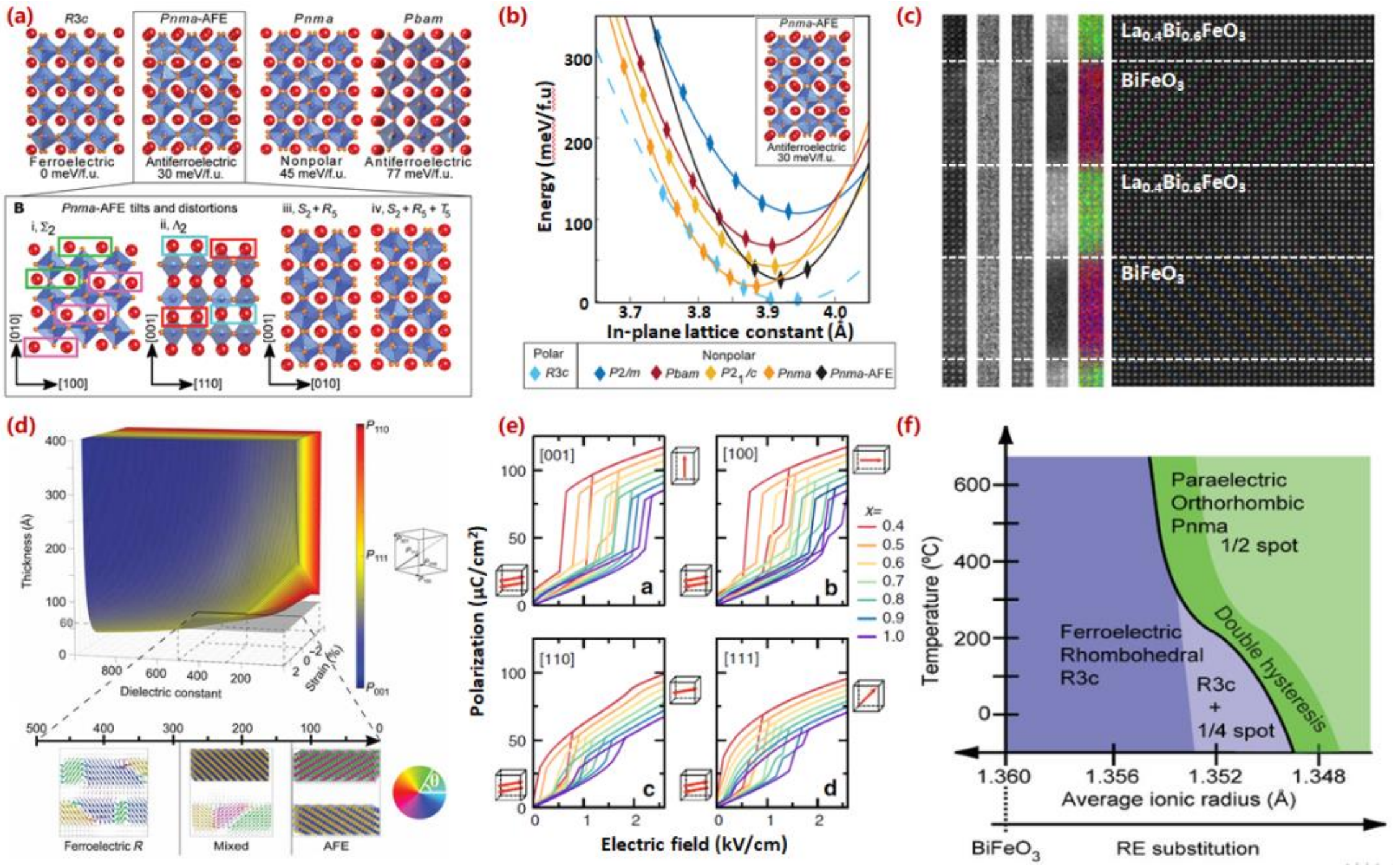

**Fig. 9.** Antiferroelectricity in $BiFeO_3$-based thin films. (a) Energetics of $BiFeO_3$ ground states, (b) energetics of $BiFeO_3$ ground states with various symmetry in (a), (c) annular dark field and EELS spectroscopic imaging reveal the antipolar feature and the atomic concentrations of bismuth, iron, and lanthanum are in red, blue, and green, respectively, (d) phase stability of $BiFeO_3$ heterostructures. Reproduced with permission from [188]. (e) calculated *P-E* hysteresis curves of $Bi_{1-x}Nd_xFeO_3$ solid solutions. Reproduced with permission from [257]. (f) phase diagram for rare element substituted $BiFeO_3$. Reproduced with permission from [261].

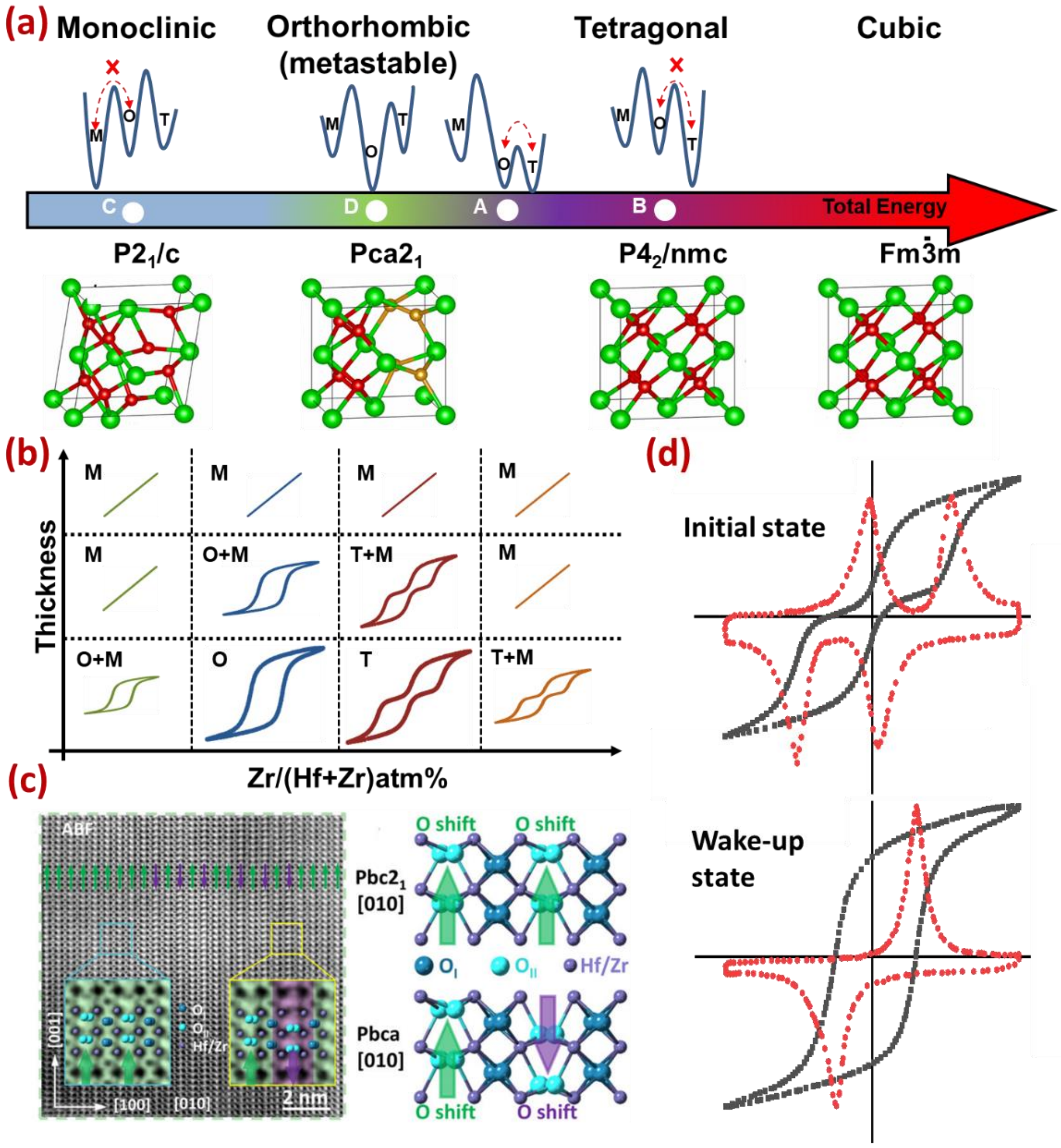

**Fig. 10.** Ferroelectricity and antiferroelectricity in $HfO_2$-based thin films. (a) Phase equilibria and properties of $HfO_2$. Reproduced with permission from [268], (b) The transition of $ZrO_2$ and $HfO_2$ solid solution and their size effect. Reproduced with permission from [28]. (c) Polar $O_{FE}$ *Pbc2₁* to antipolar $O_{AFE}$ *Pbca* phase in a fatigued $(Hf,Zr)O_2$ thin film. Reproduced with permission from [276]. (d) Wake-up effect in $(Hf,Zr)O_2$ thin films, the antiferroelectric behavior exist only a few switching cycles and transform into ferroelectric behavior. Reproduced with permission from [278].

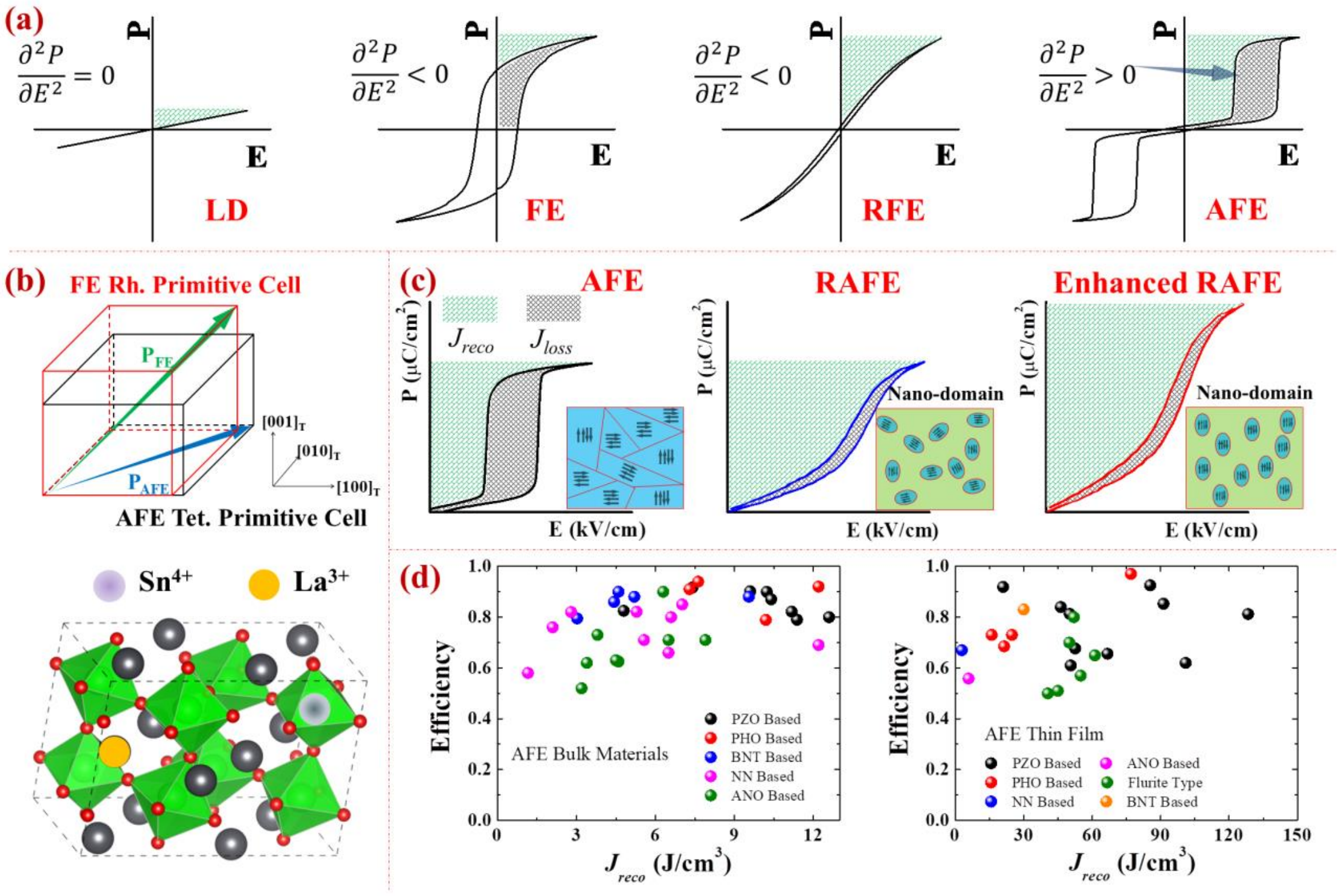

**Fig. 11.** Characterization of energy storage performance in AFEs. (a) Four distinctive *P*-*E* hysteresis loops and their energy storage behavior: linear dielectric, ferroelectric, relaxor ferroelectric, antiferroelectric, and relaxor antiferroelectric, (b) Schematic illustration of polarization enhancement via orientation control (upper panel) and structure heterogeneity with doping (lower panel). (c) The modification of *P*-*E* loops with the combination of structure heterogeneity and orientation control. (d) A comparison of energy storage characteristics in various AFE bulk systems (left panel) [172, 234, 237, 252, 296, 302, 309-338], and a comparison of energy storage characteristics in various antiferroelectric thin film systems (right panel) [21, 90, 100, 145, 182, 235, 339-355].

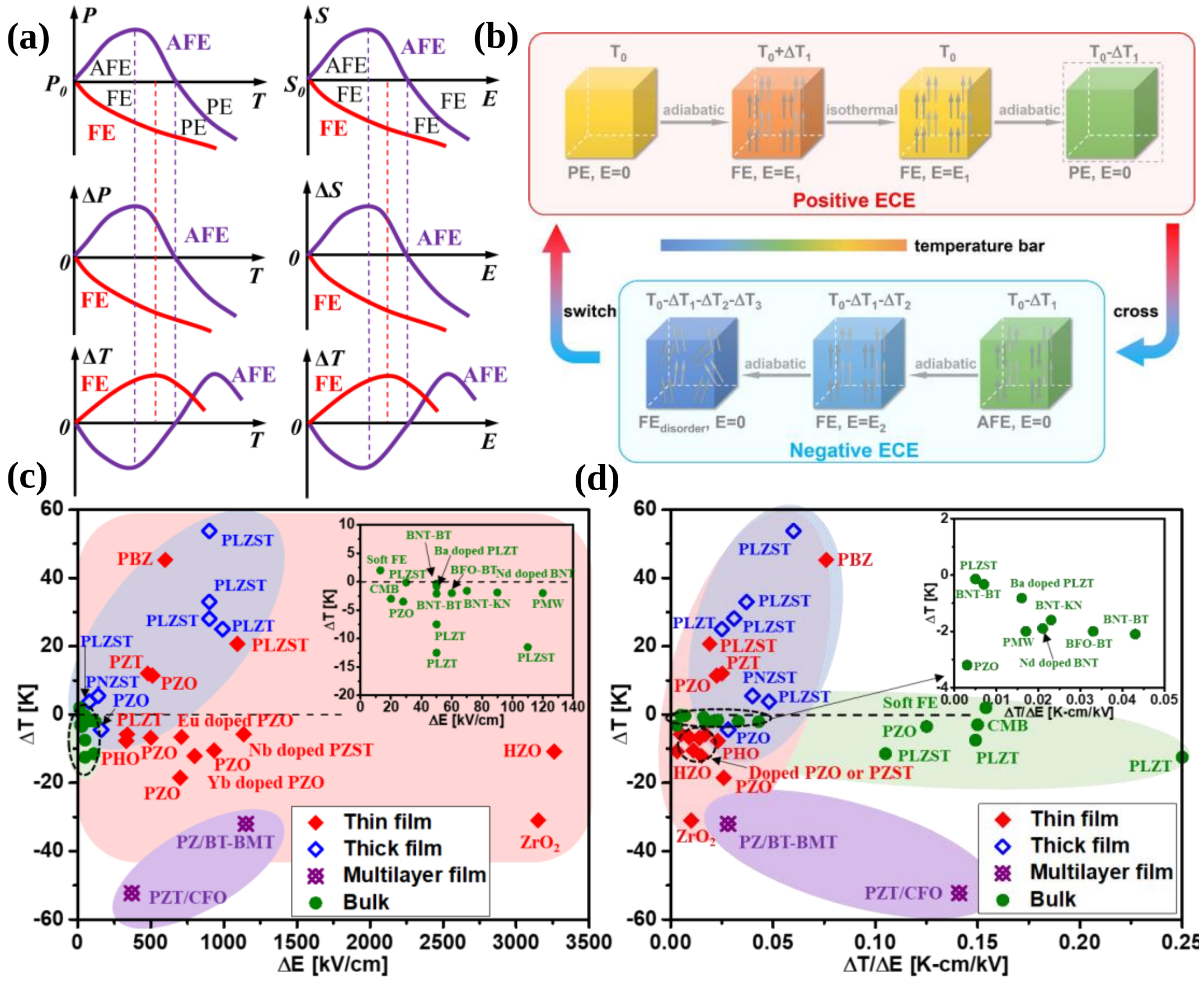

Fig. 12. Electrocaloric effect in AFEs. (a) The temperature (electric field) dependent evolution of polarization $P$ (entropy $S$), polarization change $\Delta P$ (entropy change $\Delta S$) and adiabatic temperature change $\Delta T$ for AFE (purple lines) and FE (red lines) materials respectively. (b) The EC cycle based on the combination of positive and negative ECE. Reproduced with permission from [371]. The ECE strength $\Delta T/\Delta E$ (c) and $\Delta E$ (d) dependent $\Delta T$ [6, 40, 165, 290, 366-369, 371, 373, 375-402].

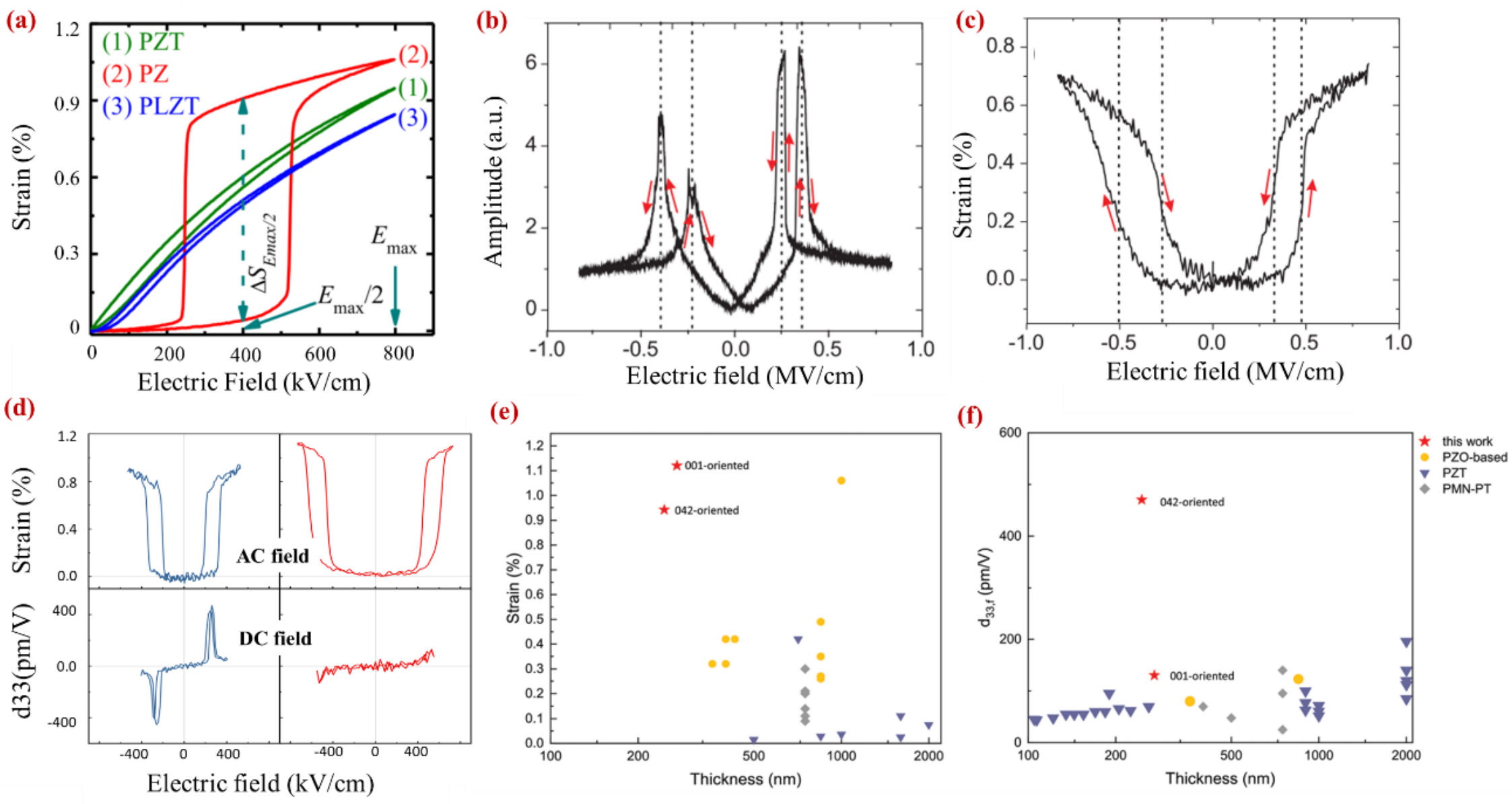

Fig. 13. Strain related effects in AFE materials. (a) Unipolar piezoelectric strain versus electric field (*S*-*E*) loops of $PbZrO_3$, PZT and PLZT. Reproduced with permission from [404]. (b) a quasi-static strain-field hysteresis loop and (c) a local piezoresponse force microscopy (PFM) amplitude loop measured at 1 Hz in epitaxial $PbZrO_3$ films on $SrTiO_3$ substrate. Reproduced with permission from [408]. (d-f) Electromechanical properties of textured $PbZrO_3$ films: (d) (042) orientation (left) and (001) orientation (right) electromechanical response of strain and effective longitudinal piezoelectric coefficient as a function of AC field and DC field, respectively, Comparison of electromechanical properties of $PbZrO_3$ films and other antiferroelectric and ferroelectric materials (e) strain and (f) effective piezoelectric coefficient ($d_{33}$,f). Reproduced with permission from [7].

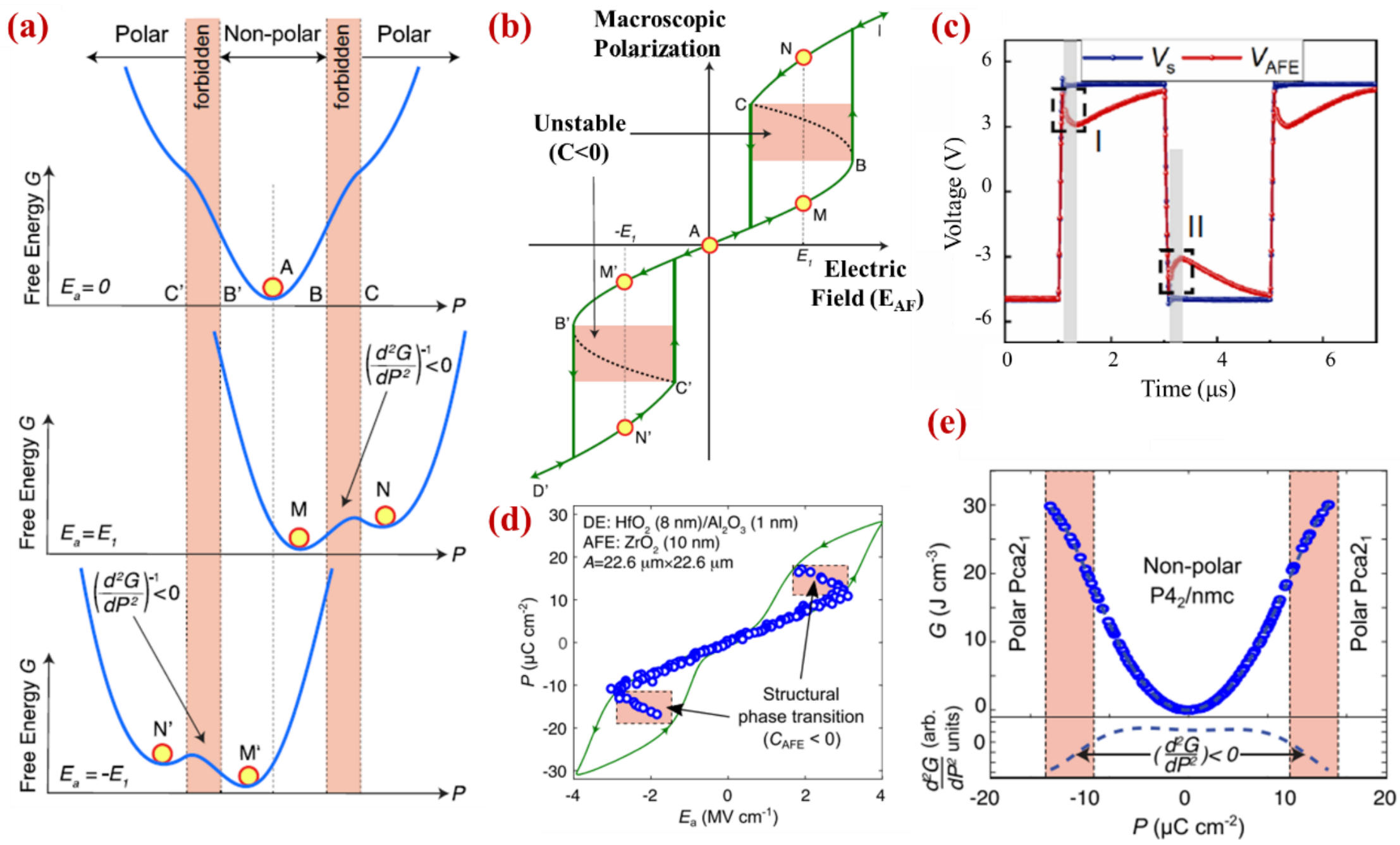

**Fig. 14.** Negative capacitance in AFE thin films. (a) The AFE free energy landscape at $E_a = 0$, $E_1$ and $-E_1$ and (b) The *P-E* characteristics of an AFE material, and the segments BC and B'C' represent the negative capacitance (C < 0) regions. Reproduced with permission from [62]. (c) Source voltage $V_s$ and the AFE voltage $V_{AFE}$ in AFE $PbZrO_3$ films. Reproduced with permission from [45]. (d) Polarization *P* as a function of extracted electric field $E_a$ across the $ZrO_2$ layer in a $HfO_2$ (8nm)/$Al_2O_3$ (~1nm)/$ZrO_2$ (10nm) heterostructure capacitor as compared with the double hysteresis loop measured from an equivalent stand-alone $ZrO_2$ capacitor. Reproduced with permission from [62]. (e) Extracted energy landscape of $ZrO_2$. Second derivative of the free energy *G* with respect to *P* based on a polynomial fit is shown below. Reproduced with permission from [62].

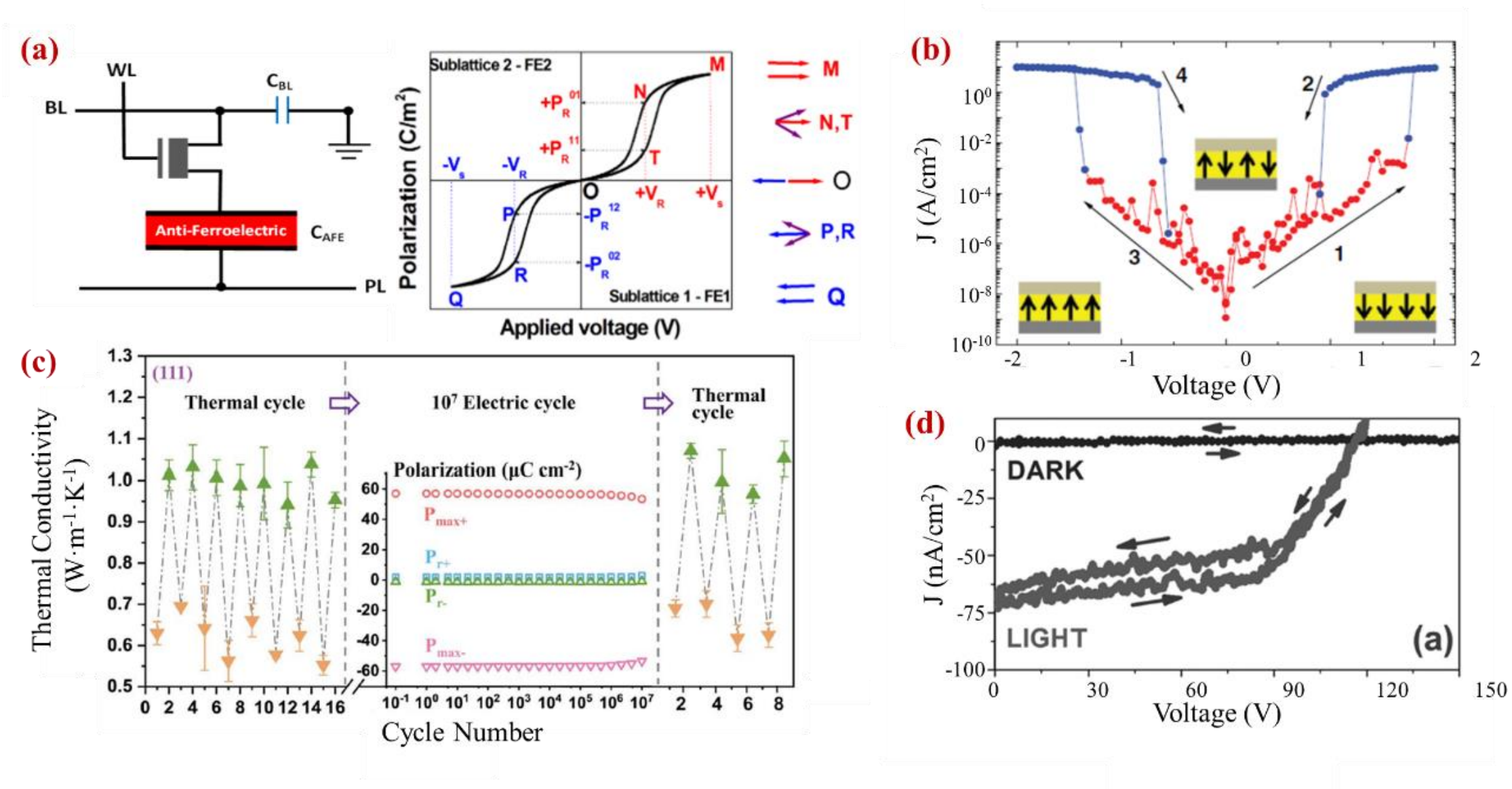

**Fig. 15.** Demonstration for novel devices application in AFE thin films. (a) AFE memory cell architecture and polarization versus applied voltage response of a typical AFE material, Key points on the double hysteresis curve are marked, as well as the saturating voltages, reversal voltages and quasi-remanent polarization memory states. The vectors represent diagrammatically the polarization states of the two sublattices at different stages on the hysteresis curve. Red corresponds to positive and blue to negative polarization, respectively. Reproduced with permission from [9]. (b) Current-voltage characteristics of $Co/PbZrO_3/(La,Sr)MnO_3$ tunnel junctions. Reproduced with permission from [426]. (c) Repeatability of switching thermal conductivity after $10^7$ electric cycles. Reproduced with permission from [10]. (d) Steady-state photovoltaic response of the AFE capacitor, showing an open-circuit photovoltage more than 100 V. Reproduced with permission from [11].

**Table 1** Key parameters in typical AFEs. The values are at room temperature.

| **Materials** | **Phase Symmetry** | **Lattice Parameter (Å)** | **Pesudo-tetragonal axis (Å)** | **Glazer notation** | **Dielectric constant $\varepsilon_r$** | **$T_C$ ($^oC$)** |
|---|---|---|---|---|---|---|
| $NaNbO_3$ | *Pbcm* [220] | $a$=5.506, $b$=5.566, $c$=15.520 [438] | $a_0$=$b_0$=3.915 $c_0$=3.880 | $(a^-a^-b^+)/(a^-a^-b^-)$[439] | ≈270 [440] | ≈482 [218] |
| $AgNbO_3$ | $Pmc2_1$ [233] | $a$=15.648, $b$=5.552, $c$=5.609 [233] | $a_0$=$b_0$=3.944 $c_0$=3.915 | $(a^-b^-c^+)/(a^-b^-c^-)$ [233] | 335 [441] | 353 [441] |
| | *Pbcm* [442, 443] | $a$=5.547, $b$=5.604, $c$=15.642 [443] | $a_0$=$b_0$=3.942 $c_0$=3.911 | $(a^-b^-c^-)/(a^-b^-c^+)$ [443] | | |
| $(Bi_{0.5}Na_{0.5})TiO_3$ | *P4bm* [444, 445] | $a$= 5.519, $b$=5.519, $c$=3.909 [444] | $a_0$=$b_0$=3.942 $c_0$=3.911 | $a^0a^0c^+$[444] | 220 [446] | 320 [447] |
| $PbZrO_3$ | *Pbam* [287] | $a$=5.888, $b$=11.771, $c$=8.226 [448] | $a_0$=$b_0$=4.161 $c_0$=4.113 | $a^-a^-c^0$[448] | 150-180[39, 449, 450] | ≈233 [50] |
| $PbHfO_3$ | *Pbam* [451] | $a$=5.856, $b$=11.729, $c$=8.212 [452] | $a_0$=$b_0$=4.140 $c_0$=4.110 | $a^-a^-c^0$[453] | ≈250 [454] | 205 [454] |
| $PbSnO_3$ | / | $a$=5.795, $b$=11.591, $c$=8.368[60] | $a_0$=$b_0$=4.098 $c_0$=4.184 | $a^-a^-c^+$[60] | 90-130[60] | 120-150 [60] |
| $Bi_{0.7}La_{0.3}FeO_3$ | *Pbam*[455] | $a$=5.572, $b$=5.540, $c$=7.864 [455] | $a_0$=$b_0$=3.940 $c_0$=3.932 | | 120-140[455] | / |
| $ZrO_2$ | $P4_2/nmc$ [97] | $a$=$b$=3.59, $c$=5.15 [97] | / | | 30-40[97] | / |
| $Hf_{0.25}Zr_{0.75}O_2$ | $P4_2/nmc$ [28] | / | / | | 40-45[427] | / |